\documentclass{jfm}
\usepackage{graphicx}
\usepackage{epstopdf, epsfig}
\usepackage{xcolor}
\usepackage{amsmath}

\shorttitle{Resolvent-based modelling of an oscillator vs. an amplifier}
\shortauthor{S. Symon, D. Sipp and B. J. McKeon}

\title{A tale of two airfoils: resolvent-based modelling of an oscillator vs. an amplifier from an experimental mean}

\author{Sean Symon\aff{1,3}
  \corresp{\email{ssymon@unimelb.edu.au}},
  Denis Sipp\aff{2}
 \and Beverley J. McKeon\aff{3}}

\affiliation{\aff{1}Department of Mechanical Engineering, University of Melbourne,
VIC 3010, Australia
\aff{2}ONERA DAAA, 8 rue des Vertugadins, 92190 Meudon, France
\aff{3}Graduate Aerospace Laboratories, California Institute of Technology, Pasadena, CA 91125, USA}

\begin{document}

\maketitle

\begin{abstract}
The flows around a NACA 0018 airfoil at a chord-based Reynolds number of $Re = 10250$ and angles of attack of $\alpha = 0^{\circ}$ and $\alpha = 10^{\circ}$ are modelled using resolvent analysis and limited experimental measurements obtained from particle image velocimetry. The experimental mean velocity profiles are data-assimilated so that they are solutions of the incompressible Reynolds-averaged Navier-Stokes equations forced by Reynolds stress terms which are derived from experimental data. Spectral proper orthogonal decompositions of the velocity fluctuations and nonlinear forcing suggest different modelling approaches should be taken based on the angle of attack under consideration. For the $\alpha = 0^{\circ}$ case, the cross-spectral density tensors of both the velocity fluctuations and nonlinear forcing are low-rank at the shedding frequency and its higher harmonics. In the $\alpha = 10^{\circ}$ case, low-rank behaviour is observed for the velocity fluctuations in two bands of frequencies. Resolvent analysis of the data-assimilated means identifies low-rank behaviour only in the vicinity of the shedding frequency for $\alpha = 0^{\circ}$ and none of its harmonics. The resolvent operator for the $\alpha = 10^{\circ}$ case, on the other hand, identifies two linear mechanisms whose frequencies are a close match with those identified by spectral proper orthogonal decomposition. It is also shown that the second linear mechanism, corresponding to the Kelvin-Helmholtz instability in the shear layer, cannot be identified just by considering the time-averaged experimental measurements as a mean flow for resolvent analysis. This is due to the fact that experimental data are missing near the leading edge of the airfoil. The $\alpha = 0^{\circ}$ case is classified as an oscillator where the flow is organized around an intrinsic instability mechanism while the $\alpha = 10^{\circ}$ case behaves like an amplifier whose forcing is unstructured. For both cases, resolvent modes resemble those from spectral proper orthogonal decomposition when the operator is low-rank. To model the higher harmonics where this is not the case, we add parasitic resolvent modes, as opposed to classical resolvent modes which are the most amplified, by approximating the nonlinear forcing from limited triadic interactions of known resolvent modes. The amplifier case is modelled without parasitic modes at frequencies where the resolvent is low-rank. The two cases suggest that resolvent-based modelling can be achieved for more complex flows with limited experimental measurements and the nonlinear forcing need not be approximated unless the flow behaves like an oscillator.
\end{abstract}


\section{Introduction}
Recent studies, e.g. \cite{Gomez16a, Beneddine17, Symon18b, Symon18c, He19}, have shown that it is possible to construct a reduced-order model of a flow using the time-averaged (mean) flow field and a small set of time-resolved velocity data at isolated measurement points. For low Reynolds number bluff body wakes, this could be as few as a single point \citep[see][]{Gomez16b, Symon19}. In a similar vein, \cite{Beneddine16} and \cite{Thomareis18} have been able to estimate the spectra at various points of the flow using only the aforementioned measurements. The success of the method relies on the low-rank nature of the (linear) resolvent operator, which one obtains after linearising the Navier-Stokes equations around the (turbulent) mean flow, and treating the nonlinear terms as a source of intrinsic forcing \citep{McKeon10}. The exact form of the forcing can often be neglected due to a dominant amplification mechanism at particular spatial wavenumbers and frequencies. Linear analysis alone does not reveal the amplitude and phase of the most amplified disturbances yet they can be calibrated using limited unsteady data \citep{Gomez16a, Beneddine16}. 

One of the major advantages of this approach is that time-resolved data at all locations in the flow field is not required. From the governing equations, it is possible to obtain a prediction of the temporal Fourier mode in the actual fluctuation field if the dominant singular value of the resolvent operator is much larger than the others. The nonlinear forcing field, furthermore, must not be preferentially biased towards the forcing mode corresponding to lower singular values and this was demonstrated by \cite{Beneddine16} for turbulent flow over a backward-facing step. The goal of Dynamic Mode Decomposition (DMD) \citep{Schmid10, Rowley09} is also to compute modes which oscillate at a distinct frequency, suggesting that resolvent analysis is attempting to approximate DMD modes from the equations instead of time-resolved data. A striking similarity between the most amplified resolvent mode and the DMD mode was noted by \cite{Gomez14} in turbulent pipe flow and the relationship between resolvent analysis and DMD was formalised in \cite{Sharma16}.

Connections between resolvent analysis and other data-driven approaches have been discussed by \cite{Towne18}. In particular, they examined modes extracted by Spectral Proper Orthogonal Decomposition (SPOD), introduced by \cite{Lumley70}, which produces a basis for the fluctuation field at each temporal frequency and optimally represents the space-time flow statistics. \cite{Towne18} determined that SPOD modes and resolvent modes are equivalent if the nonlinear forcing field is white in space and time. In practice, this might not be the case since the resolvent framework is a closed loop \citep{McKeon13, McKeon17} where the nonlinear interactions between the velocity fluctuations constitute the forcing to the resolvent operator. Despite this fact, resolvent analysis is quite often able to predict the structure of energetic velocity fluctuations. Moreover, if the singular values of the resolvent operator are well separated, it often suggests a similar separation in the SPOD eigenspectra as demonstrated by \cite{Schmidt18} for a turbulent jet. 

Resolvent analysis may be used for predicting structures even if the singular values are not well separated. Recent work by \cite{Rosenberg19} proved that considering a small subset of triadic interactions of `correct' resolvent modes (i.e. modes computed at frequencies where the singular values are well separated) can approximate the nonlinear forcing, a process reminiscent of weakly nonlinear analysis \citep{Sipp07} without being limited to the vicinity of a critical Reynolds number. When these interactions are fed back through the resolvent operator, they produce the correct structure. These modes were labelled parasitic in the sense that they fed off nonlinear interactions of other modes rather than emerge naturally from a linear amplification mechanism embedded in the operator. 

The majority of the aforementioned studies have utilised a mean obtained from simulations although a growing number have considered experimental means \citep{Sasaki17, Beneddine17, Symon18c, He19}. The objective in this article is to obtain a reduced-order representation using a partially known mean obtained from experiments. We use data-assimilation to recover the mean profile on a larger domain and enforce no-slip boundary conditions on the airfoil surface. Data-assimilation is a process whereby experimental measurements are merged with computational fluid dynamics (CFD) \citep[see][]{Hayase15} and has been a growing field in fluid mechanics. A list of recent developments can be found in \cite{daSilva19}. The particular framework used here was developed by \cite{Foures14} and later applied to flow around an idealised airfoil by \cite{Symon17}. 

In this article, we consider the flow around a NACA 0018 airfoil at a chord-based Reynolds number of $Re = 10250$ at two angles of attack: $\alpha = 0^{\circ}$ (A0 case) and $\alpha = 10^{\circ}$ (A10 case). The flows are obtained experimentally from particle image velocimetry (PIV). As will be discussed later in the article, the A0 and A10 cases behave like an oscillator and amplifier, respectively. The former has intrinsic dynamics which are insensitive to background noise while the latter filters and amplifies upstream noise in the downstream direction \citep{Huerre98, Sipp10}. Furthermore, the resolvent norm, or largest singular value of the resolvent operator, tends to have sharp peaks isolated at resonant frequencies in the cases of oscillators, e.g. \cite{Symon18}. Amplifier flows, on the other hand, may have separation between the first and second singular values over a broad ranges of frequencies, e.g. \cite{Dergham13, Beneddine16}. Of particular interest to us is the nature of the nonlinear forcing for each type of flow. If the nonlinear forcing is structured, then it can be deduced from a limited number of nonlinear interactions between coherent structures. This was the case in the oscillator flows studied by \cite{Rosenberg19}, where the nonlinear forcing at energetic frequencies consisted of coherent structure at the oscillation frequency and its harmonics. In the case of a turbulent jet \citep{Towne15}, which is more representative of an amplifier flow, the nonlinear forcing field resembles incoherent turbulent fluctuations.

The principal objectives of this article can be summarized as follows: (1) does data-assimilation of the mean velocity profile make a measurable difference on predicting the velocity fluctuations? (2) what does SPOD reveal when applied to the nonlinear forcing?  (3) can additional structure be extracted from resolvent analysis if the singular values are not separated but the nonlinear forcing is approximated? (4) is there a fundamental difference between oscillator and amplifier type flows? (5) how far can we go in modelling with just a few mean flow measurements and some unsteady point measurements? These questions are addressed in the rest of the paper which is organised as follows. In \S\ref{sec:experiment}, we describe the PIV experiments and in \S\ref{sec:tools} we discuss the mathematical background of the analysis tools used to construct the resolvent-based model. In \S\ref{sec:assimilation}, the mean velocity profiles are data-assimilated onto a larger, more resolved mesh. In \S\ref{sec:SPOD}, we apply SPOD to the velocity fluctuations and nonlinear forcing of both airfoils to highlight the low-rank behaviour in each flow. In \S\ref{sec:resolvent analysis}, the data-assimilated mean profiles are used as an input to resolvent analysis and the results are compared to those of SPOD. We also try to establish new links between resolvent analysis and SPOD which depend on whether the flow is an oscillator or an amplifier. In \S\ref{sec:modeling}, we use resolvent modes and, where appropriate, parasitic modes to construct a model of the coherent fluctuations in the two airfoil flows before concluding in \S\ref{sec:conclusion}. 

\section{Experimental setup}\label{sec:experiment}

The flow around a NACA 0018 airfoil at a Reynolds number of $Re = 10250$ and two angles of attack, $\alpha = 0^{\circ}$ (A0) and $\alpha = 10^{\circ}$ (A10), is measured using PIV. The experiments are performed in a free-surface water facility with a test section measuring 1.6 m in length, 0.46 m in width and 0.5 m in height. The NACA 0018 airfoil is mounted vertically in the tunnel, so that its span is parallel to the test section height, and has a chord length of 10 cm. Its spanwise length is 48 cm, resulting in an aspect ratio of $AR = 4.8$. The laser sheet is provided by a YLF dual cavity solid-state laser and is centred at a height of 220 mm. \cite{Dunne16} measured the out-of-plane velocity component to be less than 4\% of the free-stream velocity, allowing the mean flow at this height to be considered two-dimensional. 

The PIV setup consists of two Phantom Miro 320 cameras with 50 mm focal length Nikon lenses and 1:1.8 aperture. They have an overlap of 18\% in the streamwise direction and sample the flow at a frequency of 125 Hz. The camera resolution is 1920 $\times$ 1200 pixels and they are calibrated at 8.2 px/mm. The seeding particles are hollow glass spheres (reference 110P8 with an average diameter of 11.7 $\mu$m and a specific gravity of 1.1) and the seeding density is approximately 0.1 particles per square pixel. To reduce the impact of surface reflections, the image intensity is calibrated using white-image subtraction and background-image subtraction, which are taken before each run and averaged over 100 snapshots. 

\begin{figure}
	\centering
	\includegraphics[scale=0.29]{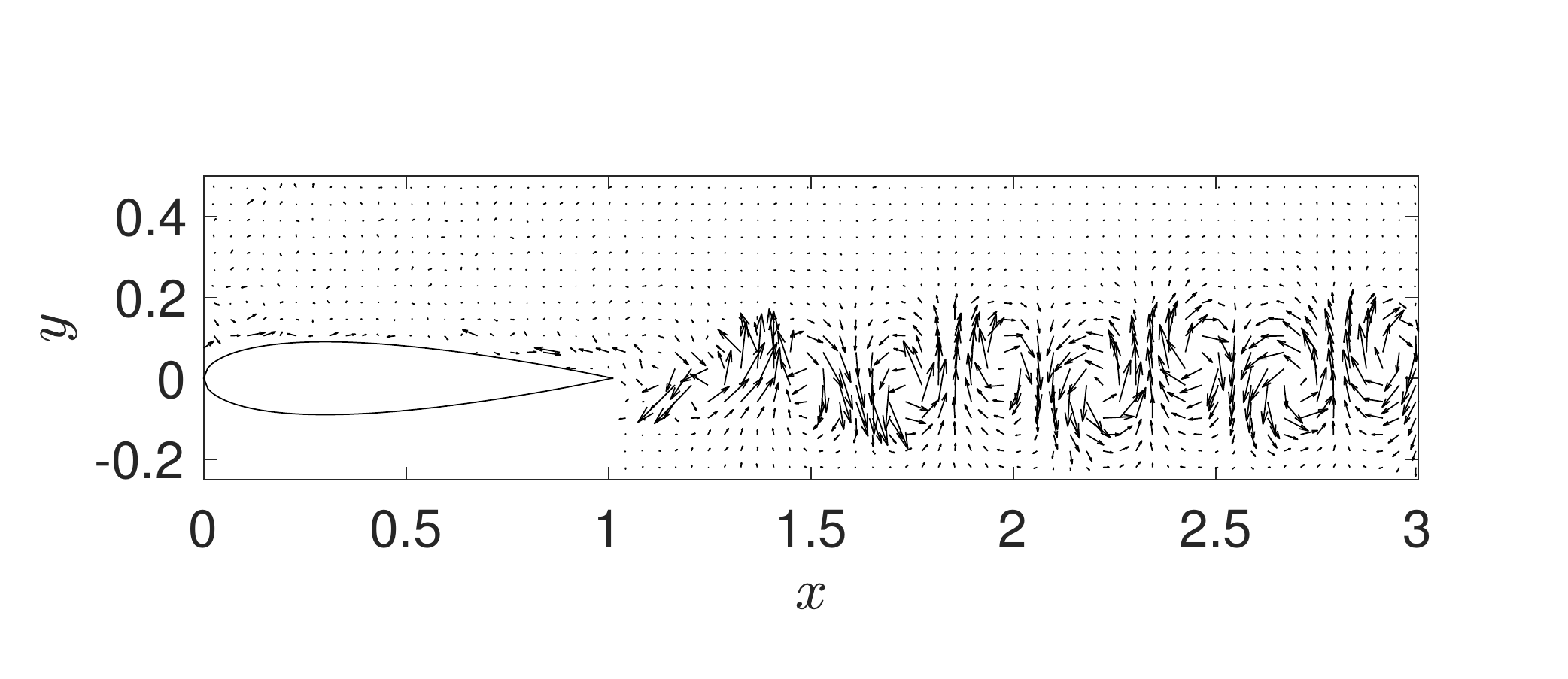}
	\includegraphics[scale=0.29]{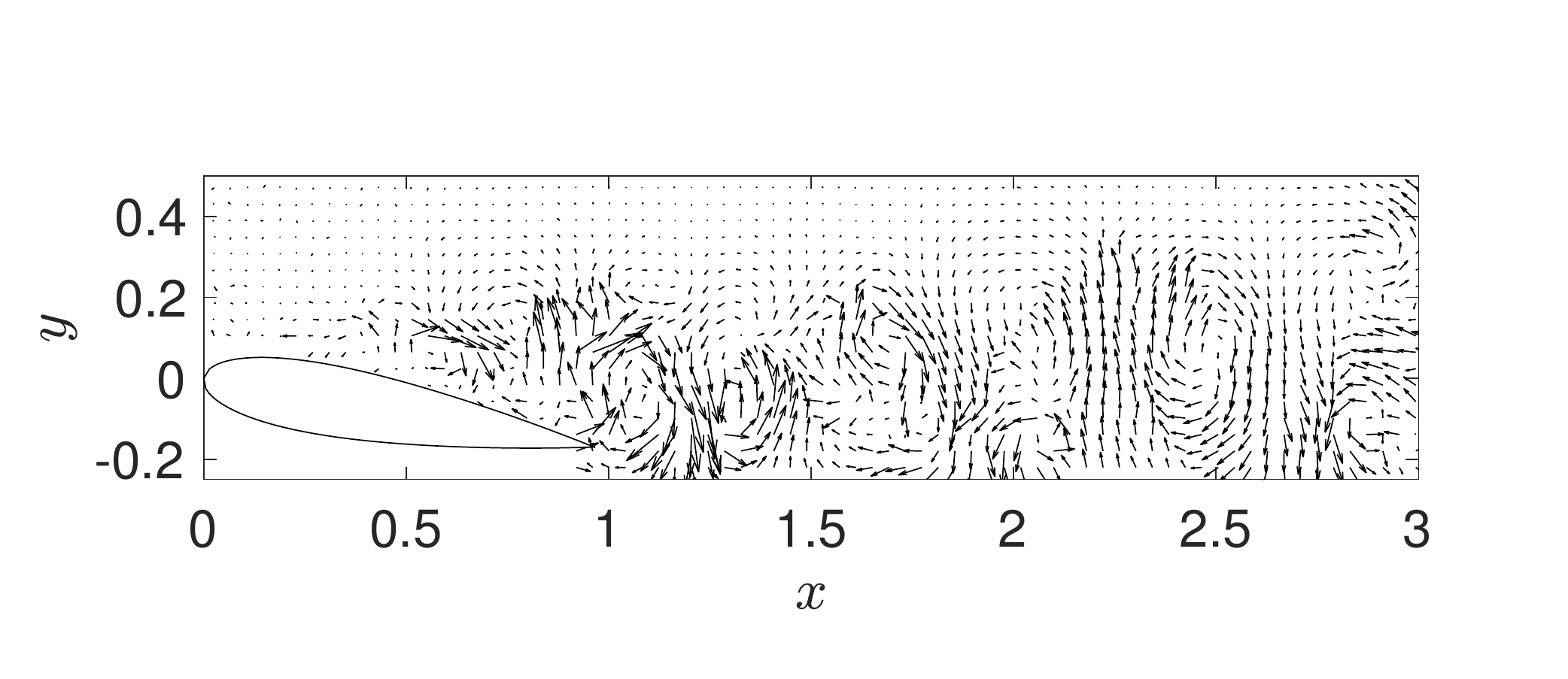}
	
	\caption{Sample vector plots of the instantaneous velocity for the A0 case (left) and A10 case (right). The mean velocity has been subtracted and 1 out of every 2 vectors has been removed in $x$ and $y$ to facilitate visualization of coherent structure.}\label{fig:PIV examples}
\end{figure}

LaVision's software package DaVis computes the velocity vectors where a standard cross-correlation technique via Fast Fourier Transformation is applied to each sequential image. The window-size is reduced from 32 $\times$ 32 px$^2$ to 16 $\times$ 16 px$^2$ over three passes. Once the data are post-processed, missing vectors are interpolated using an average of all non-zero neighbourhood vectors and a median filter \citep{Westerweel05} is used for outlier detection. A sample of the post-processed data for both cases is displayed in figure \ref{fig:PIV examples}. The mean profile, obtained after time-averaging two runs of 3,499 snapshots, is subtracted so that the coherent structures can be more readily identified. As seen in figure \ref{fig:PIV examples}, velocity vectors below the airfoil cannot be obtained since the laser sheet is obstructed in this region. For the A0 case, the mean velocity profile is symmetric so the mean profile can be reflected over the centreline (the transverse velocity changes sign) to obtain the mean below the centreline and in the shadow region. This is not possible for the A10 case, so the mean profiles for $\alpha = \pm 10^{\circ}$, which are obtained separately during different runs, are stitched together to obtain data around the entire airfoil.  

\section{Analysis tools}\label{sec:tools}

\subsection{Data-assimilation of the mean velocity profiles}

The experimental profiles are data-assimilated using the framework devised by \cite{Foures14} so that they can later be used as an input to resolvent analysis. Only the equations are reviewed here as its application to experimental data is discussed by \cite{Symon17}. The flow is governed by the incompressible Navier-Stokes equations, which are non-dimensionalised by the free-stream velocity $U_{\infty}$ and the airfoil chord $c$:
\begin{subequations} \label{eq:NS}
	\begin{equation}
		\partial_t \boldsymbol{u} + \boldsymbol{u} \cdot \nabla \boldsymbol{u} = -\nabla p + Re^{-1}\nabla^2 \boldsymbol{u},
	\end{equation}
	\begin{equation}
		\nabla \cdot \boldsymbol{u} = 0,
	\end{equation}
\end{subequations}
where $\boldsymbol{u}$ and $p$ are the velocity and pressure, respectively. Substituting $\boldsymbol{u} = \overline{\boldsymbol{u}} + \boldsymbol{u}'$, where an overbar denotes a mean and a prime denotes a fluctuation, into (\ref{eq:NS}) and averaging in time yields the incompressible Reynolds-averaged Navier-Stokes (RANS) equations:
\begin{subequations} \label{eq:RANS}
	\begin{equation} \label{eq:RANSm}
		\overline{\boldsymbol{u}} \cdot \nabla \overline{\boldsymbol{u}} + \nabla \overline{p} - Re^{-1} \nabla^2 \overline{\boldsymbol{u}} = \boldsymbol{f} = - \nabla \cdot \boldsymbol{R},
	\end{equation}
	\begin{equation}
		\nabla \cdot \overline{\boldsymbol{u}} = 0.
	\end{equation}
\end{subequations}
The right-hand side of (\ref{eq:RANSm}), or $\boldsymbol{f}$, represents the divergence of the Reynolds stress tensor $\boldsymbol{R}$. 

The data-assimilation algorithm treats $\boldsymbol{f}$ as an unknown and leverages partial knowledge of $\overline{\boldsymbol{u}}$ to reconstruct not only $\boldsymbol{f}$ but also $\overline{\boldsymbol{u}}$ on a larger domain which obeys prescribed boundary conditions. A constrained optimisation problem is formulated as follows
\begin{equation}\label{eq:lagrangian}
	\mathcal{L}(\overline{\boldsymbol{u}}^{\dagger}, \overline{p}^{\dagger}, \overline{\boldsymbol{u}},\overline{p},\boldsymbol{f}) = \mathcal{E}(\overline{\boldsymbol{u}}) - \left<\overline{\boldsymbol{u}}^{\dagger},\overline{\boldsymbol{u}}\cdot \nabla \overline{\boldsymbol{u}} + \nabla \overline{p} - Re^{-1}\nabla^2 \overline{\boldsymbol{u}} - \boldsymbol{f} \right> - \left<\overline{p}^{\dagger},\nabla \cdot \overline{\boldsymbol{u}} \right>,
\end{equation} 
to minimise the difference between a measurement of $\overline{\boldsymbol{u}}$, which satisfies  (\ref{eq:RANS}), and the experimental mean profile $\overline{\boldsymbol{u}}_{exp}$ measurement. In (\ref{eq:lagrangian}), $(\cdot)^{\dagger}$ signifies an adjoint quantity and $\left<\boldsymbol{a}\cdot \boldsymbol{b}\right> = \int_{\Omega}\boldsymbol{a} \cdot \boldsymbol{b} d \Omega$. The scalar $\mathcal{E}$ quantifies the discrepancy and is given by 
\begin{equation} \label{eq:E}
	\mathcal{E}(\overline{\boldsymbol{u}}) = \frac{1}{2} \| \overline{\boldsymbol{m}} - \mathcal{M}(\overline{\boldsymbol{u}})\|^2_M.
\end{equation}
The operator $\mathcal{M}$ projects the numerical data to the subspace spanned by the known PIV measurements $\overline{\boldsymbol{m}}$. 

The variations of $\mathcal{L}$ with respect to each dependent variable are set to zero to obtain the equations which are solved iteratively until $\mathcal{E}$ is minimised. Since $\mathcal{E} \neq \mathcal{E}(\overline{p})$, only the solenoidal component of $\boldsymbol{f}$ is recovered \citep[see][for details]{Symon19}. To remove the irrotational component and facilitate comparisons to the experimental Reynolds stresses, only $\nabla \times \boldsymbol{f}$ is reported in \S\ref{sec:assimilation}. 

The direct and adjoint equations are solved in FreeFem++ \citep{Hecht12} on the computational domain $\Omega \in -5 \leq x \leq 15 \cup -5 \leq y \leq 5 $ with the leading edge of the airfoil being centred at the origin. The equations are spatially discretised using quadratic basis functions for the velocity and linear basis functions for the pressure resulting in approximately 360,000 and 1,000,000 degrees of freedom for the A0 and A10 cases, respectively. The reader is referred to \cite{Foures14, Symon17} for more details about the algorithm and its implementation. 

\subsection{Spectral proper orthogonal decomposition}

The velocity fluctuations obtained from the processed PIV datasets are analysed using two modal decomposition techniques. The first, introduced by \cite{Lumley67, Lumley70}, is spectral proper orthogonal decomposition (SPOD). From the raw, experimental data, SPOD calculates energy-ranked modes which are orthogonal to one another and oscillate at a single frequency. The SPOD modes are computed using the procedure described in \cite{Towne18} and the code developed by \cite{Schmidt18b}. A brief summary of the steps is presented below.

The PIV data are rearranged into the matrix 
\begin{equation}
	 \boldsymbol{Q} = \left[\begin{array}{cccc} \boldsymbol{q}_1 & \boldsymbol{q}_2 & \cdots & \boldsymbol{q}_n  \end{array} \right]  \in \mathbb{R}^{m \times n}, 
\end{equation}
where $m$ represents the number of states, i.e. the streamwise and transverse velocity of all PIV vectors, and $n$ is the total number of snapshots. Thus, the overall size of $\boldsymbol{Q}$ is 49,104 total states by 6,800 snapshots. Following Welch's method \citep{Welch67}, $\boldsymbol{Q}$ is split into $p$ smaller, overlapping segments, or blocks, containing $n_b$ snapshots where $n_b < n$
\begin{equation}
	\boldsymbol{Q}^{(p)} = \left[\begin{array}{cccc} \boldsymbol{q}_1^{(p)} & \boldsymbol{q}_2^{(p)} & \cdots & \boldsymbol{q}^{(p)}_{n_{b}}  \end{array} \right]  \in \mathbb{R}^{m \times n_b}. 
\end{equation}
Each row of $\boldsymbol{Q}^{(p)}$ is then Fourier-transformed in time using the discrete Fourier transform (DFT), yielding an ensemble of Fourier realizations
\begin{equation}
	\hat{\boldsymbol{Q}}^{(p)} = \left[\begin{array}{cccc} \hat{\boldsymbol{q}}_{\omega_1}^{(p)} & \hat{\boldsymbol{q}}_{\omega_2}^{(p)} & \cdots & \hat{\boldsymbol{q}}_{\omega_{n_b}}^{(p)}  \end{array} \right]  \in \mathbb{C}^{m \times n_b}.
\end{equation}
The columns of $\hat{\boldsymbol{Q}}^{(p)}$ represent a Fourier mode instead of a snapshot in time as they did in $\boldsymbol{Q}^{(p)}$. The Fourier modes for a specific frequency $\omega$ can be arranged into a new data matrix 
\begin{equation}
	\hat{\boldsymbol{Q}}(\omega) = \left[\begin{array}{cccc} \hat{\boldsymbol{q}}_{\omega}^{(1)} & \hat{\boldsymbol{q}}_{\omega}^{(2)} & \cdots & \hat{\boldsymbol{q}}_{\omega}^{(b)}  \end{array} \right]  \in \mathbb{C}^{m \times b}, 
\end{equation}
where $b$ is the number of blocks.

The cross-spectral density matrix for a specific temporal frequency $\hat{\boldsymbol{S}}(\omega)$ is 
\begin{equation}
	\hat{\boldsymbol{S}}(\omega) = \hat{\boldsymbol{Q}}(\omega)\hat{\boldsymbol{Q}}^*(\omega) \in \mathbb{R}^{m \times m} .
\end{equation} 
The SPOD eigenvectors $\hat{\boldsymbol{V}}(\omega)$ and eigenvalues $\boldsymbol{\Lambda}(\omega)$ can be obtained via an eigenvalue decomposition of the cross-spectral density matrix
\begin{equation}
	\hat{\boldsymbol{S}}(\omega)\hat{\boldsymbol{V}}(\omega) = \hat{\boldsymbol{V}}(\omega)\boldsymbol{\Lambda}(\omega).
\end{equation}
Since the number of blocks is much smaller than the number of states in the problem, i.e. $b \ll m$, it is computationally less demanding to consider the following eigenvalue problem instead
\begin{equation}
	\hat{\boldsymbol{Q}}^*(\omega)\hat{\boldsymbol{Q}}(\omega)\hat{\boldsymbol{\Theta}}(\omega) = \hat{\boldsymbol{\Theta}}(\omega)\boldsymbol{\Lambda}(\omega), 
\end{equation}
as $\hat{\boldsymbol{Q}}^*(\omega)\hat{\boldsymbol{Q}}(\omega) \in \mathbb{R}^{b \times b}$, and compute the eigenvector of the full system using
\begin{equation}
	\hat{\boldsymbol{V}}(\omega) = \hat{\boldsymbol{Q}}(\omega)\hat{\boldsymbol{\Theta}}(\omega)\boldsymbol{\Lambda}(\omega)^{-1/2}.
\end{equation}
It should be noted that the choice of $\omega$ is limited by $n$ and the sampling frequency which, for both experiments, is 125Hz. As will be seen in the next section, this is a major disadvantage of SPOD, which requires a lot of data to obtain converged modes and high frequency resolution, in comparison to resolvent analysis. 

The PIV snapshots are divided into blocks containing $n = 1700$ snapshots with 50\% overlap, resulting in $b = 7$ realisations of the flow. The resulting frequency resolution is $\Delta \omega = 0.46$. SPOD is performed using a rectangular windowing function (nearly identical results are obtained using a Hamming window). The choice of $n$ and $b$ is a compromise between suitable resolution of the lower frequencies, where the energy is most concentrated, and obtaining a sufficient number of realisations to identify low-rank behaviour. We find that the trends are unaffected as long as we consider at least 3 realisations of the flow.

\subsection{Resolvent analysis} \label{sec:resolvent}

Once the mean profile has been data-assimilated, it can be used as an input to resolvent analysis, which has been shown to provide an efficient basis for velocity fluctuations by \cite{McKeon10} and many others. Subtracting (\ref{eq:RANS}) from (\ref{eq:NS}) results in the equations governing the fluctuations
\begin{subequations}\label{eq:fluctuations}
	\begin{equation} \label{eq:fluctuatingmomentum}
		\partial_t \boldsymbol{u}' + \overline{\boldsymbol{u}} \cdot \nabla \boldsymbol{u}' + \boldsymbol{u}' \cdot \nabla \overline{\boldsymbol{u}} + \nabla p' - Re^{-1}\nabla^2 \boldsymbol{u}' = -\boldsymbol{u}'\cdot \nabla \boldsymbol{u}' + \overline{\boldsymbol{u}'\cdot \nabla \boldsymbol{u}'} = \boldsymbol{f}',
	\end{equation}
	\begin{equation}
		\nabla \cdot \boldsymbol{u}' = 0.
	\end{equation}
\end{subequations}
The left-hand side of (\ref{eq:fluctuatingmomentum}) contains the linear terms and the right-hand side is lumped into a single forcing term $\boldsymbol{f}'$ to account for the nonlinear terms. Substituting $\boldsymbol{f}' = \hat{\boldsymbol{f}}e^{i\omega t}$, where $\omega$ is a real-valued frequency, implies $\boldsymbol{u}' = \hat{\boldsymbol{u}} e^{i\omega t}$ and leads to the following input-output form of the linearised Navier-Stokes equations
\begin{equation}\label{eq:resolvent}
	\hat{\boldsymbol{u}} = \boldsymbol{C}^T(i\omega \boldsymbol{B} - \boldsymbol{L})^{-1} \boldsymbol{C} \hat{\boldsymbol{f}} = \mathcal{H}(\omega)\hat{\boldsymbol{f}}.
\end{equation}
The matrix $\mathcal{H}(\omega)$ is the resolvent operator and the linear operators $\boldsymbol{C}$, $\boldsymbol{B}$, and $\boldsymbol{L}$ are defined as 
\begin{equation}
	\boldsymbol{C} = \left( \begin{array}{c} 1 \\ 0 \end{array} \right), ~~~ \boldsymbol{B} = \left(\begin{array}{cc} 1 & 0 \\ 0 & 0 \end{array} \right), ~~~ \boldsymbol{L} = \left(\begin{array}{cc} -\overline{\boldsymbol{u}} \cdot \nabla () - () \cdot \nabla \overline{\boldsymbol{u}} + Re^{-1} \nabla^2 ()  & -\nabla () \\ \nabla \cdot () & 0  \end{array} \right).
\end{equation}
$\boldsymbol{L}$ is the linear Navier-Stokes (LNS) operator, $\boldsymbol{B}$ restricts the nonlinear forcing to the velocity subspace, and $\boldsymbol{C}$ can be adjusted to restrict the spatial domain of the input and output vectors. 

The singular value decomposition (SVD) of the resolvent operator yields basis functions for the nonlinear term and velocity fluctuations at each temporal frequency $\omega$
\begin{equation}\label{eq:SVD}
	\mathcal{H}(\omega) = \sum_{j=1}^{\infty} \hat{\boldsymbol{\Psi}}(\omega) \boldsymbol{\Sigma}(\omega) \hat{\boldsymbol{\Phi}}^*(\omega).
\end{equation}
The matrices $\hat{\boldsymbol{\Phi}}(\omega)$ and $\hat{\boldsymbol{\Psi}}(\omega)$ contain the optimal forcing and response modes, respectively, which are ranked by their kinetic energy gains contained in the diagonal matrix $\boldsymbol{\Sigma} (\omega)$. Following \cite{Sipp13}, the singular values $\sigma_i(\omega)$ are computed by generating the relevant operators in FreeFem++ and reformulating (\ref{eq:SVD}) into the following eigenvalue problem:
\begin{equation}
	\mathcal{H}^*(\omega)\mathcal{H}(\omega)\hat{\boldsymbol{\phi}}_i(\omega) = \sigma_i^2(\omega)\hat{\boldsymbol{\phi}}_i(\omega).
\end{equation}
The largest eigenvalues of the Hermitian operator $\mathcal{H}^*(\omega)\mathcal{H}(\omega)$ are computed using the ARPACK library \citep{Lehoucq96} and the MUMPS parallel solver \citep{Amestoy01}. 

\subsection{Reduced-order modelling of the fluctuation fields}

In this section, we discuss how an efficient basis for the fluctuations can be educed from resolvent analysis by approximating the linear operator $\mathcal{H}(\omega)$, if it is low-rank, or the nonlinear forcing $\hat{\boldsymbol{f}}(\omega)$. We also explain how time-resolved probe points are used to calibrate the amplitude and phase of the resolvent modes.

\subsubsection{Approximation of the (linear) resolvent operator}

The resolvent operator's singular values are said to be well separated if $\sigma_1(\omega) \gg \sigma_2(\omega)$, in which case the following approximation holds
\begin{equation} \label{eq:approximate operator}
	\mathcal{H}(\omega) \approx \hat{\boldsymbol{\psi}}_1(\omega) \sigma_1(\omega)\hat{\boldsymbol{\phi}}_1(\omega). 
\end{equation}
(\ref{eq:approximate operator}) can be interpreted as an approximation of the (linear) resolvent operator and signifies that the first response mode $\hat{\boldsymbol{\psi}}_1(\omega)$ is sufficient to represent the velocity fluctuations at a frequency $\omega$. In other words, it is possible to write
\begin{equation} \label{eq:rank1}
	\hat{\boldsymbol{u}}(\omega) \approx  \hat{\boldsymbol{\psi}}_1(\omega) \sigma_1 (\omega) \hat{\boldsymbol{\phi}}_1^*(\omega) \hat{\boldsymbol{f}}(\omega) = \hat{\boldsymbol{\psi}}_1(\omega) \chi_1(\omega),
\end{equation}
where the procedure for computing the complex weight $\chi_1(\omega)$ is described in \S\ref{sec:weighting}. It should be noted that, in general, the Fourier mode $\hat{\boldsymbol{u}}(\omega)$ is not well defined unless the flow is truly periodic like cylinder flow. Hence, we refer to $\hat{\boldsymbol{u}}(\omega)$ and $\hat{\boldsymbol{f}}(\omega)$ as an arbitrary sample related to a specific bin.

\subsubsection{Approximation of the (nonlinear) forcing}

When there is an energetic frequency in the flow where the singular values are not well separated, the resolvent modes are no longer an efficient basis to represent the velocity fluctuations. An alternative approach is to approximate the nonlinear forcing $\hat{\boldsymbol{f}}(\omega)$ instead of the linear operator itself \citep{Rosenberg19}. (\ref{eq:rank1}) is modified into the following form
\begin{equation} \label{eq:driven}
	\hat{\boldsymbol{u}}(\omega) \approx \mathcal{H}(\omega)\hat{\boldsymbol{f}}_a(\omega),
\end{equation}
where $\hat{\boldsymbol{f}}_a(\omega)$ is the approximated nonlinear forcing that yields the parasitic mode. As stated in \cite{McKeon13}, the full expression for $\hat{\boldsymbol{f}}(\omega)$ is a convolution over all triadically consistent Fourier modes
\begin{equation} \label{eq:full forcing}
	\hat{\boldsymbol{f}}(\omega_c) = \sum_{ \substack{ \omega_a + \omega_b = \omega_c \\ \omega_a,\omega_b \neq 0  }} - \nabla \cdot (\hat{\boldsymbol{u}}(\omega_a)\hat{\boldsymbol{u}}^T(\omega_b)).  
\end{equation}
In this article, we will explain the conditions under which the sum in (\ref{eq:full forcing}) can be severely truncated to a few key triadic interactions and for which types of flows this procedure is applicable.

\subsubsection{Weighting the modes}\label{sec:weighting}

Once a resolvent or parasitic mode for each frequency of interest has been computed using one of the two methods described above, its complex weight $\chi_1(\omega)$ is calibrated using a time-resolved probe $\check{u}(t,\boldsymbol{x}_0)$ in a region of the flow where $\hat{\boldsymbol{\psi}}_1(\omega)$ is energetic as demonstrated by \cite{Gomez16a} and \cite{Beneddine16}. The signal is Fourier-transformed in time so that $\chi_1(\omega)$ can be computed in the frequency domain through the following expression
\begin{equation} \label{eq:calibrate}
	\chi_1(\omega) = \hat{\boldsymbol{\psi}}_1(\omega,\boldsymbol{x}_0)/\check{u}(\omega,\boldsymbol{x}_0).
\end{equation}
It should be noted that while $\hat{\boldsymbol{\psi}}_1$ has been used in (\ref{eq:calibrate}), the equation is still applicable to parasitic modes once their shapes have been computed using (\ref{eq:driven}). After the complex amplitude of each mode has been determined, the linear superposition of all modes can be inverse Fourier-transformed to obtain a reduced-order model of the velocity fluctuations in the time domain, i.e.
\begin{equation}
	\boldsymbol{u}'(t) = \sum_{j=1}^{N_\omega}\hat{\boldsymbol{\psi}}_1(\omega_j)\chi_1(\omega_j)e^{i(\omega_jt)}.
\end{equation}

\subsection{Summary}

\begin{figure} 
	\centering
	\includegraphics[scale=0.45]{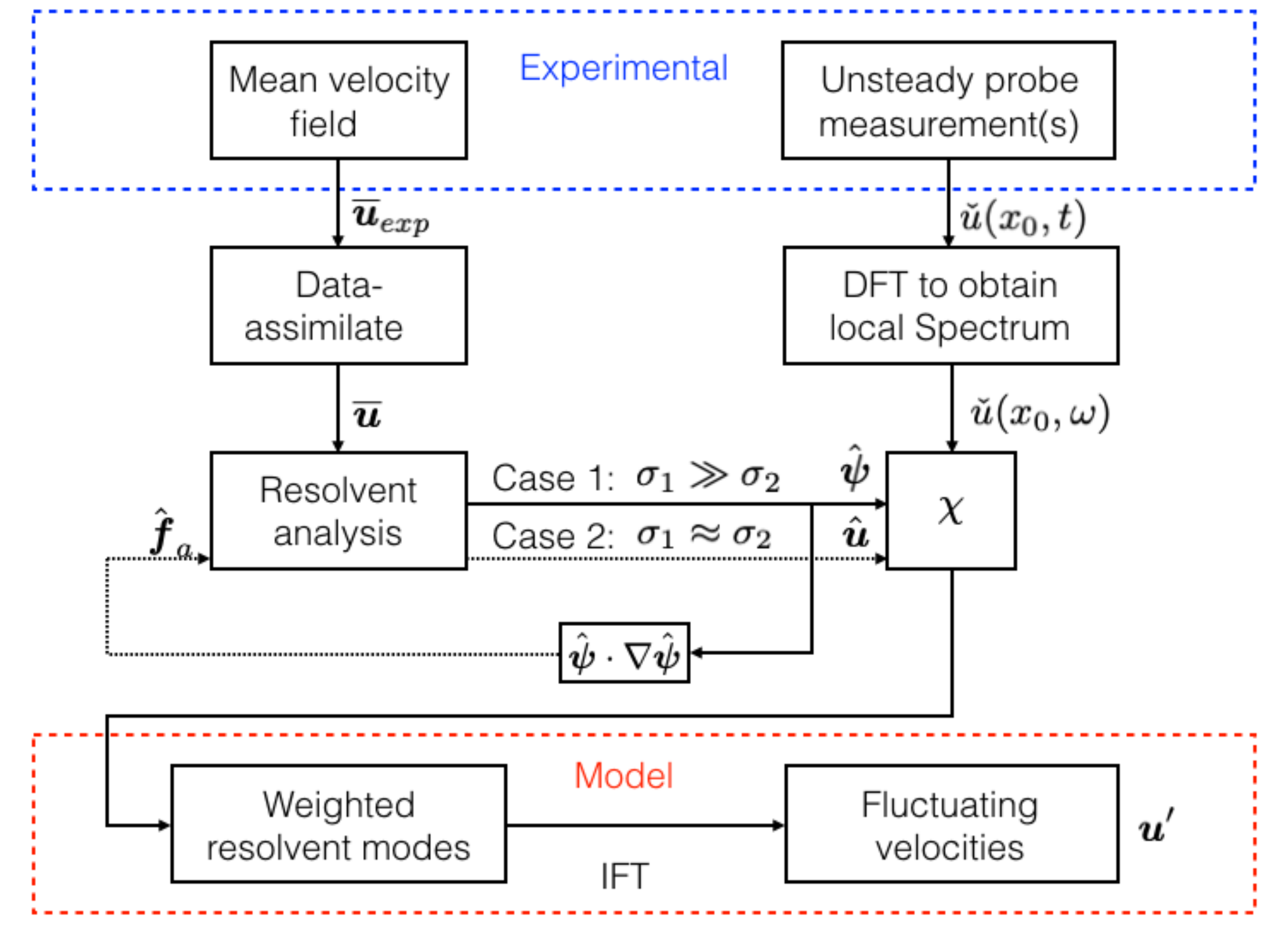}
	\caption{Steps to obtain the resolvent-based model of the velocity fluctuations starting from the experimental measurements, adapted from \cite{Symon18b}.}\label{fig:flowchart}
\end{figure}

The resolvent-based modelling procedure is summarised in figure \ref{fig:flowchart}, which contains the relevant steps starting with the acquisition of the experimental measurements (top row). The mean profile is data-assimilated onto a FreeFem mesh with the correct boundary conditions and the time-resolved probe measurements are Fourier-transformed in time (row 2). The structure associated with frequency $\omega$ is obtained from resolvent analysis of the data-assimilated mean profile in one of two ways: (1) approximating the operator via the singular value decomposition if there is a large separation of singular values ($\sigma_1 \gg \sigma_2$) or (2) approximating the nonlinear forcing if the frequency is energetic but there is no separation of singular values. These two methods are demarcated by solid and dotted lines, respectively, in row 3 of figure \ref{fig:flowchart}. The resolvent modes are weighted from the Fourier-transformed probe points and inverse Fourier-transformed to obtain a model of the velocity fluctuations in time (bottom row).

\section{Assimilated flow fields} \label{sec:assimilation}

In this section, the mean profiles obtained from the time-averaged PIV are data-assimilated using the method introduced by \citet{Foures14}. Since data-assimilation attempts to reconstruct the optimal forcing $\boldsymbol{f}$ in the RANS equations, the quantity $\nabla \times \boldsymbol{f}$, in addition to the assimilated mean fields, are compared to their experimental counterparts. 

\subsection{A0 Case}

\begin{figure}
	\centering
	\includegraphics[scale=0.3]{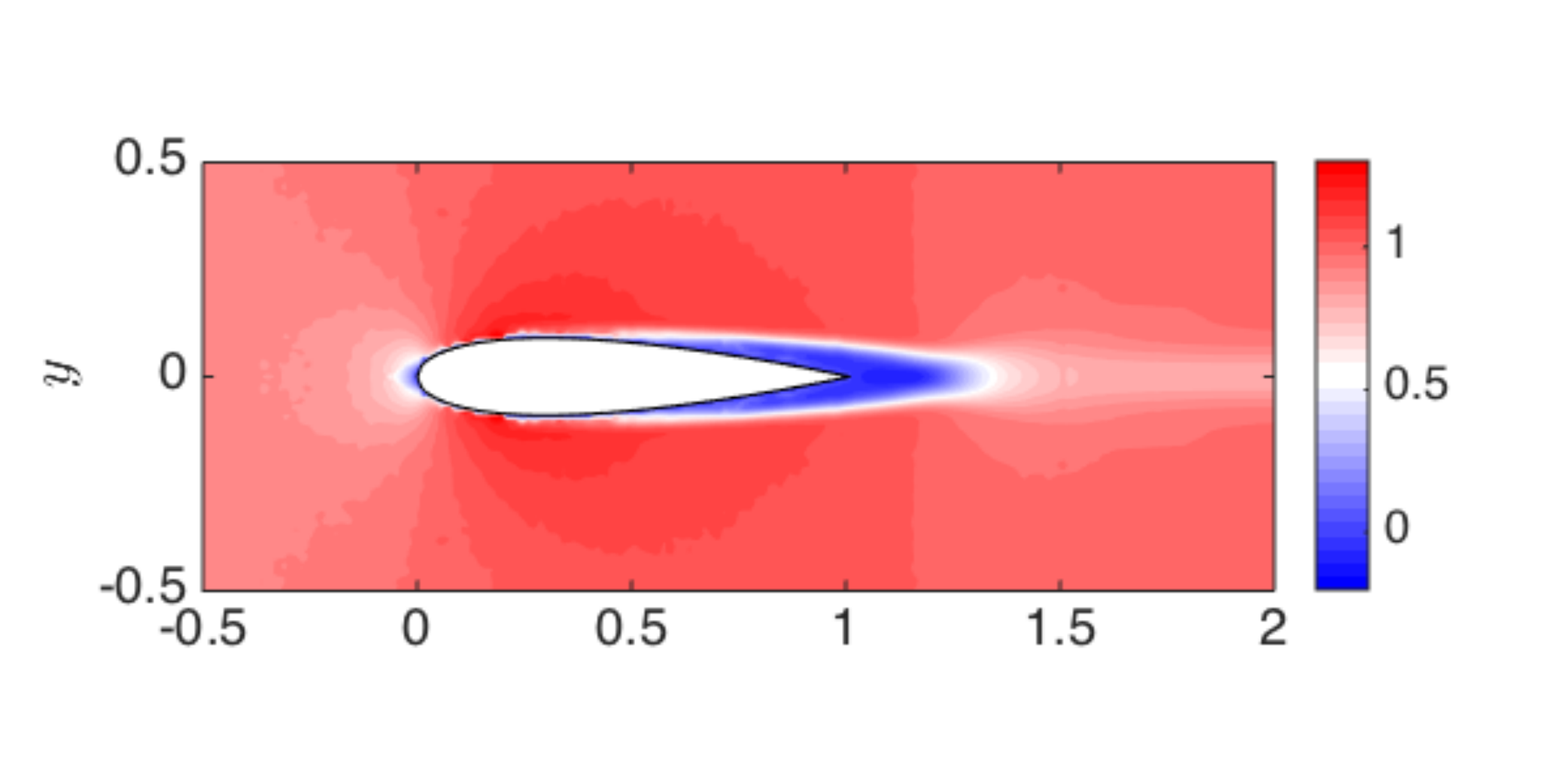}
	\includegraphics[scale=0.3]{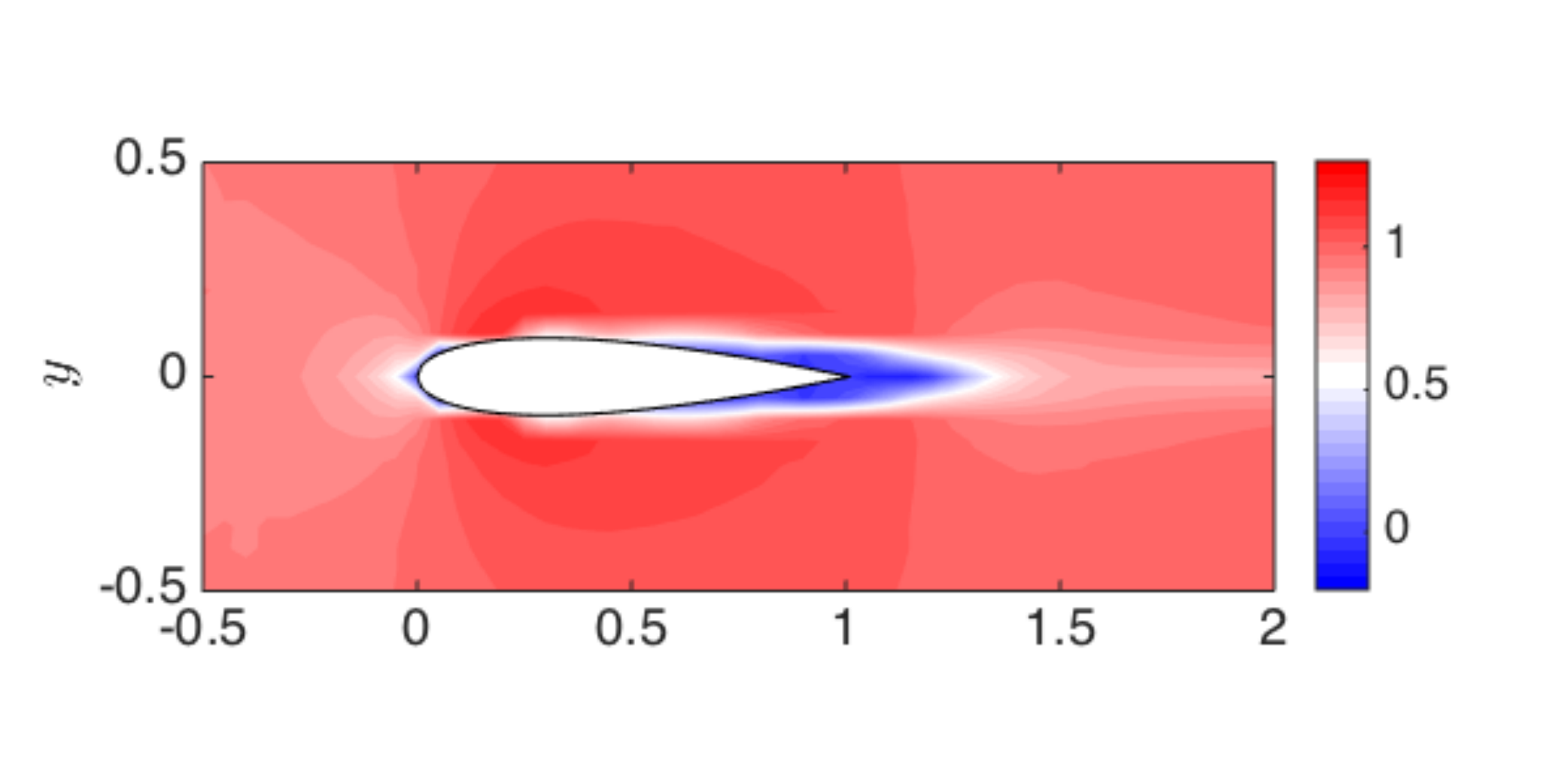}
	\includegraphics[scale=0.3]{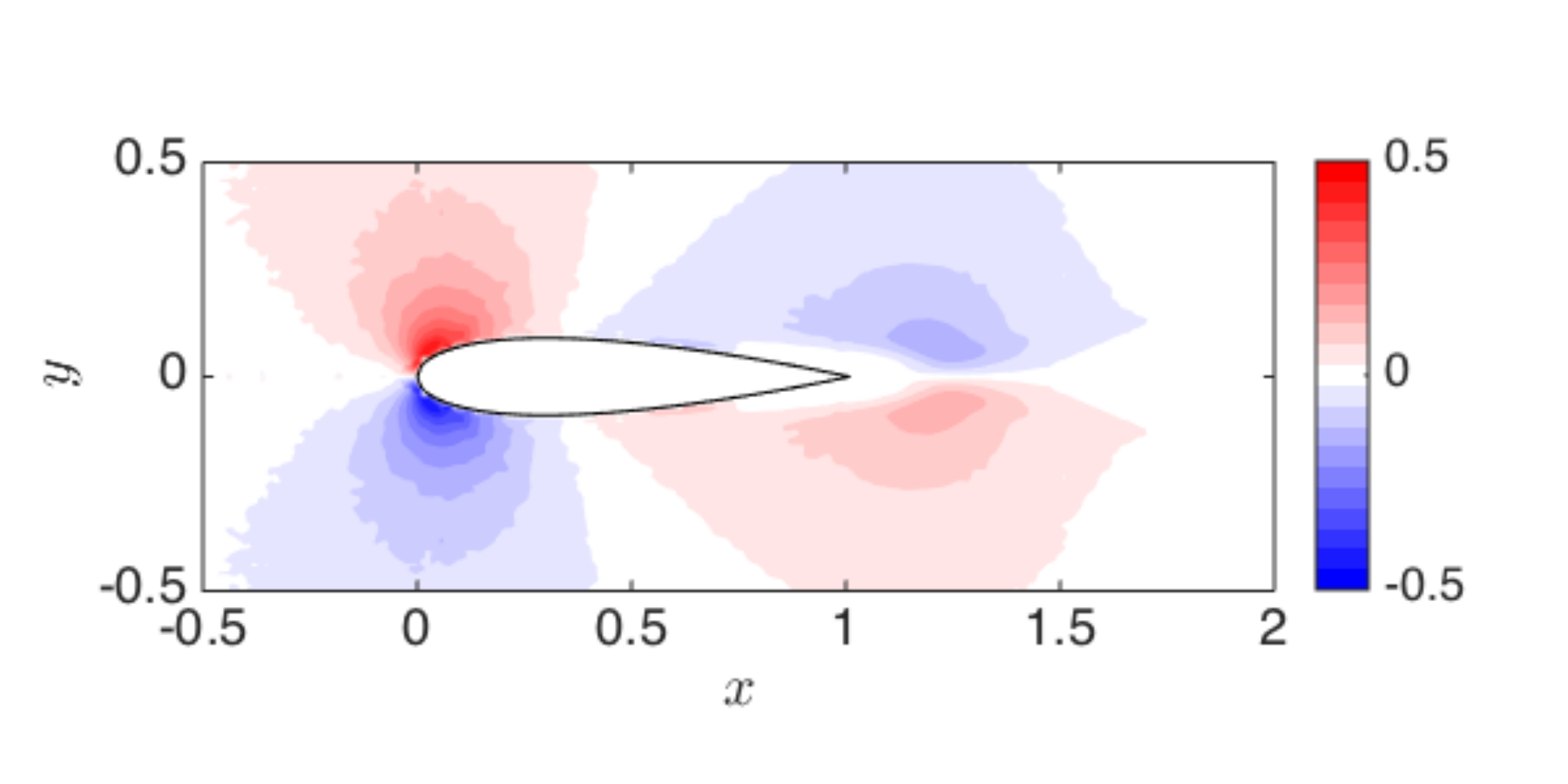}
	\includegraphics[scale=0.3]{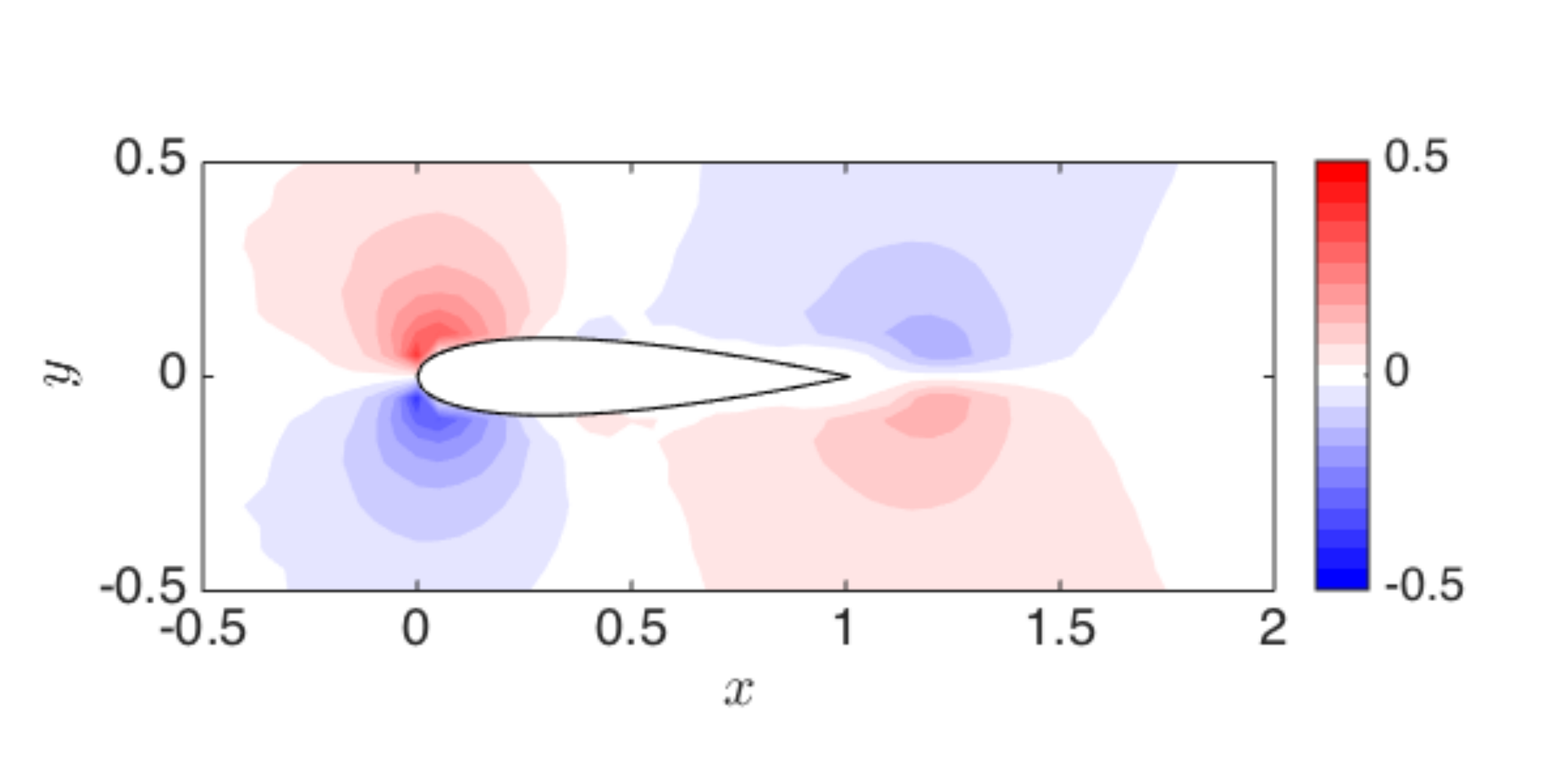}
	
	\caption{Data-assimilation results for the A0 case. The assimilated velocity fields (right) are compared to the experiment (left). The streamwise velocity $\overline{u}$ is in the upper row while the transverse velocity $\overline{v}$ is in the lower row.}\label{fig:assimilatedA0}
\end{figure}

As mentioned in \S\ref{sec:experiment}, the data below the airfoil are obstructed by the airfoil's shadow. To overcome this problem, the data above the centreline are reflected across, keeping the streamwise velocity constant and reversing the sign of the transverse velocity. Both the streamwise and transverse component of the data-assimilated fields are compared to their experimental counterparts in figure \ref{fig:assimilatedA0}. The fields are nearly indistinguishable although the data-assimilated contours are slightly more smooth than those in the experiment. 

\begin{figure}
	\includegraphics[scale=0.3]{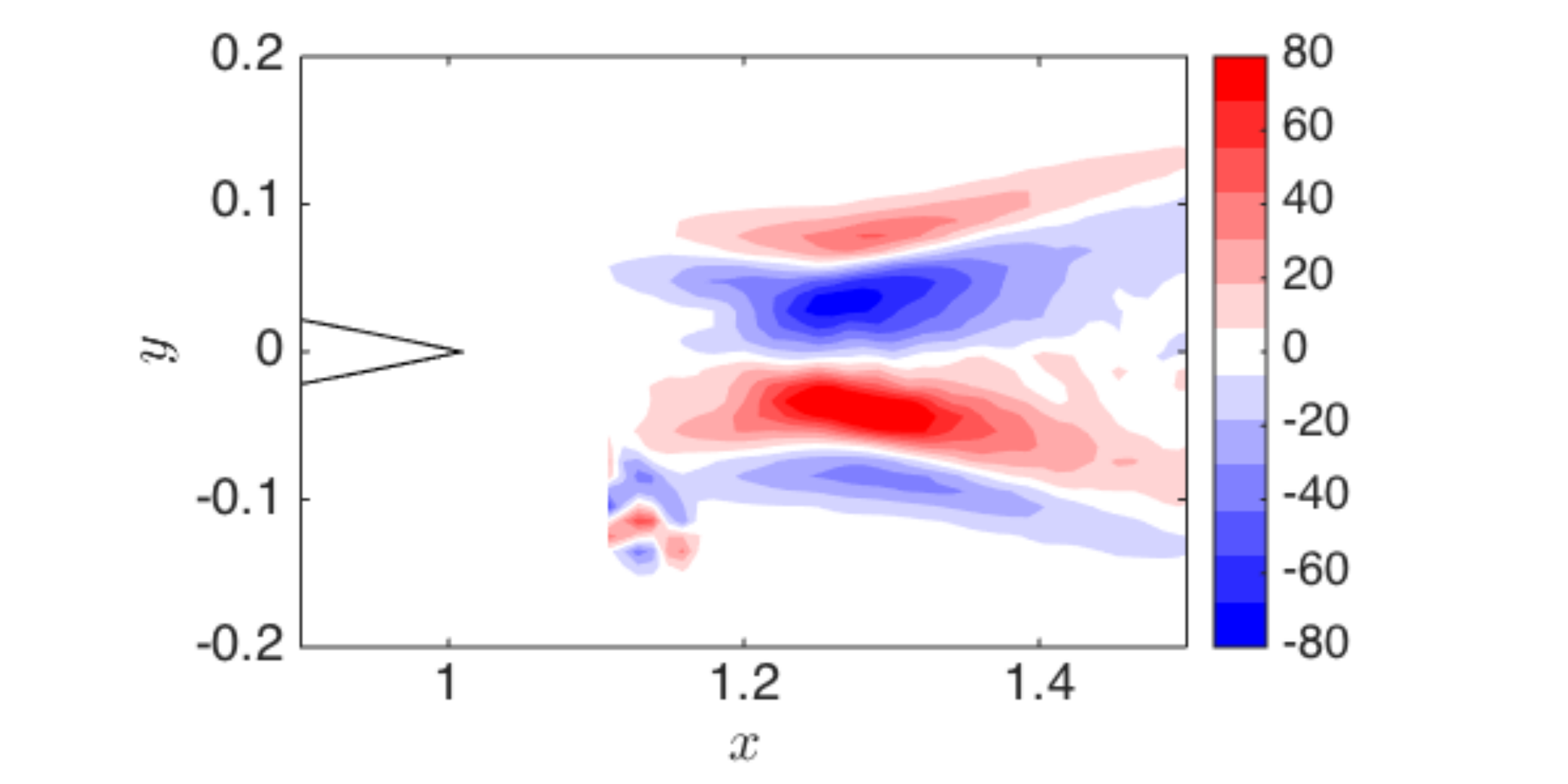}
	\includegraphics[scale=0.3]{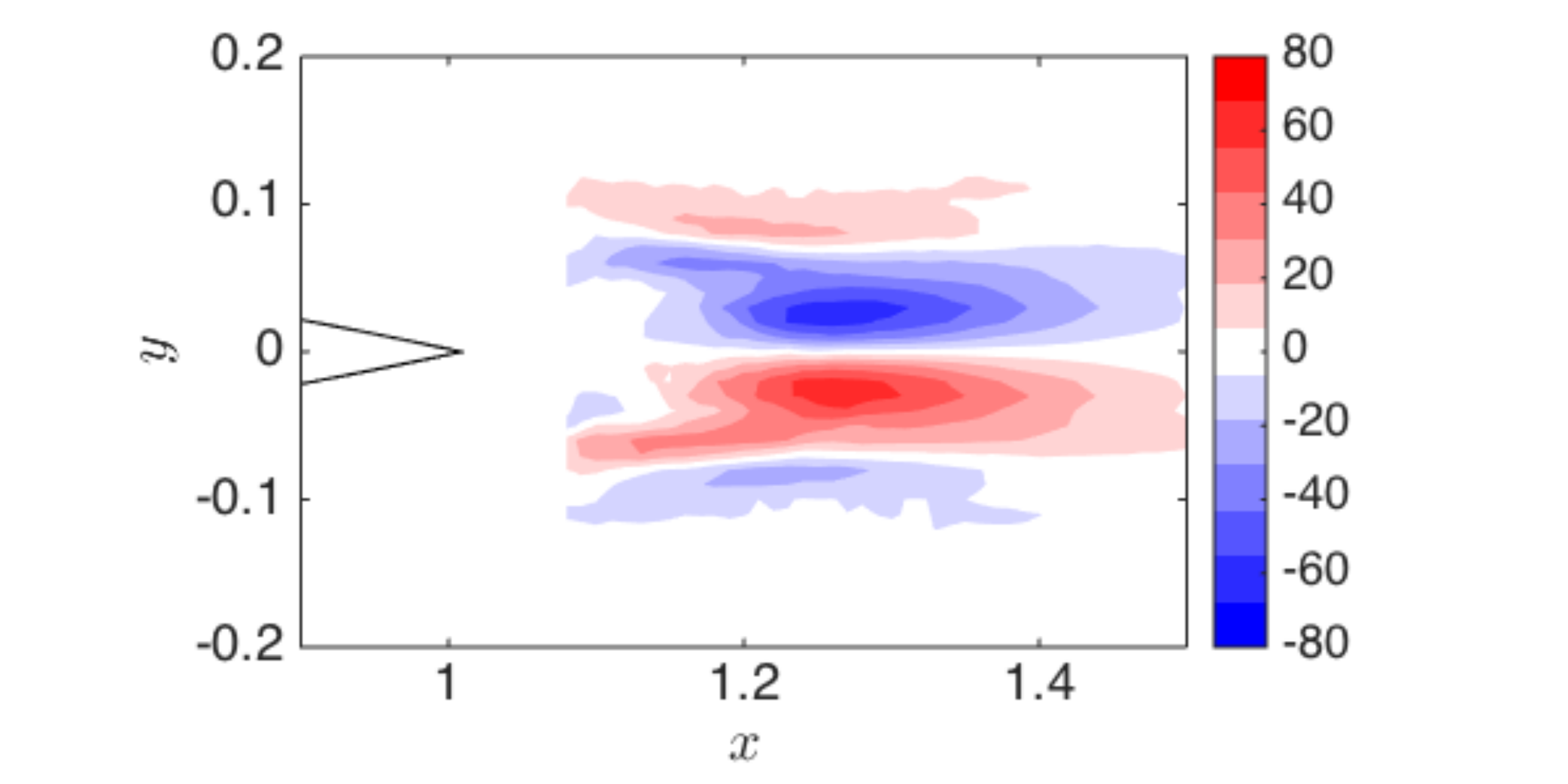}
	
	\caption{The curl of the optimal forcing $\nabla \times \boldsymbol{f}$ from the experiment (left) compared to the data-assimilation (right) in the A0 case.}\label{fig:fA0}
\end{figure}

The curl of the forcing $\nabla \times \boldsymbol{f}$, rather than $\boldsymbol{f}$ itself, is shown in figure \ref{fig:fA0} to eliminate the irrotational component which cannot be captured. The agreement between the experimental field, obtained after computing two gradients of the velocity fluctuations, and the data-assimilated field is good despite some minor discrepancies. One notable aspect about figure \ref{fig:fA0} is that $\nabla \times \boldsymbol{f}$ for this flow closely resembles that of cylinder flow (see figures 4 and 5.9 of \cite{Foures14} and \cite{Symon18}, respectively). 

\subsection{A10 Case}

Data-assimilation of the A10 case is slightly more difficult since the flow is not symmetric with respect to the centreline. Two separate experiments are conducted at both $\alpha = -10^{\circ}$ and $\alpha = 10^{\circ}$. In the first experiment, the data below the airfoil (suction side), are obstructed by the airfoil shadow so only the pressure side can be measured. In the second experiment, the pressure side is obstructed by the shadow, so only the suction side is measured. The final mean velocity field can be obtained by stitching together the two mean profiles such that the data around the entire airfoil are known. The stitching process, if imperfect, may result in non-physical measurements although the data-assimilation algorithm is able to compensate for this. 

\begin{figure}
	\centering
	\includegraphics[scale=0.3]{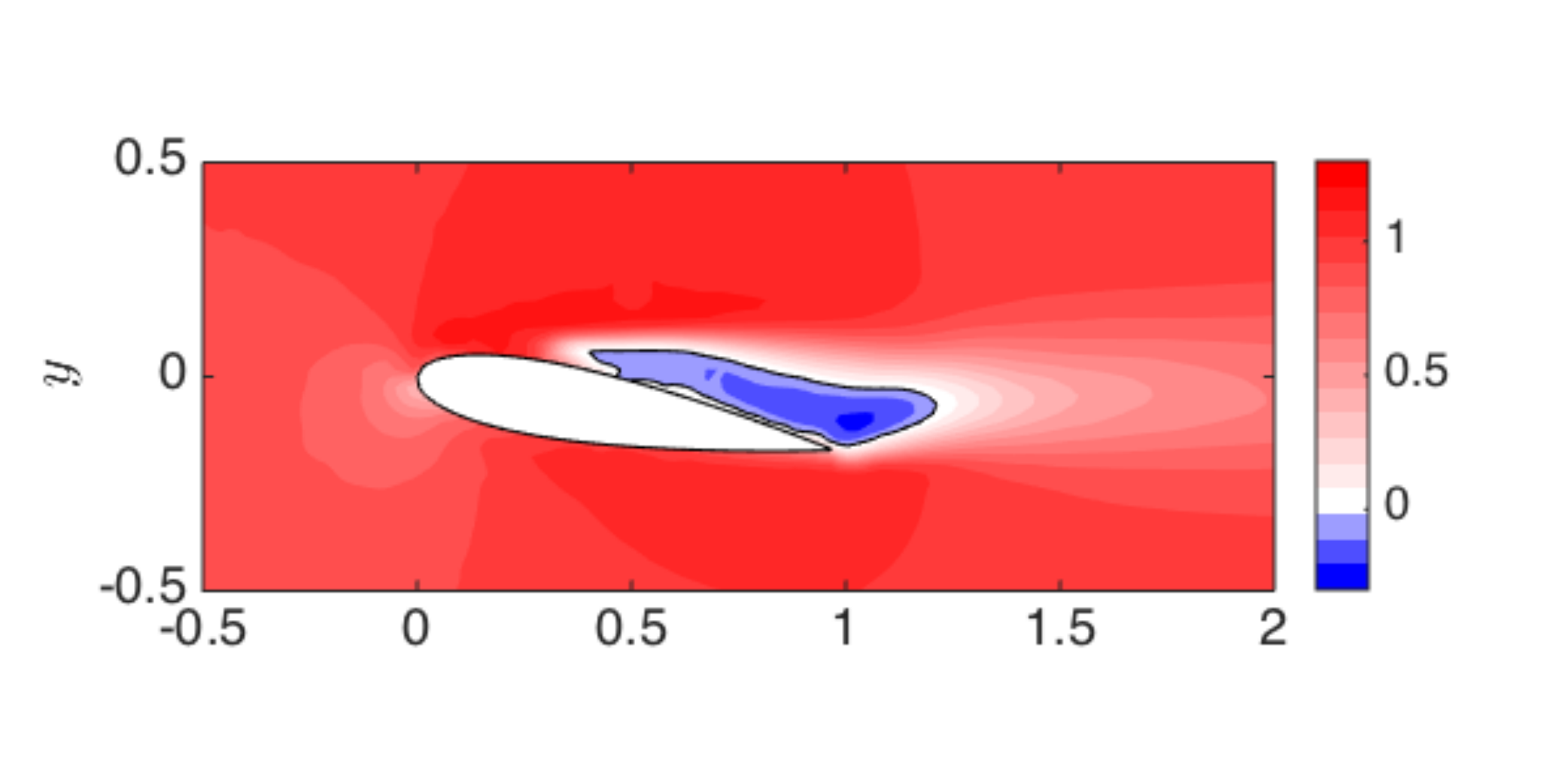}
	\includegraphics[scale=0.3]{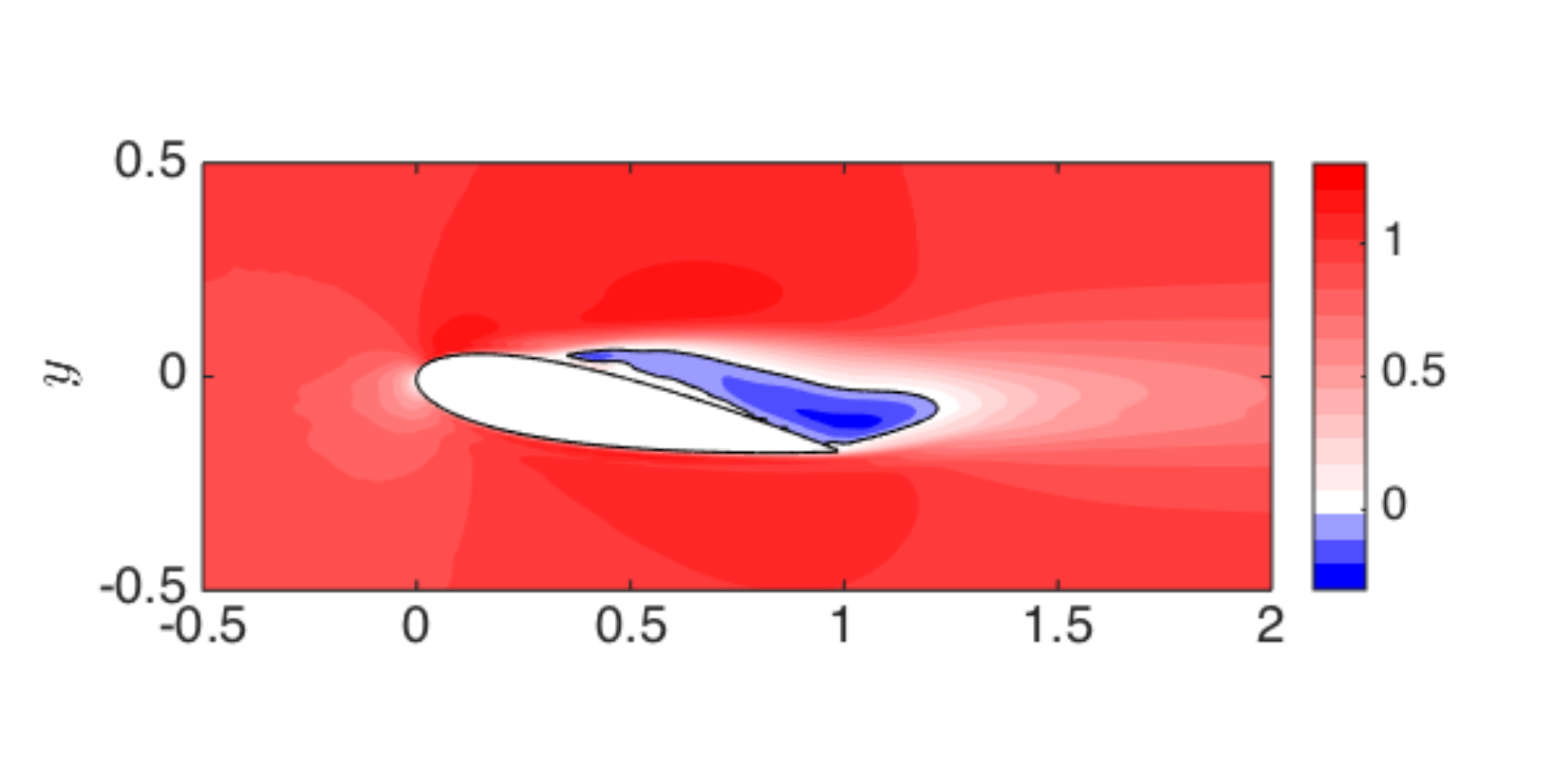}
	
	\includegraphics[scale=0.3]{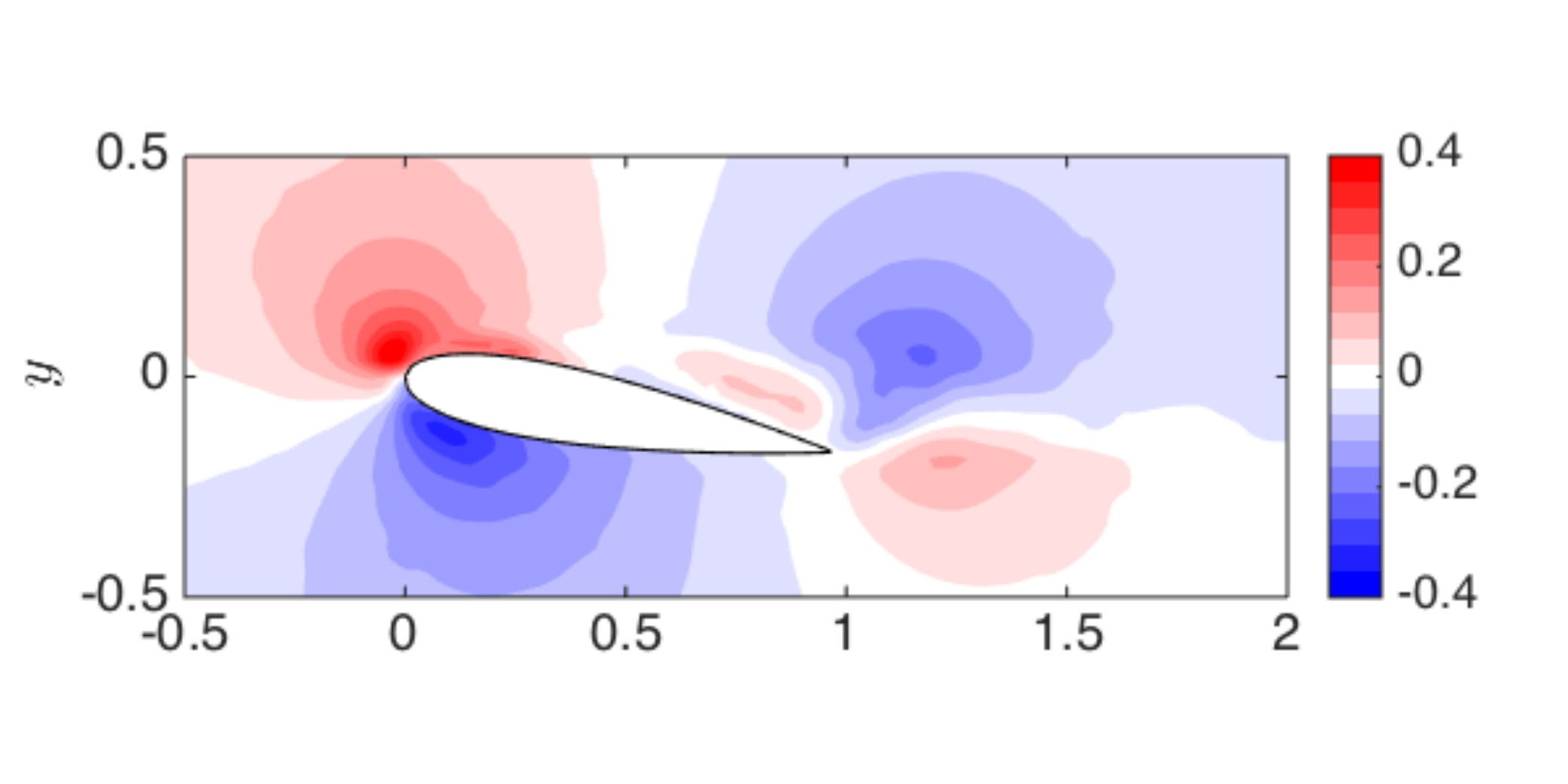}
	\includegraphics[scale=0.3]{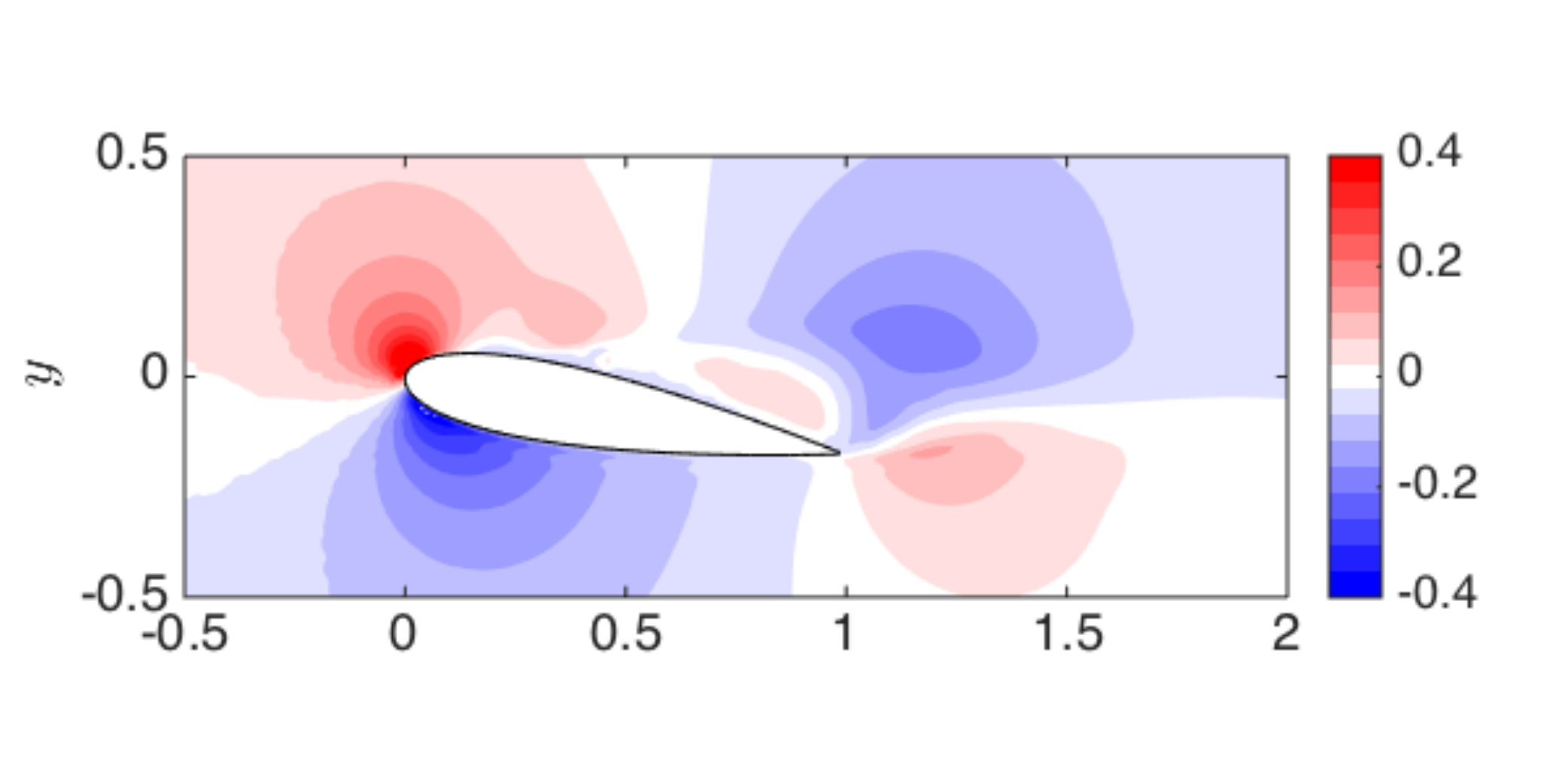}
	
	\includegraphics[scale=0.3]{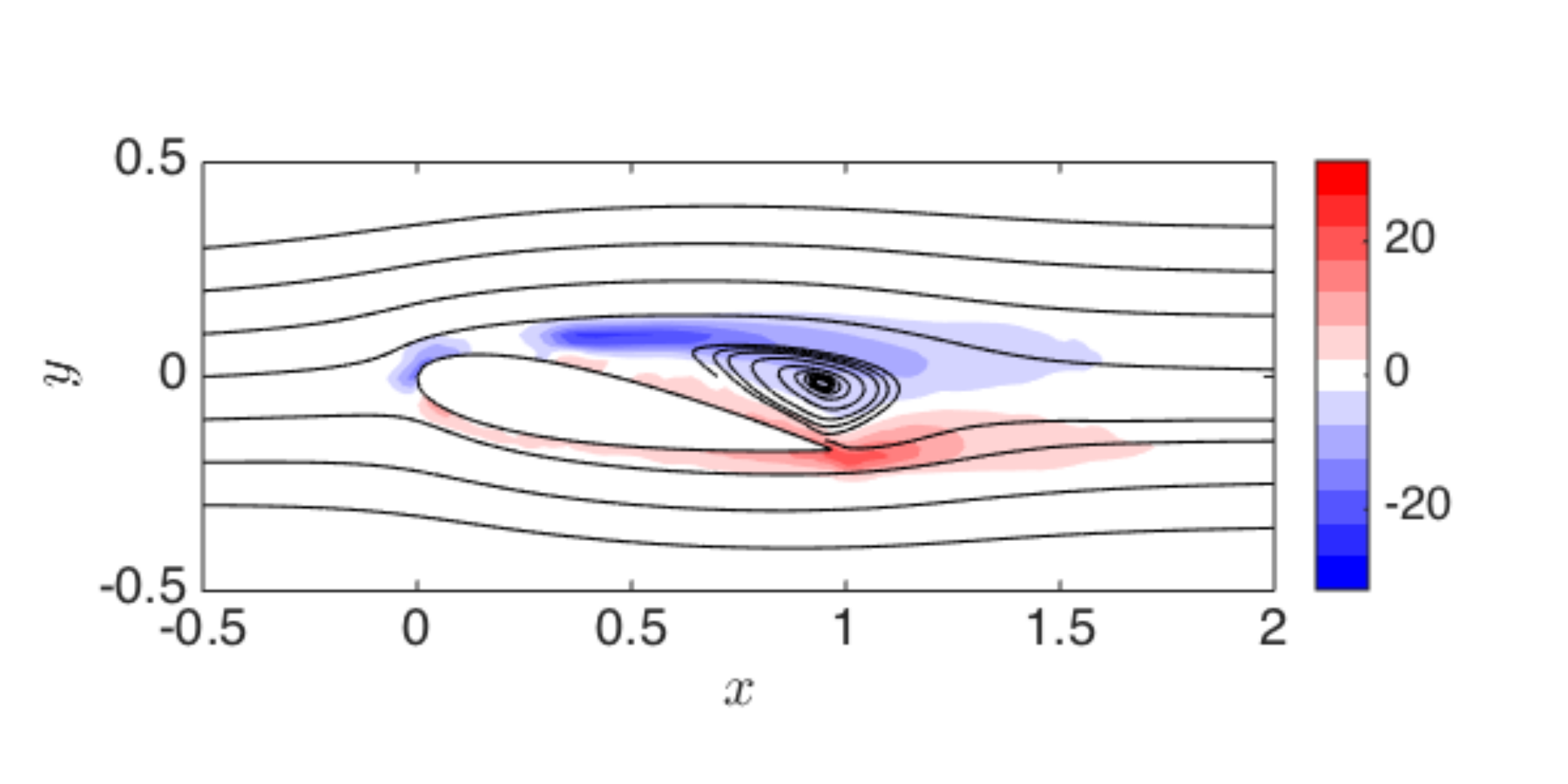} 
	\includegraphics[scale=0.3]{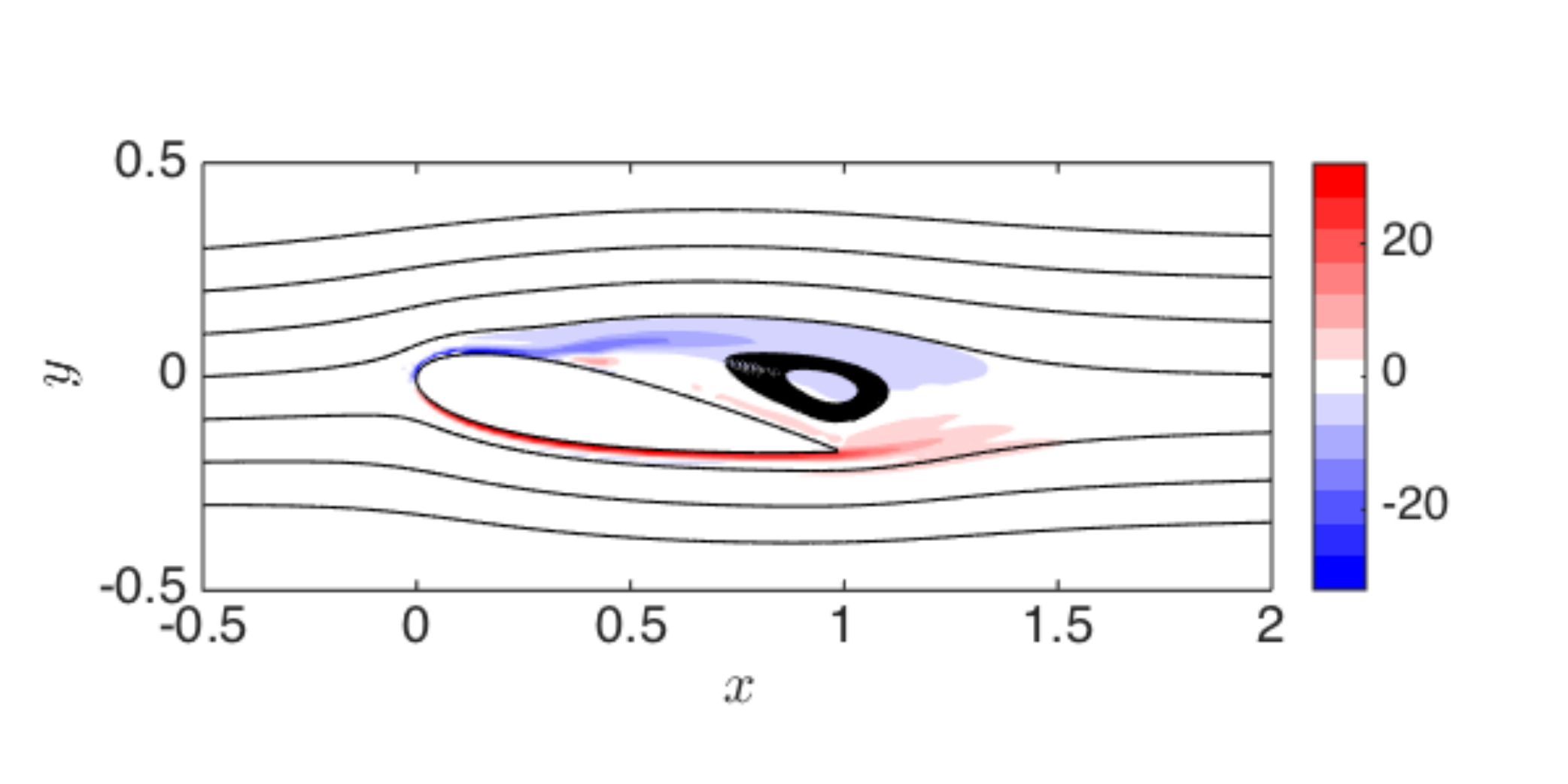} 
	
	\caption{Data-assimilation results for the A10 case. The assimilated fields (right) are compared to the experiment (left): streamwise velocity $\overline{u}$ (top row), transverse velocity $\overline{v}$ (middle row) and spanwise vorticity $\overline{\omega}_z$ (bottom row).}\label{fig:assimilatedA10}
\end{figure}

Data-assimilation is performed using the stitched profile, i.e. from measurements at $\alpha = -10^{\circ}$ and $\alpha = 10^{\circ}$, and the resulting fields are compared to their experimental counterparts in figure \ref{fig:assimilatedA10}. In addition to comparing the two components of the mean velocity field, the mean spanwise vorticity $\overline{\omega}_z$ is plotted in the bottom row of figure \ref{fig:assimilatedA10}. Streamlines of the mean velocity field indicate the size and location of the recirculation bubble. The streamwise and transverse velocities match very well, particularly with respect to the reverse flow region. The separated flow region is accurately reproduced in the data-assimilated fields.   

There are two regions of the flow field which are visibly improved by data-assimilation. The first is near the leading edge of the airfoil, where the flow separates and vortices are formed due to the rollup of the shear layer. This improvement is particularly visible in the vorticity plots since the negative contours of the data-assimilated field are continuous unlike in the experiment where there is a gap at $x \approx 0.2$. As noted by \cite{Raffel18}, there are difficulties associated with PIV for laminar flow around airfoils. The effects of strong centrifugal forces around the airfoil leading edge and strong shear result in the outward movement of tracer particles in a direction perpendicular to the curved streamlines. This loss of seeding compromises measurements in the region close to the solid body. 

\begin{figure}
	\centering
	\includegraphics[scale=0.3]{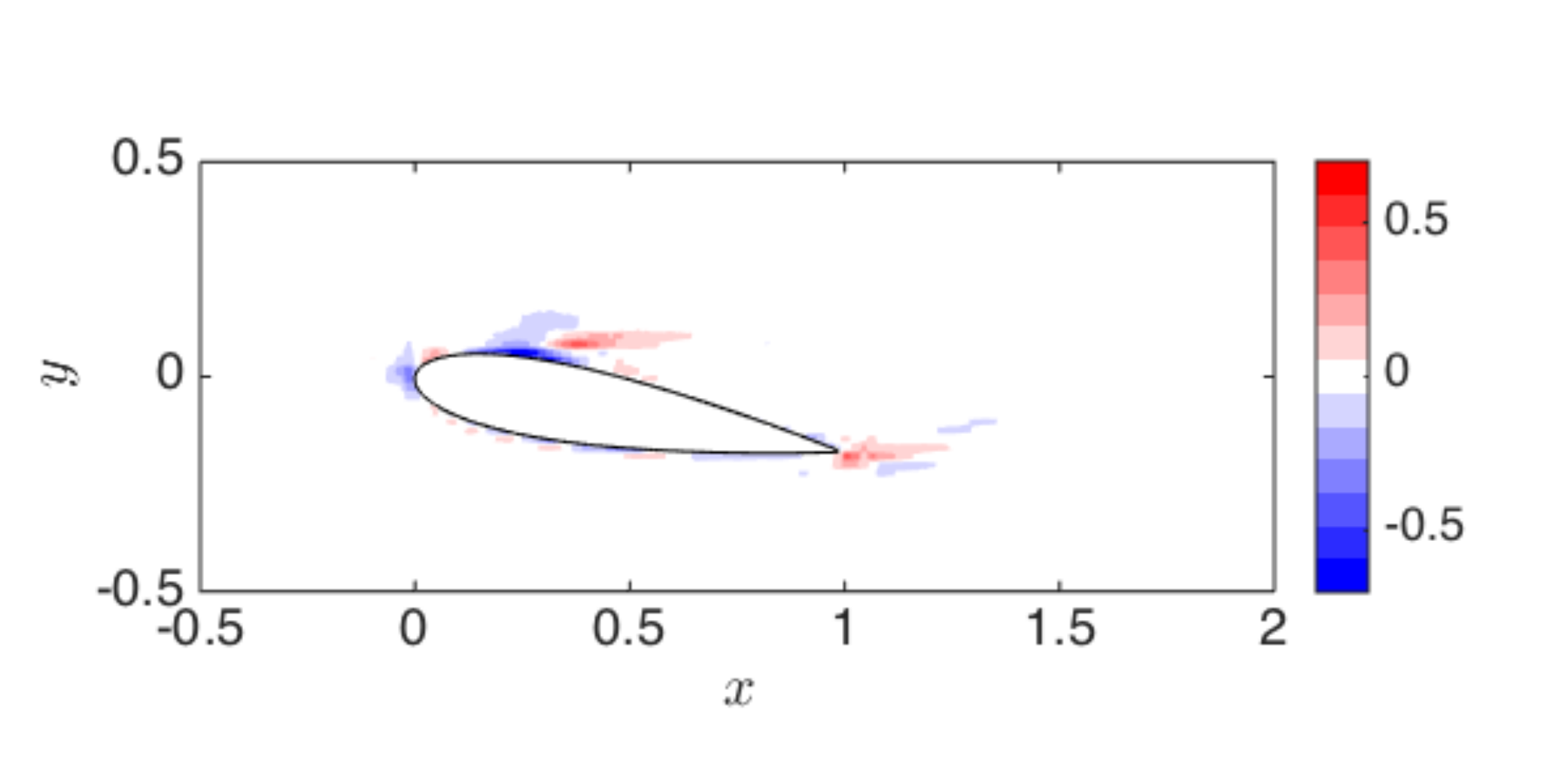}
	\includegraphics[scale=0.3]{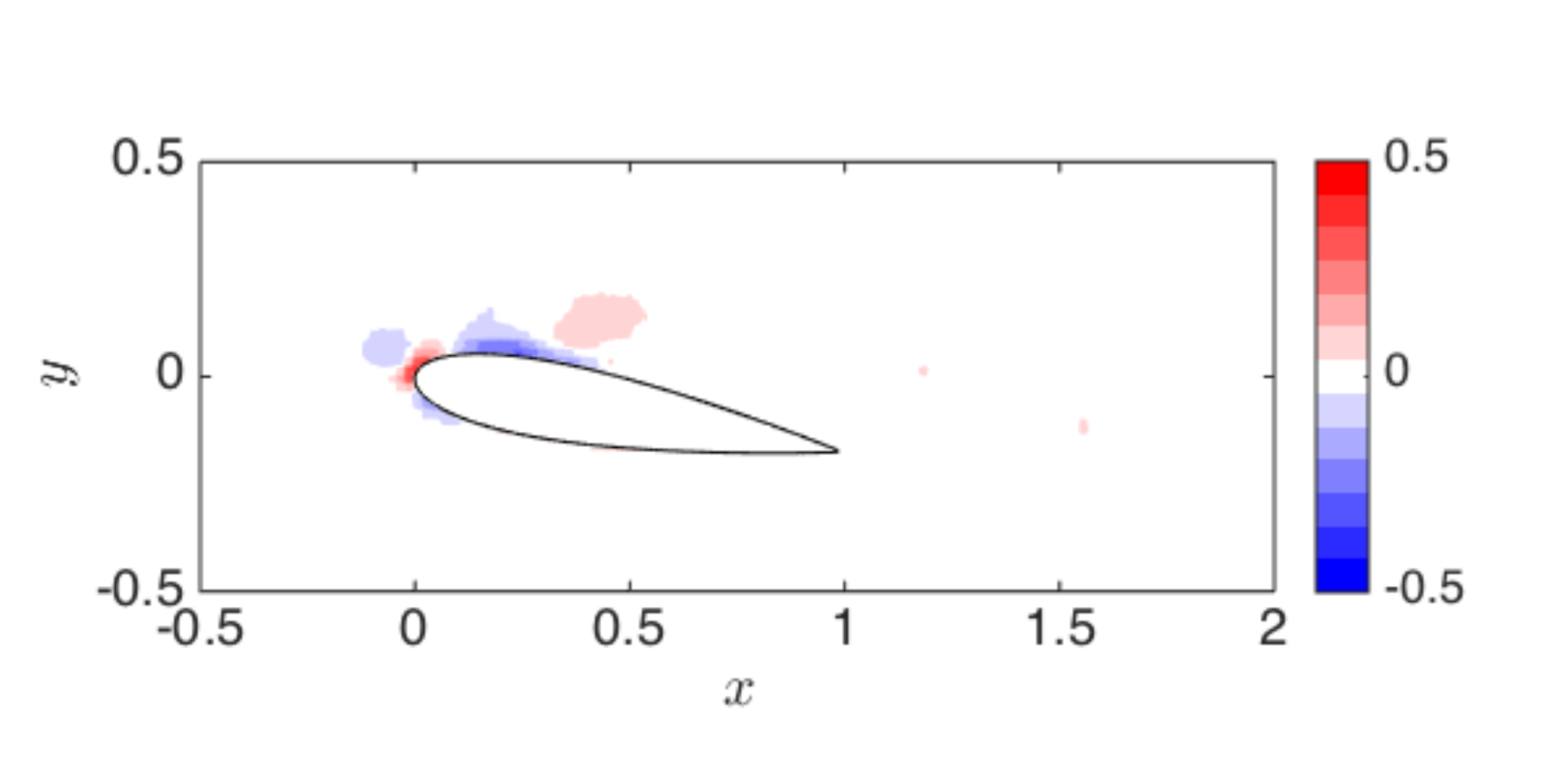}
	
	\caption{Residual discrepancy of the mean streamwise (left) and transverse (right) velocities between the experimental and data-assimilated mean profiles.}\label{fig:A10 delu}
\end{figure}

The second area of improvement is at the trailing edge where there are imperfections associated with the stitching together of the mean profiles. Contours of negative mean streamwise velocity protrude below the trailing edge which corresponds to non-physical behaviour. Data-assimilation is able to correct this problem and eliminate inaccuracies associated with the stitching process. These changes are more apparent by examining the difference between the assimilated and experimental mean velocities, i.e. $\Delta \overline{\boldsymbol{u}} = \overline{\boldsymbol{u}} - \overline{\boldsymbol{u}}_{exp}$, which is illustrated in figure \ref{fig:A10 delu} for both velocity components. The two regions of disagreement, notably, coincide with the areas of improvement discussed earlier.

\begin{figure}
	\centering
	\includegraphics[scale=0.3]{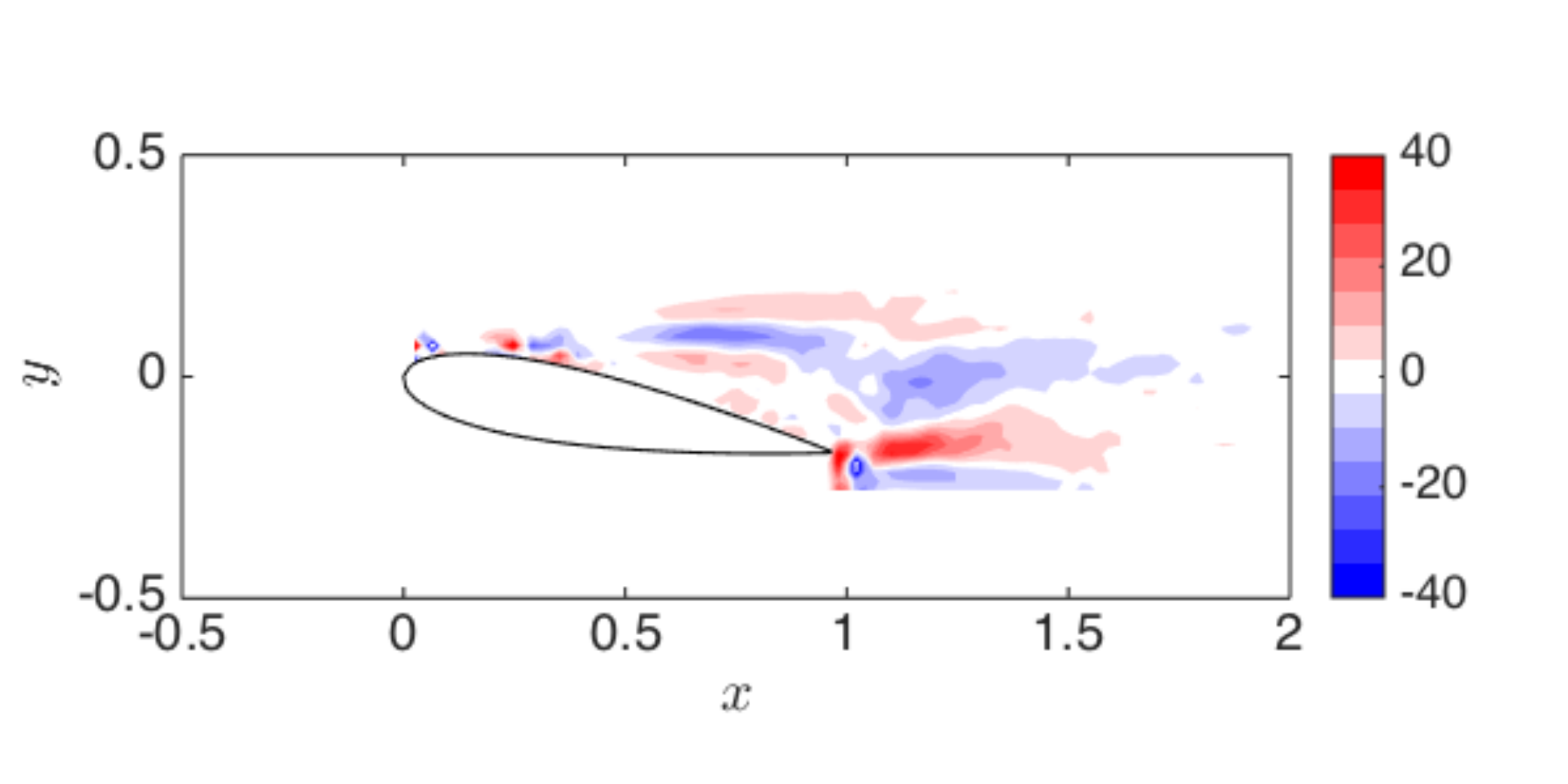}
	\includegraphics[scale=0.3]{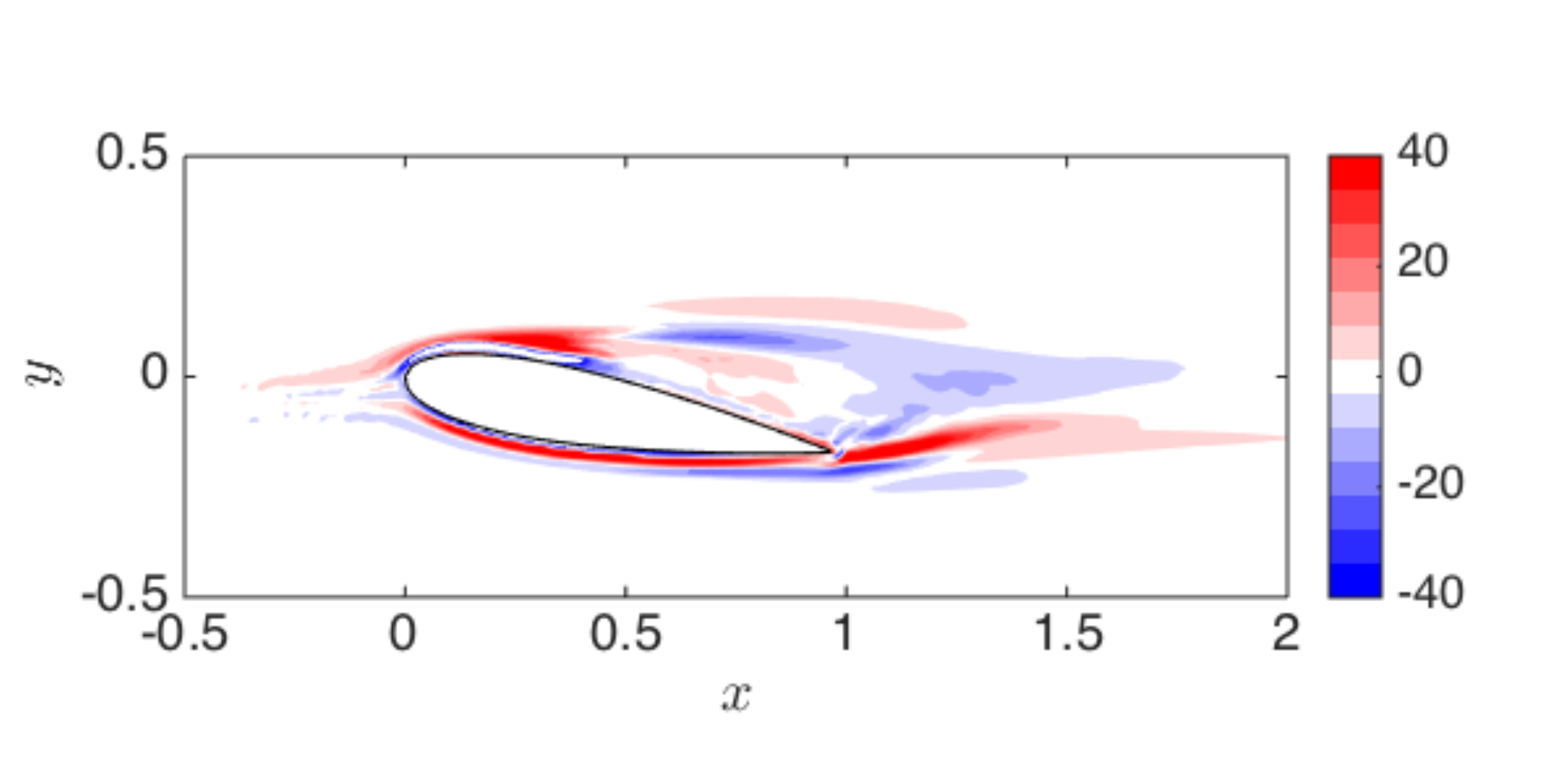}
	
	\caption{The curl of the optimal forcing $\nabla \times \boldsymbol{f}$ from the experiment (left) compared to the data-assimilation (right) in the A10 case.} \label{fig:fA10}
\end{figure}

It should be emphasised that the benefit to data-assimilating fields is not only filling in missing data, but also in obtaining gradient quantities with less noise. This is particularly noticeable when comparing $\nabla \times \boldsymbol{f}$ in figure \ref{fig:fA10}. Even though the experimental data do not have much noise to begin with, the data-assimilated field is more smooth overall. 

Data-assimilation leads to improvement for both the A0 and A10 cases although the changes for A0 are fairly unremarkable. For the A10 case, on the other hand, there are features missing from the experimental profile that data-assimilation fills in, particularly at the shear layer originating near the leading edge. As will be seen in \S\ref{sec:resolvent analysis}, this has major implications for the identification of linear amplification mechanisms in the flow. 

\section{SPOD results and the difference between an oscillator and an amplifier}\label{sec:SPOD}

In this section, SPOD is applied to the velocity fluctuations obtained from the raw PIV data. We also apply SPOD to the nonlinear term $\boldsymbol{u}'\cdot \nabla \boldsymbol{u}'$ to investigate the most energetic structures associated with the nonlinear forcing field. \citet{Towne15} and \citet{Towne16} applied SPOD to the nonlinear forcing in a turbulent jet and found that the educed structures were primarily incoherent turbulent fluctuations and not nonlinear interactions between coherent structures. We will compare the results of SPOD applied to the velocity fluctuations and nonlinear forcing for both airfoils and highlight fundamental differences between oscillator and amplifier flows. These will have consequences on when we expect SPOD and resolvent analysis to coincide in \S\ref{sec:resolvent analysis} as well as our ability to model the flows in \S\ref{sec:modeling}. 

\subsection{SPOD of the velocity fluctuations}

\begin{figure}
	\centering
	\includegraphics[scale=0.3]{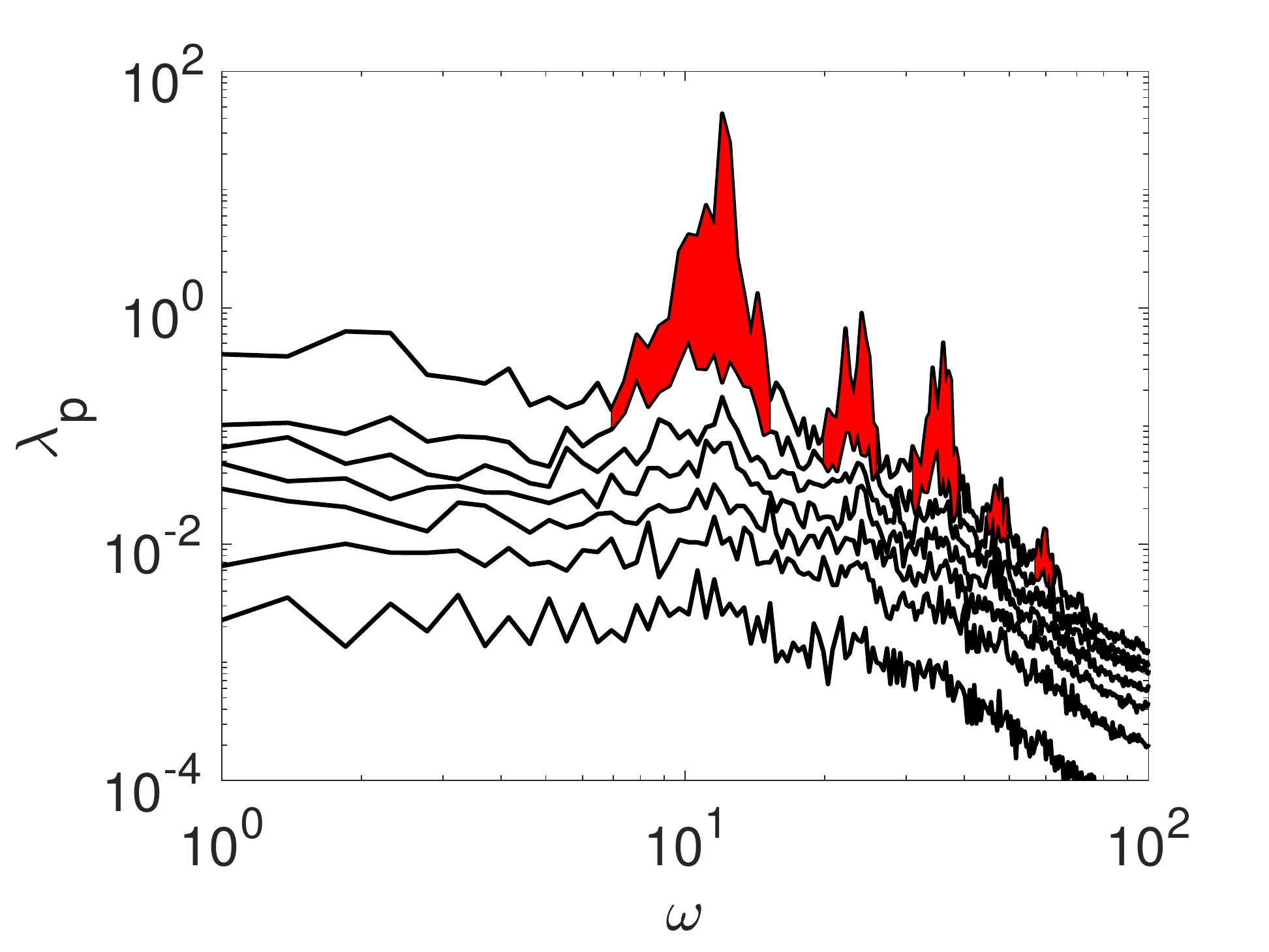}
	\includegraphics[scale=0.3]{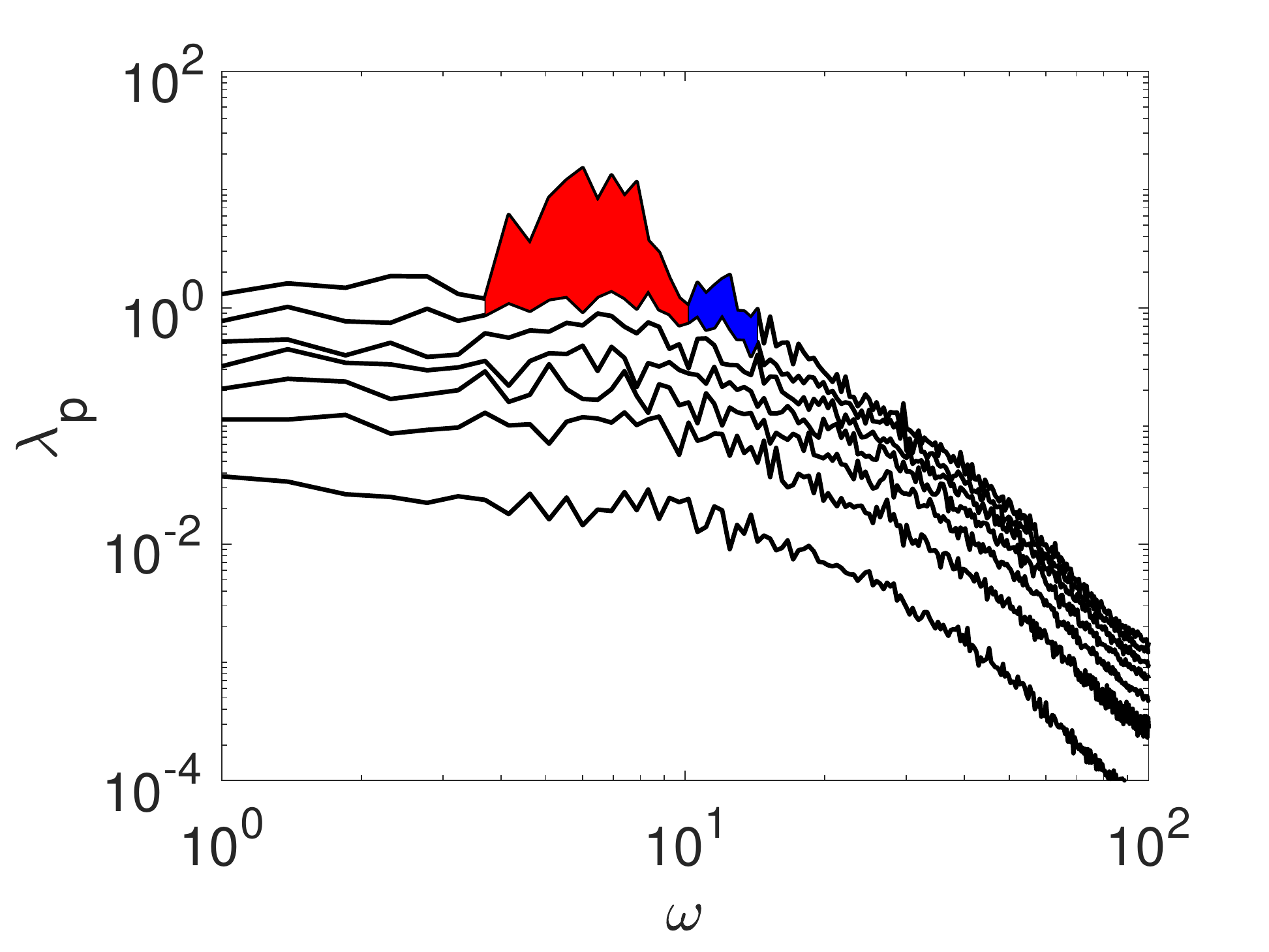}
	
	\caption{The SPOD eigenspectra of the A0 case (left) as a function of $\omega$ where red patches denote frequencies where there is significant separation between the first two eigenvalues. For the A10 case (right),  the red patch denotes low-rank behaviour corresponding to the wake dynamics and the blue patch denotes low-rank behaviour corresponding to the shear layer dynamics. } \label{fig:SPODeigs}
\end{figure}

The SPOD eigenspectra for A0 are plotted in figure \ref{fig:SPODeigs} as a function of $\omega$. The largest peak in $\lambda_1$ occurs at $\omega = 12.0$, which corresponds to the shedding frequency of the airfoil. Red patches highlight frequencies at which there is significant separation between the first and second eigenvalues, the former of which exhibits sharp peaks at the shedding frequency and its harmonics. 

\begin{figure}
	\centering
	\includegraphics[scale=0.3]{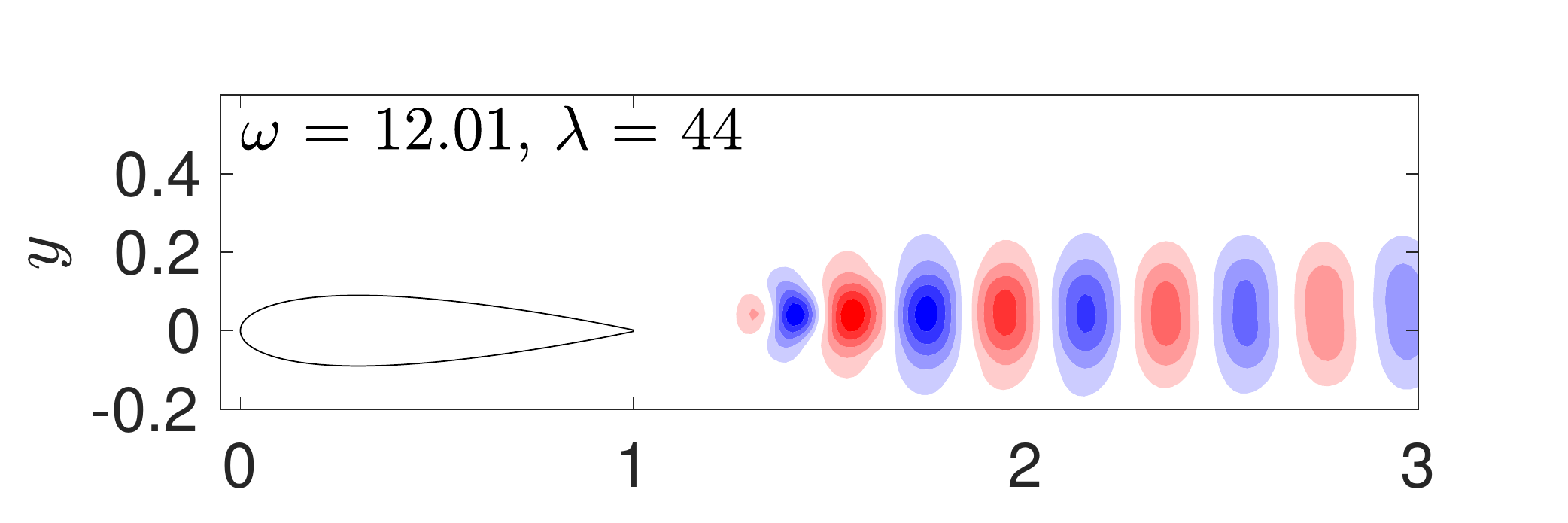}
	\includegraphics[scale=0.3]{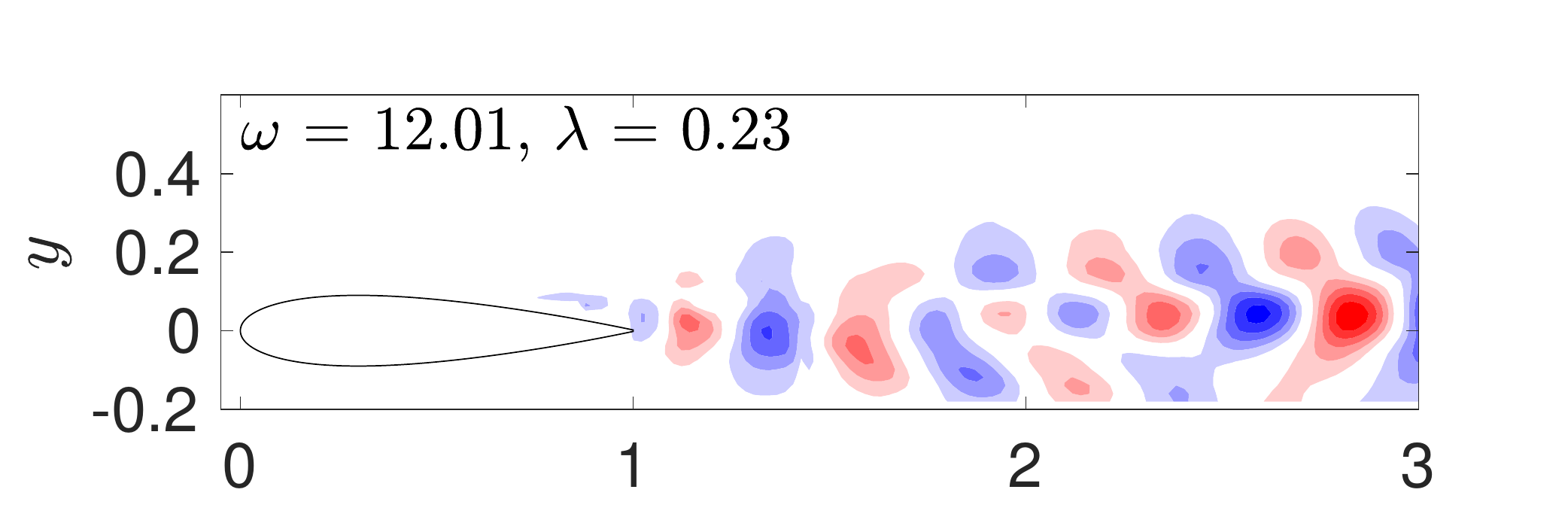} 	
	\includegraphics[scale=0.3]{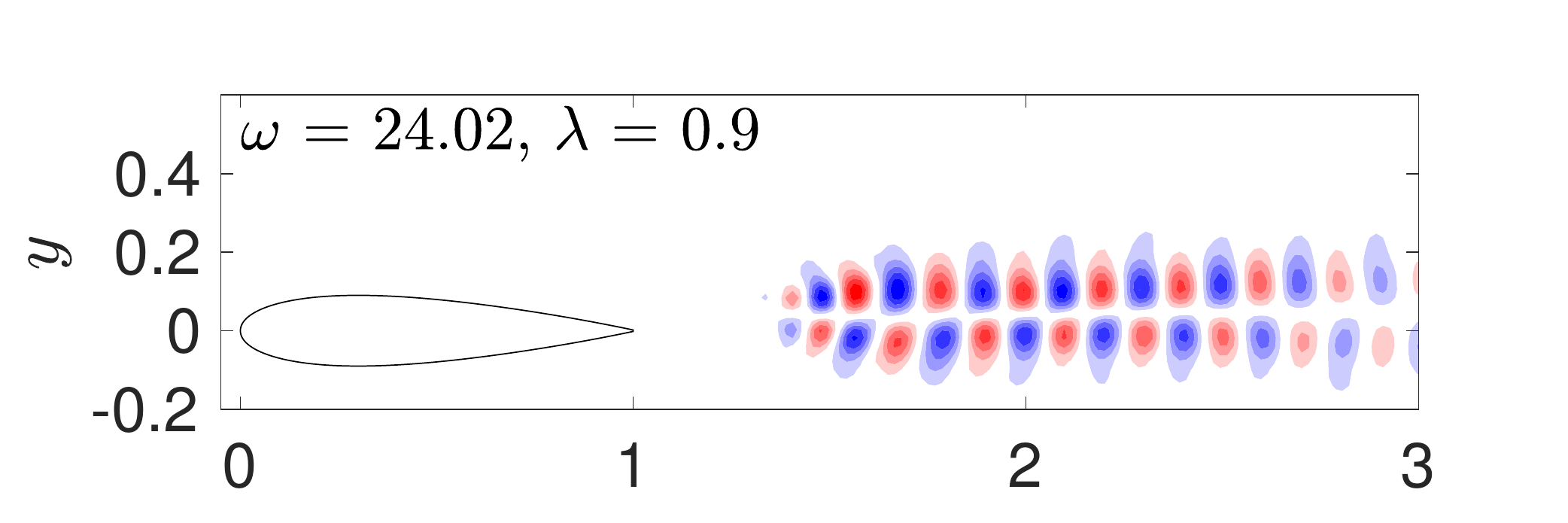}
	\includegraphics[scale=0.3]{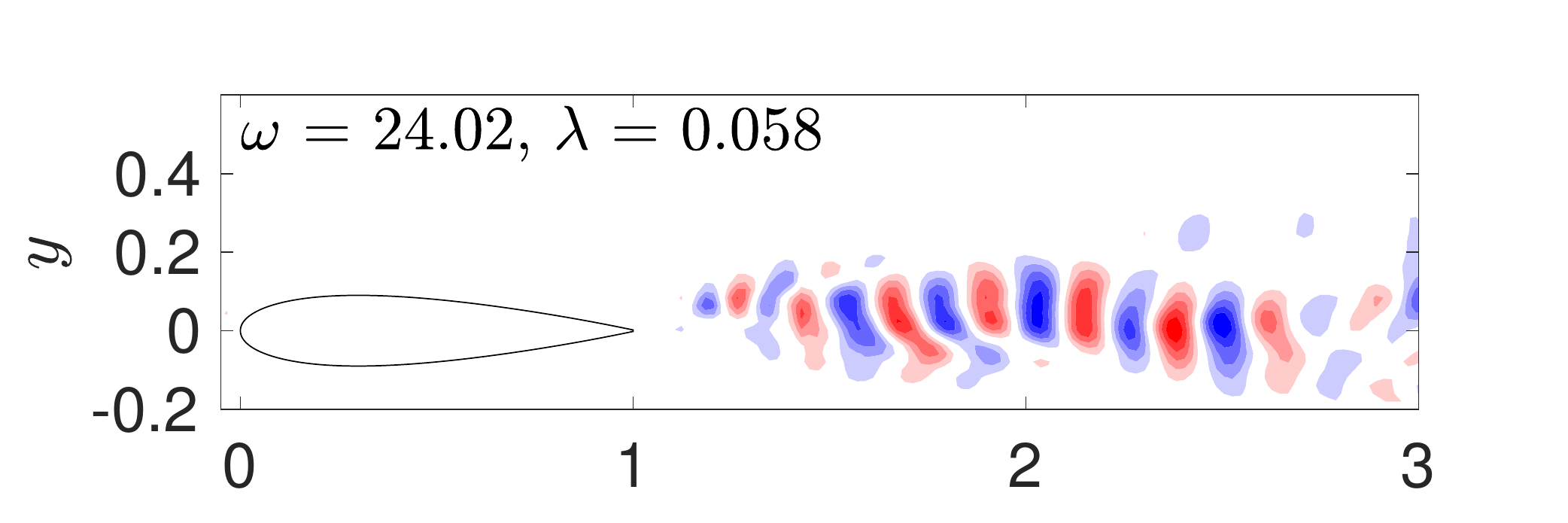}
	\includegraphics[scale=0.3]{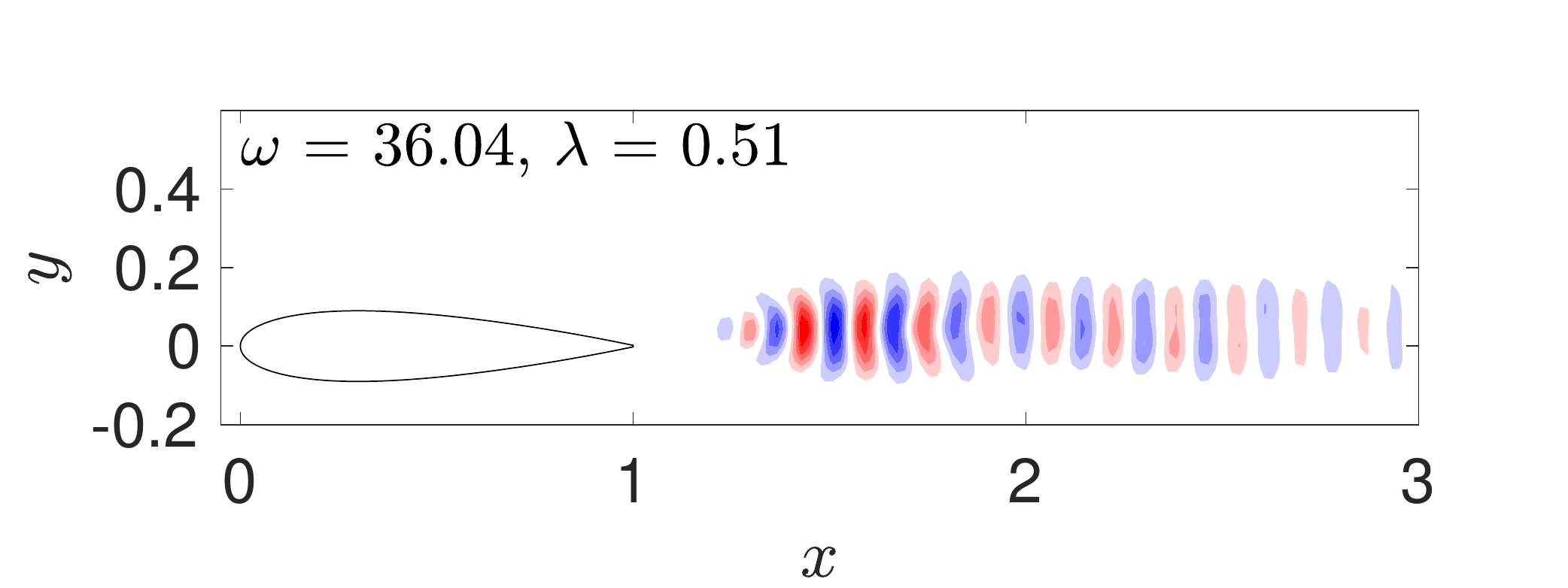}
	\includegraphics[scale=0.3]{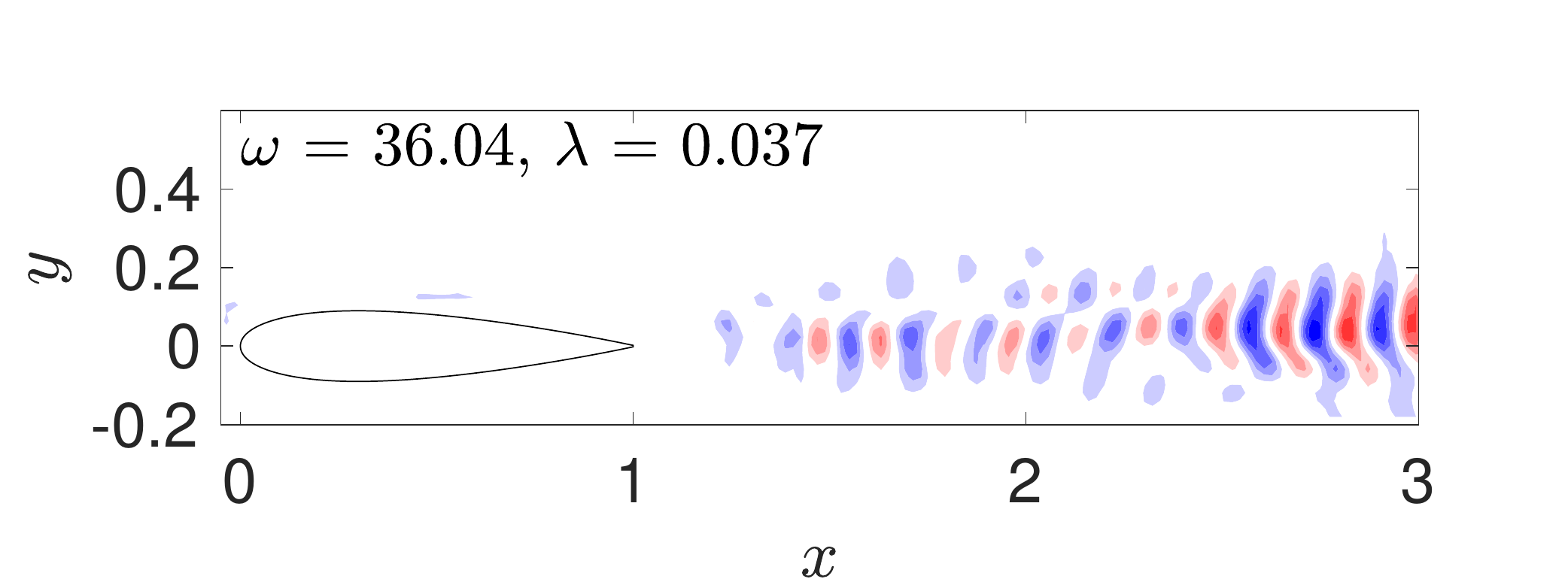}
	
	\caption{The first SPOD mode for the first three harmonics of the shedding frequency (left column) for the A0 case. The second SPOD mode for the first three harmonics of the shedding frequency (right column).}\label{fig:SPODA0}
\end{figure}

The $v$-component of the first two SPOD modes is shown in figure \ref{fig:SPODA0} for the first three harmonics. In all cases, the first eigenvalue is at least one order of magnitude larger than the second eigenvalue. The first SPOD mode is significantly more structured than the second mode for all three harmonics in figure \ref{fig:SPODA0}, an observation which is consistent with the fact that nearly all the energy at these frequencies is concentrated in the first mode. For frequencies where $\lambda_1 \approx \lambda_2$ (not shown in the interest of brevity), the mode shapes are less structured for all eigenvalues. 

Let us now consider the A10 case. The PIV data are subdivided into the same number of blocks containing the same number of snapshots as in the A0 case to allow for direct comparisons between both angles of attack. The SPOD eigenspectra are plotted in figure \ref{fig:SPODeigs}. It is immediately apparent that the SPOD eigenvalues are far less peaked than they were for the A0 case. Instead of observing a fundamental frequency and its harmonics, there are bands of frequencies where low-rank behaviour can be observed, a feature reminiscent of an amplifier flow. The first region, denoted by the red patch in figure \ref{fig:SPODeigs}, is situated around $4 < \omega < 8$ while the second one resides roughly between $10.5 < \omega < 14.5$. 

\begin{figure}
	\centering
	\includegraphics[scale=0.3]{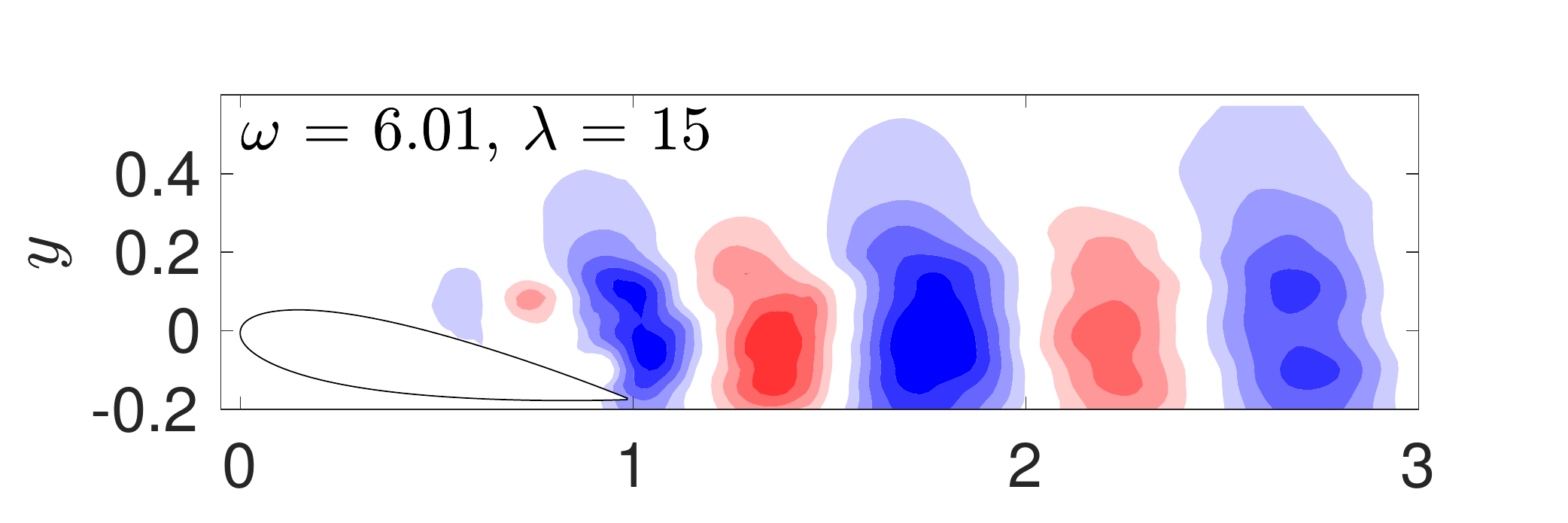}
	\includegraphics[scale=0.3]{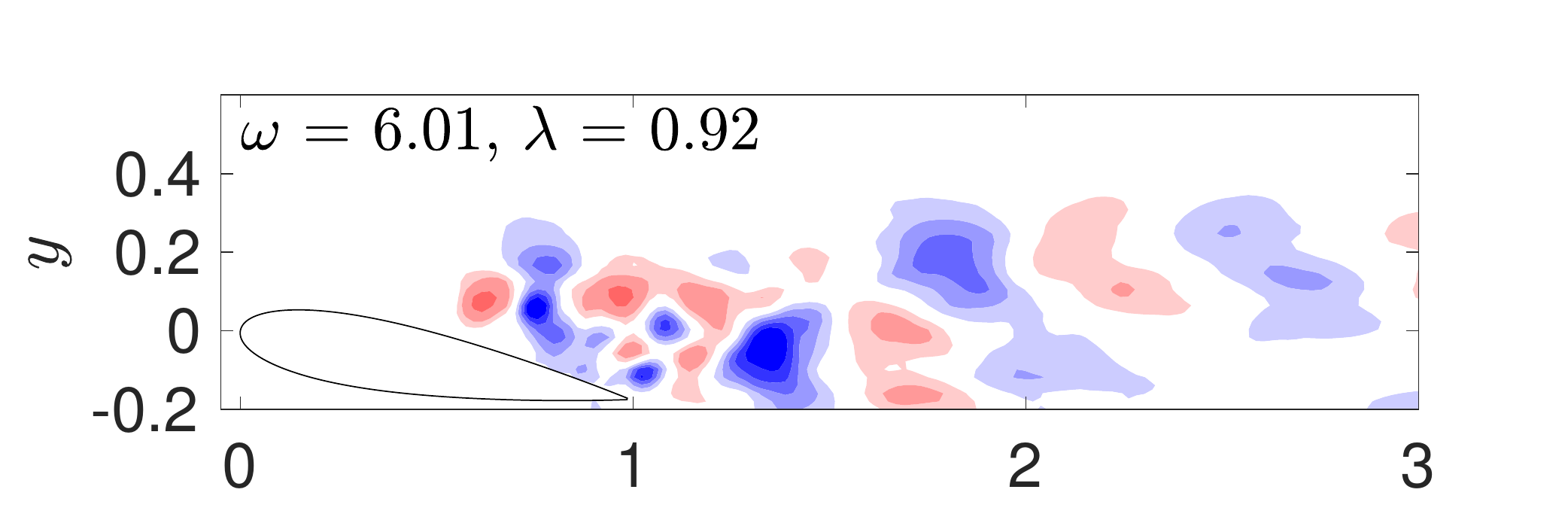}
	\includegraphics[scale=0.3]{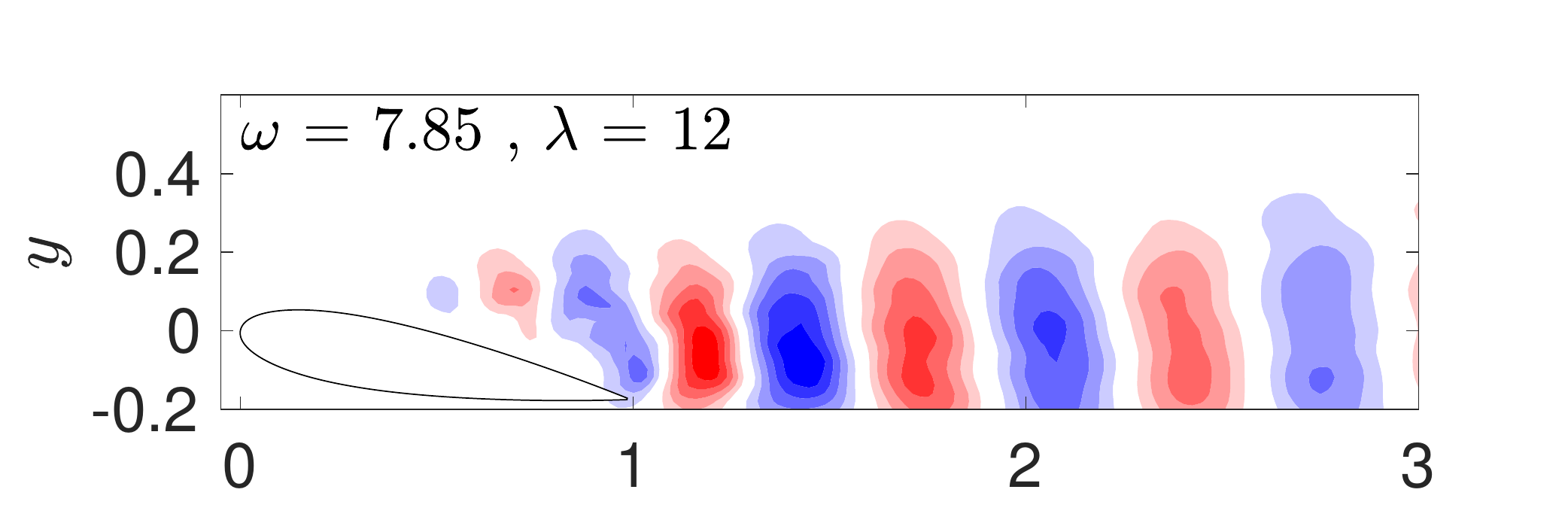}
	\includegraphics[scale=0.3]{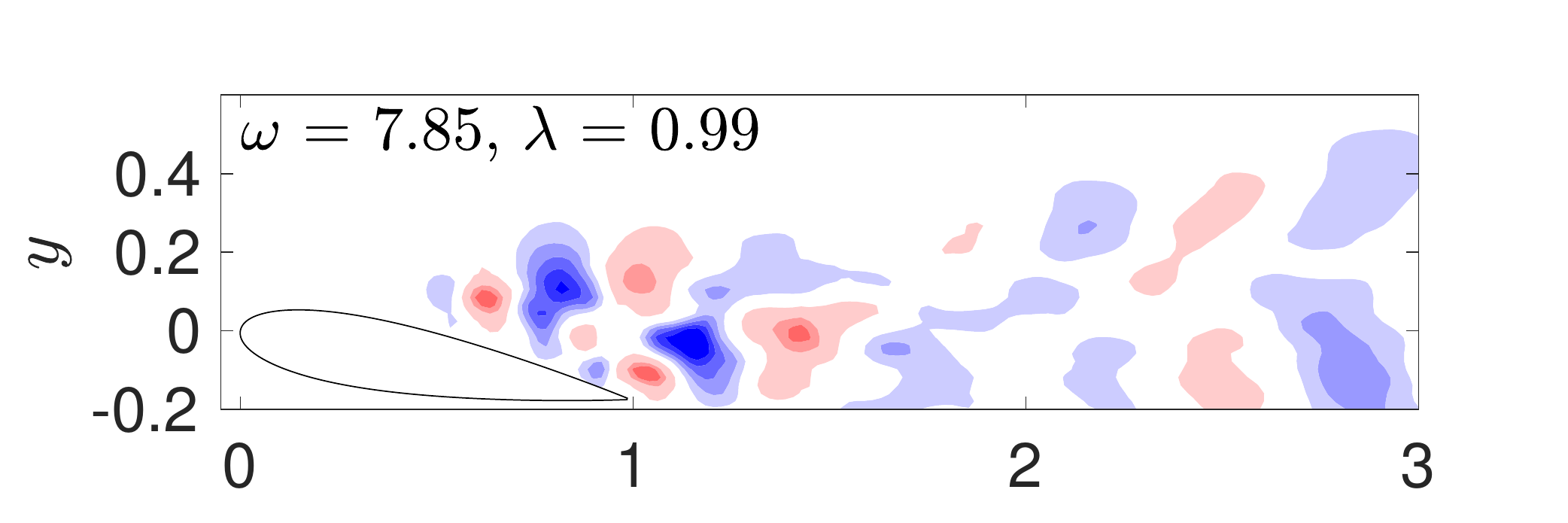}
	\includegraphics[scale=0.3]{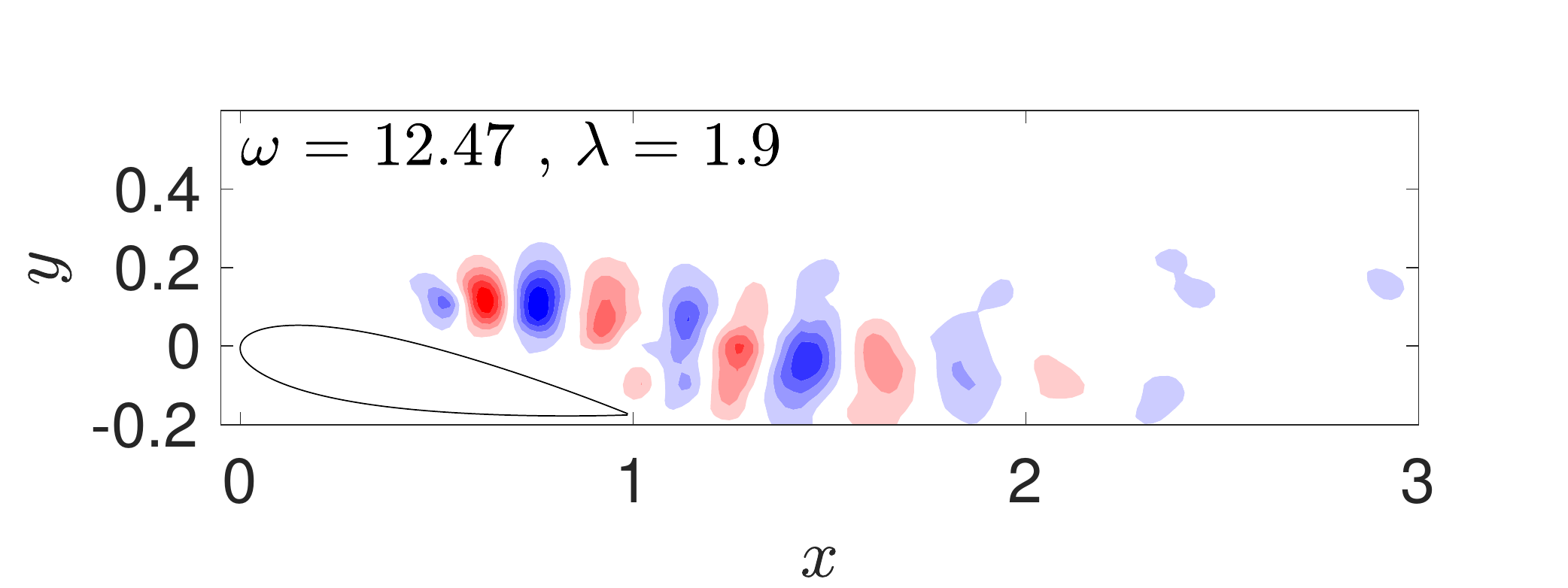}
	\includegraphics[scale=0.3]{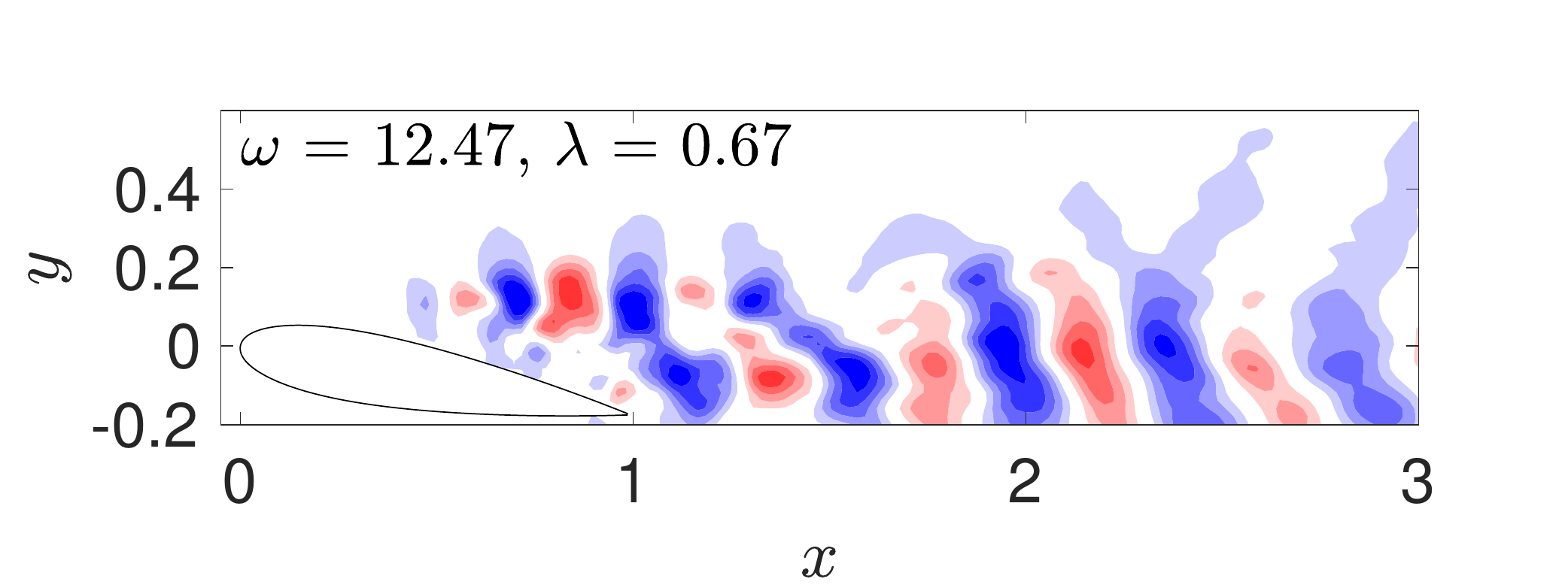}
	
	\caption{The first SPOD mode for frequencies corresponding to the wake or shear layer dynamics (left column) and the second SPOD mode for the same frequencies (right column) for the A10 case.}\label{fig:SPODA10}
\end{figure}

The modes for A10 are plotted in figure \ref{fig:SPODA10} for the highest eigenvalues in the low- and high-frequency bands. Similar to the A0 case, the leading SPOD mode is highly structured and contains about one order of magnitude more energy than the second mode for $\omega = 6.01$ and $\omega = 7.85$. Most of the energy in these modes, furthermore, resides in the wake region, or $x > 1$, although there is some structure in the shear layer. The second SPOD mode, on the other hand, is more energetic in the shear layer although it is far less structured. The final frequency considered is $\omega = 12.47$, which is roughly in the middle of the second band of frequencies identified earlier. The first mode is primarily in the shear layer while the second mode resembles a mix of both the shear layer and wake dynamics. Since SPOD takes into consideration the kinetic energy of the entire domain, the SPOD eigenvalues for frequencies corresponding to the shear layer assume lower values. They are dynamically important, nonetheless, since there is non-negligible separation between the first and second eigenvalue. They are also essential to model the fluctuations above the airfoil.  

\subsection{SPOD of the nonlinear forcing}

\begin{figure}
	\centering
	\includegraphics[scale=0.3]{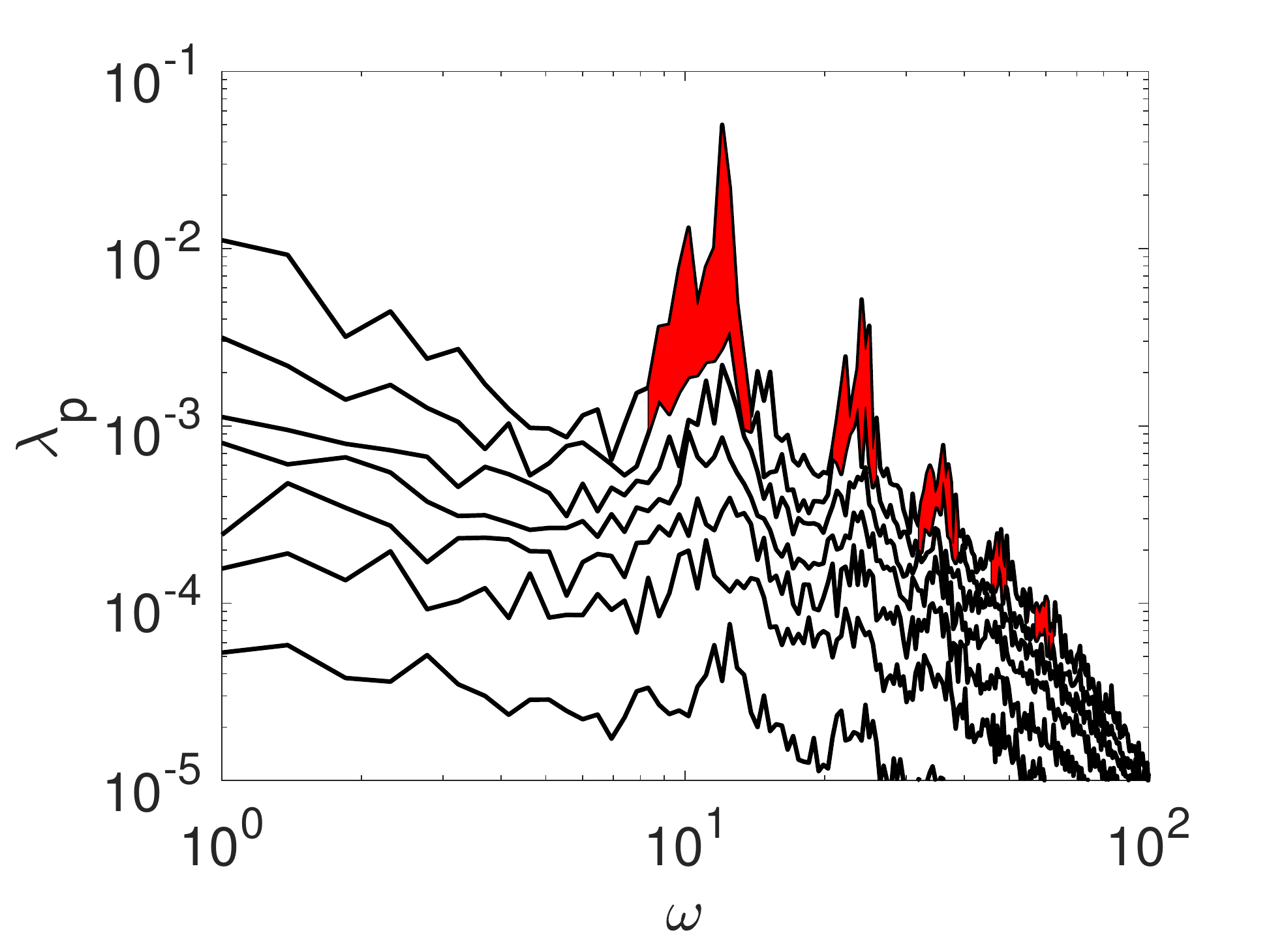}
	\includegraphics[scale=0.3]{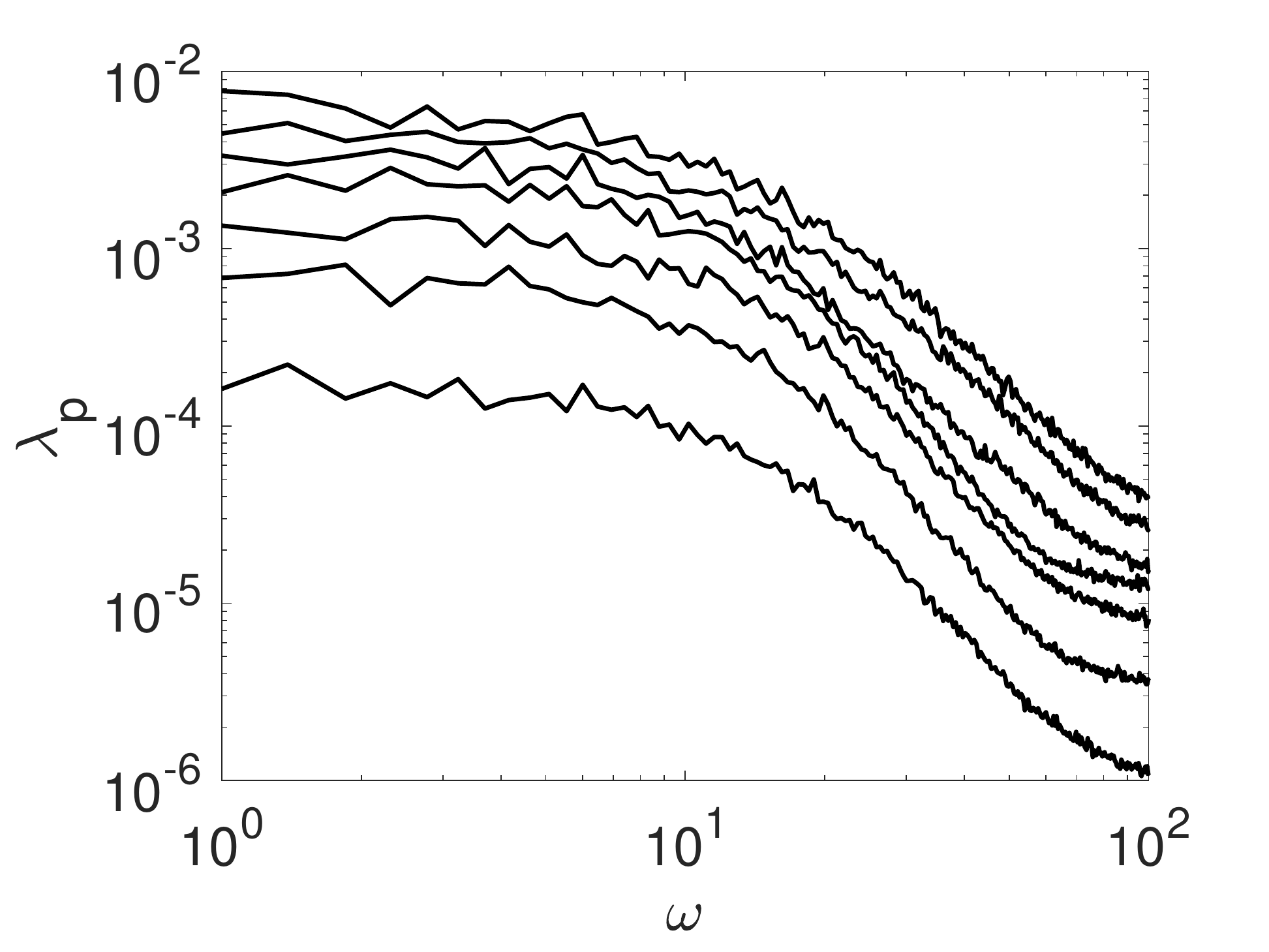}
	\caption{SPOD eigenspectra of the nonlinear forcing for the A0 (left) and A10 (right) cases. The red patches in the A0 case delineate the shedding frequency and its harmonics where there is separation between the first and second eigenvalues.} \label{fig:NLSPOD}
\end{figure}

We now utilise SPOD to characterise the structure of the nonlinear forcing, which is computed from the PIV snapshots since the data are sufficiently well-resolved in space and time. The SPOD eigenspectra for both cases (A0 \& A10) are plotted together in figure \ref{fig:NLSPOD} to underscore the contrast between the two cases. The eigenspectra for A0 are still peaked in the sense that the energy is high at the vortex shedding frequency and its harmonics. There is also a considerable gap between the first and second eigenvalues at these frequencies, similar to the trends observed for the velocity fluctuations. The spectra for the A10 case, on the other hand, are relatively flat and have no discernible peaks. The behaviour of the nonlinear forcing bears no resemblance to those of the velocity fluctuations as seen in figure \ref{fig:SPODA10}.

\begin{figure}
	\centering
	\includegraphics[scale=0.3]{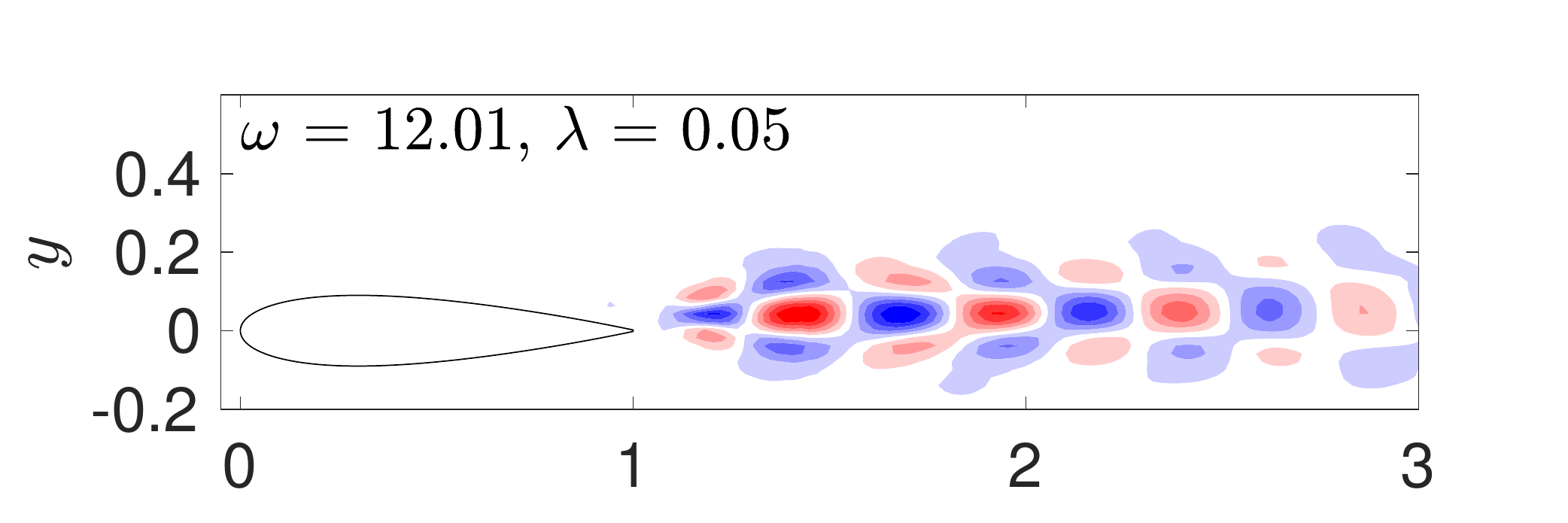}
	\includegraphics[scale=0.3]{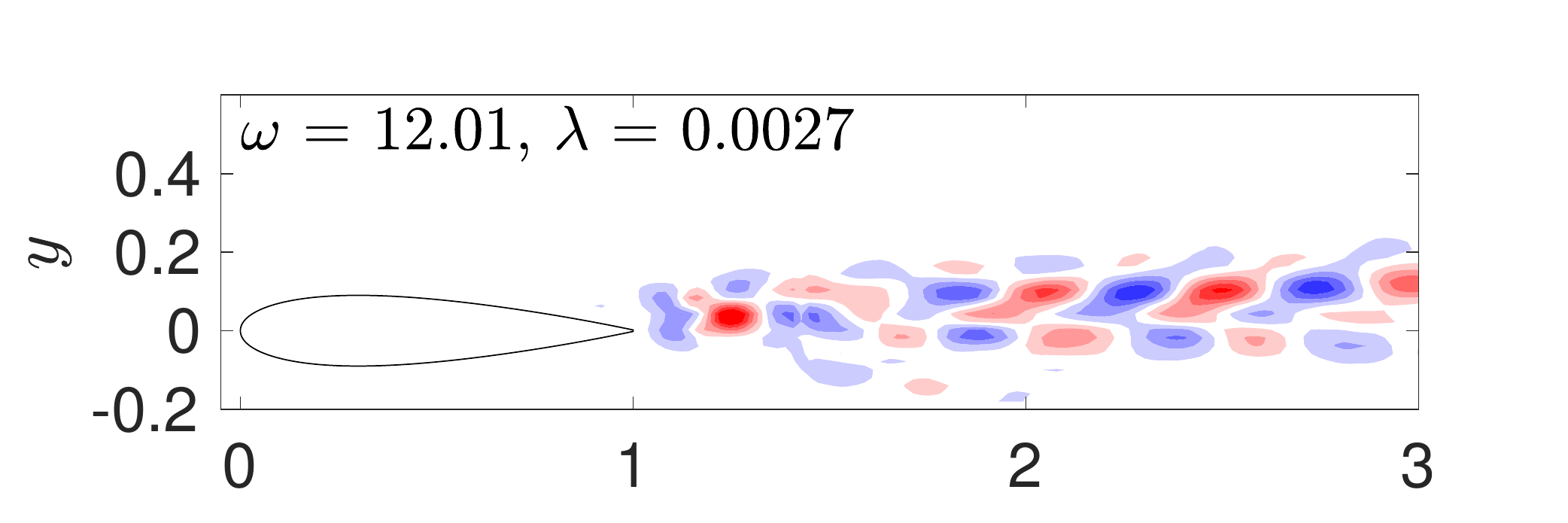} 	
	\includegraphics[scale=0.3]{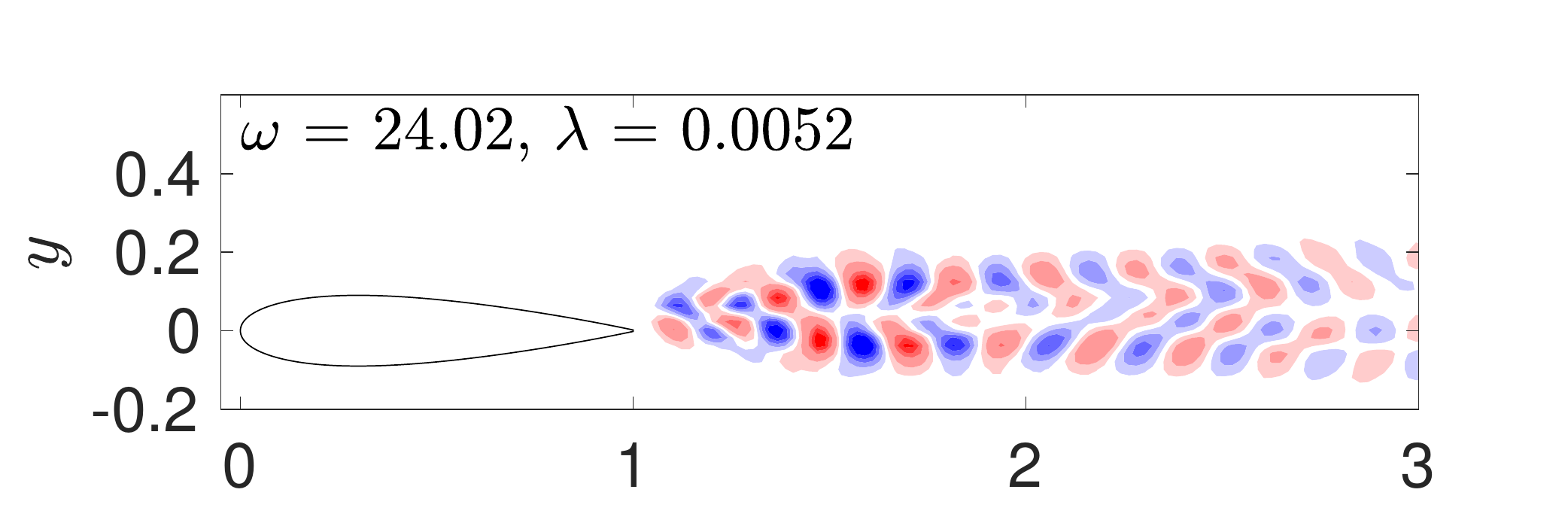}
	\includegraphics[scale=0.3]{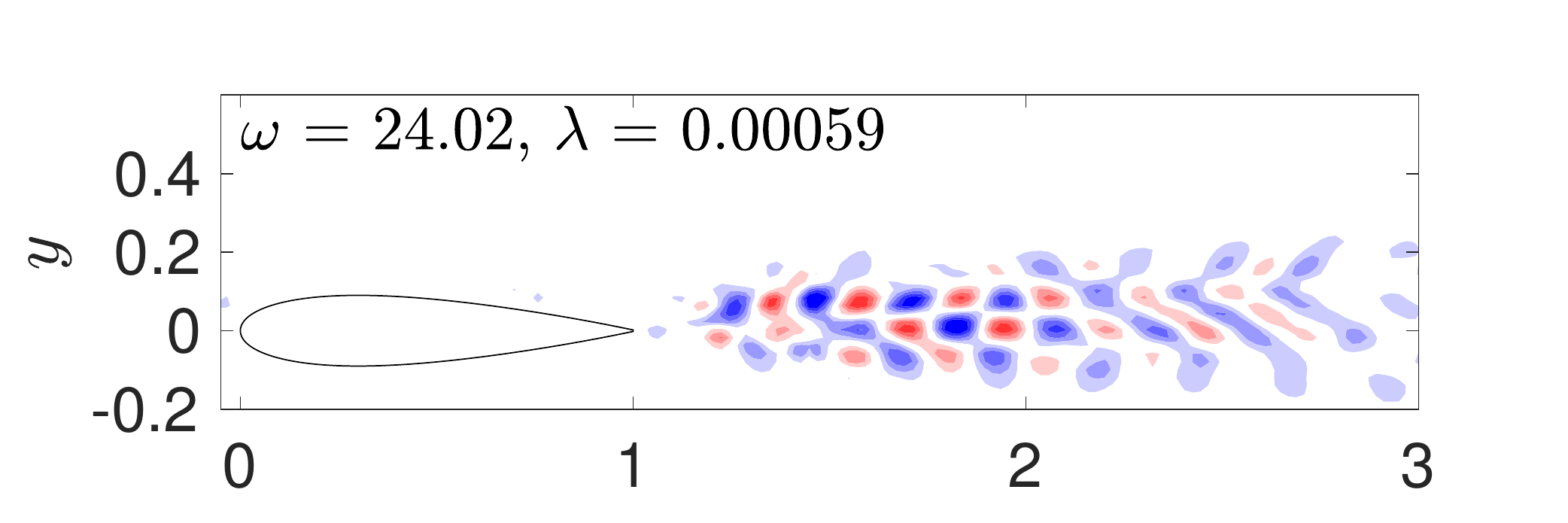} 
	\includegraphics[scale=0.3]{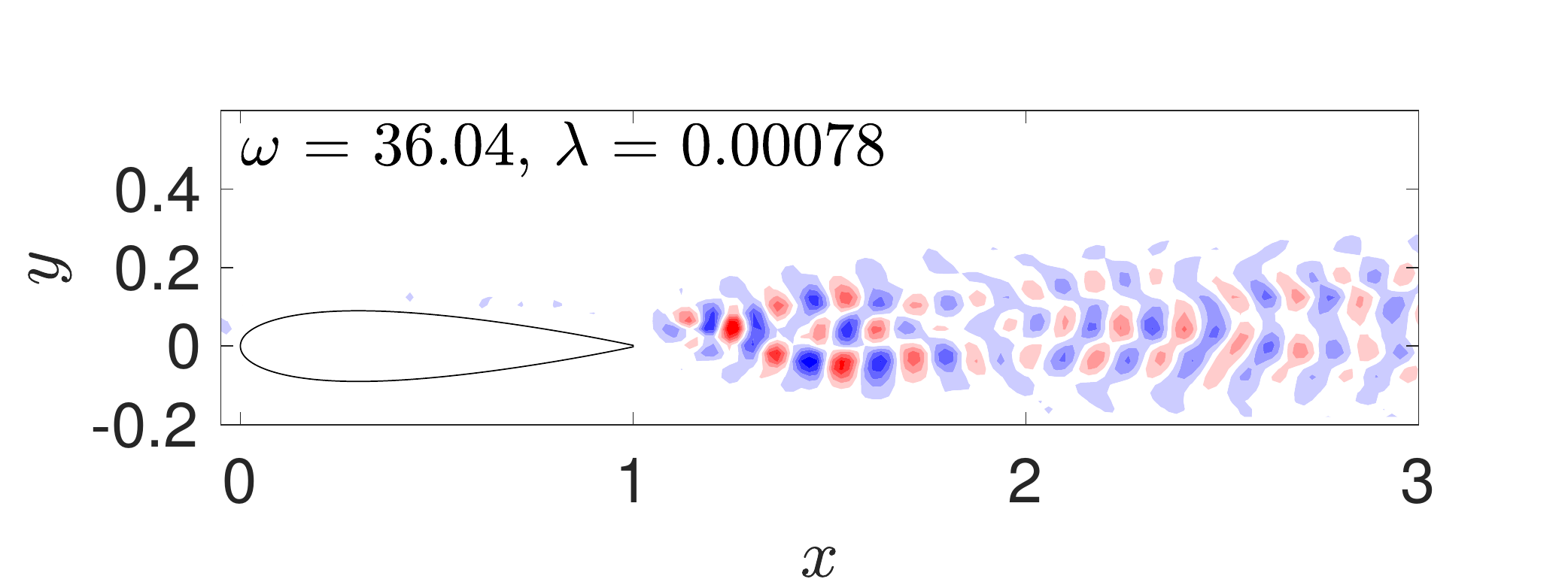}
	\includegraphics[scale=0.3]{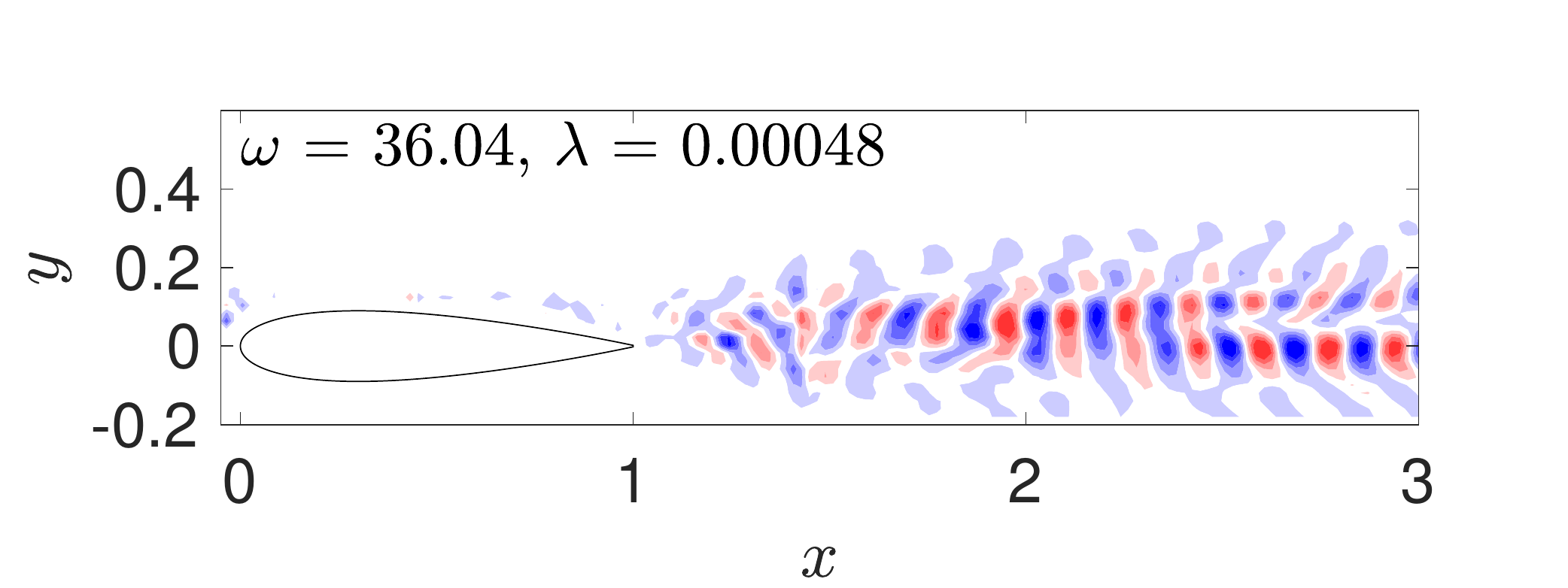} 
	
	\caption{The first (left column) and second (right column) SPOD mode for the nonlinear forcing of the A0 case.}\label{fig:NLFA0}
\end{figure}

The differences are even more evident when considering the mode shapes. For clarity, we will plot modes at the same frequencies as those in figures \ref{fig:SPODA0} and \ref{fig:SPODA10} since they are dynamically important. The modes for A0 are plotted in figure \ref{fig:NLFA0}. It is clear that the nonlinear forcing for the first two harmonics is very structured, as the contours are almost perfectly symmetric (or anti-symmetric in the case of the second harmonic) with respect to the centreline. The structure corresponding to the third harmonic looks slightly more disordered, since it is not particularly energetic, but it can be remarked that the mode shape is approximately symmetric with respect to the centreline and has a well-defined wavelength which is approximately constant, i.e. not changing as a function of $x$. 

\begin{figure}
	\centering
	\includegraphics[scale=0.3]{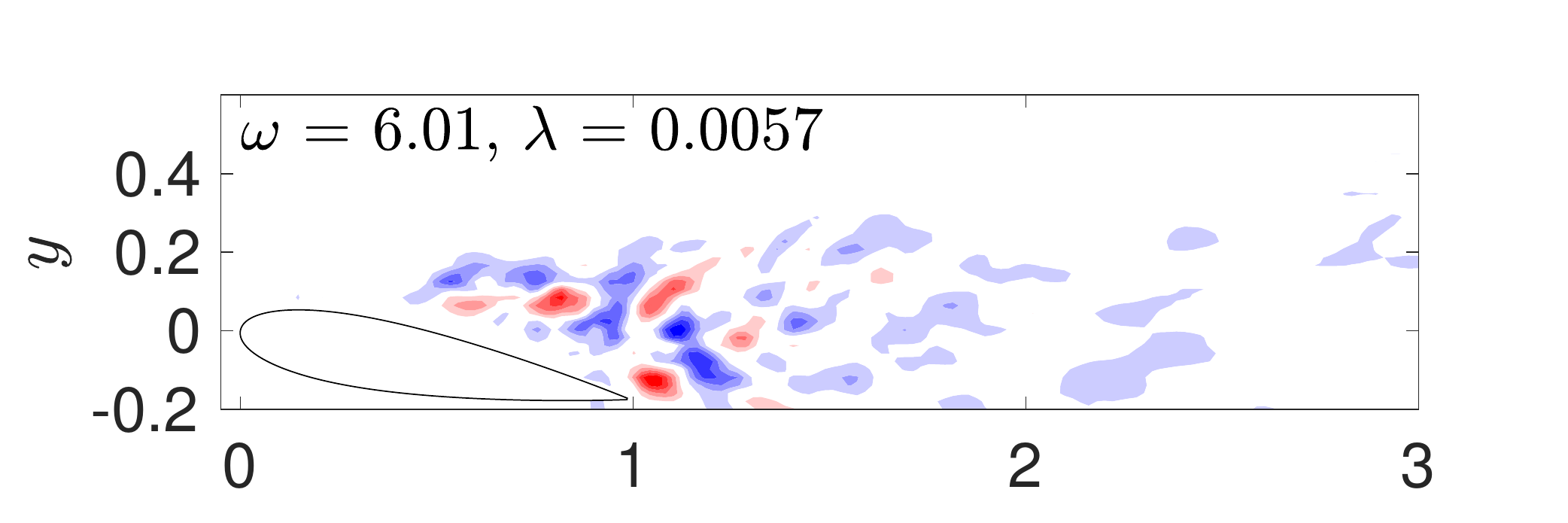}
	\includegraphics[scale=0.3]{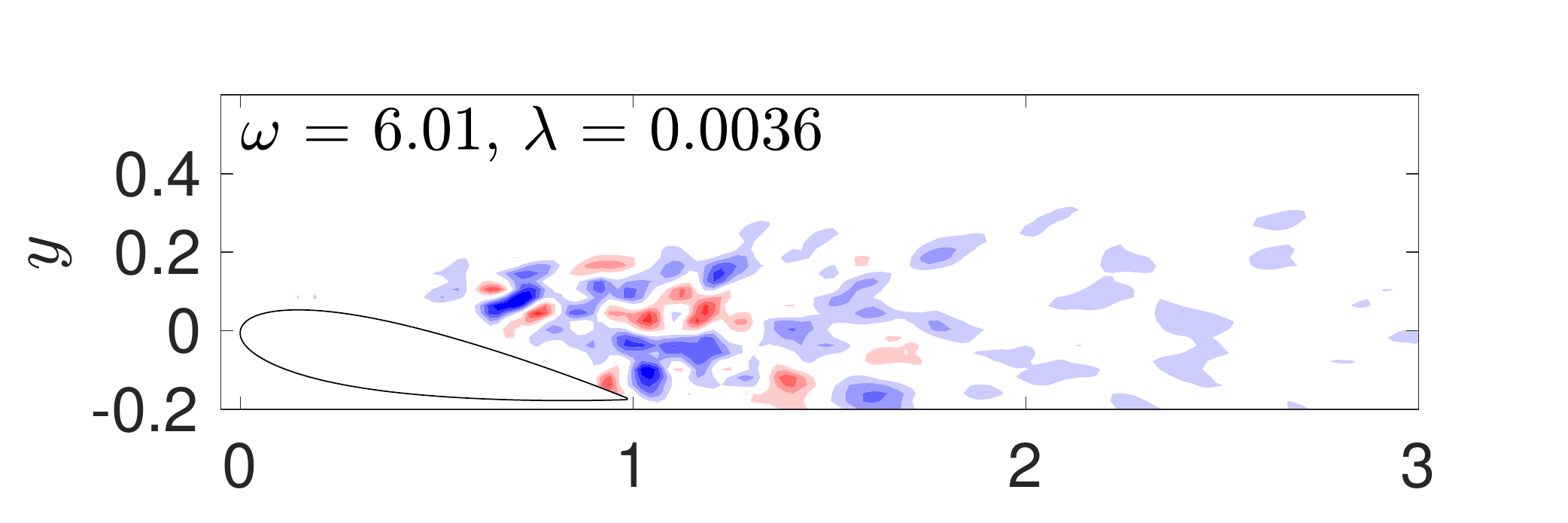} 
	\includegraphics[scale=0.3]{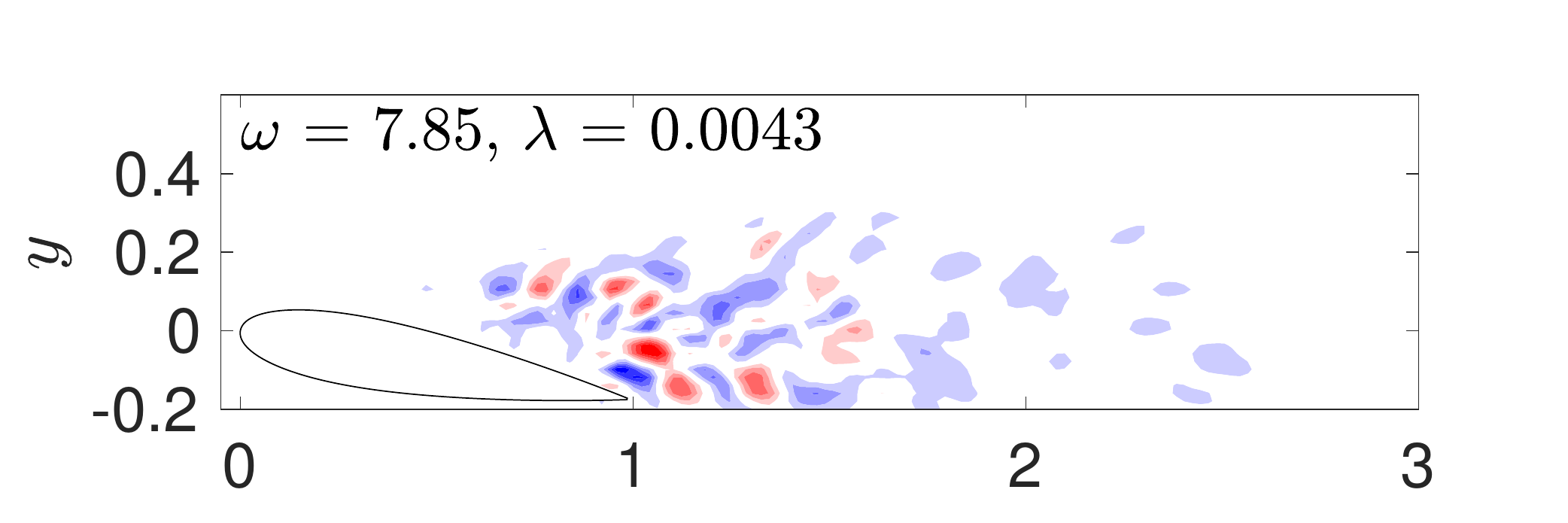}
	\includegraphics[scale=0.3]{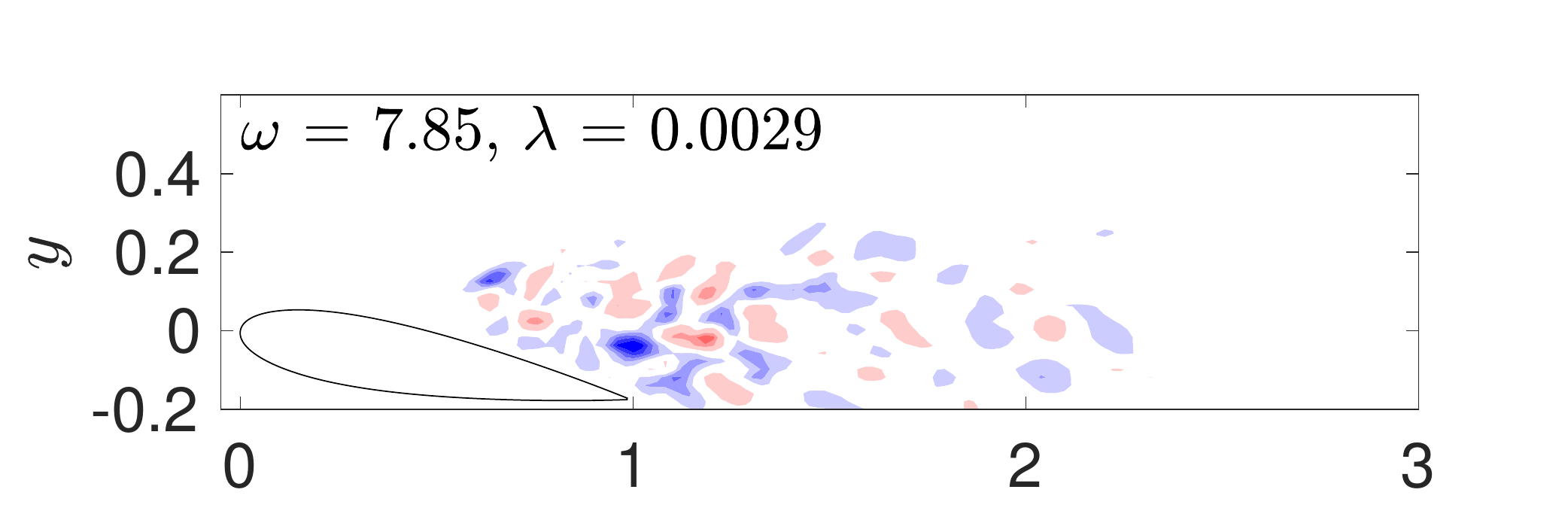} 
	\includegraphics[scale=0.3]{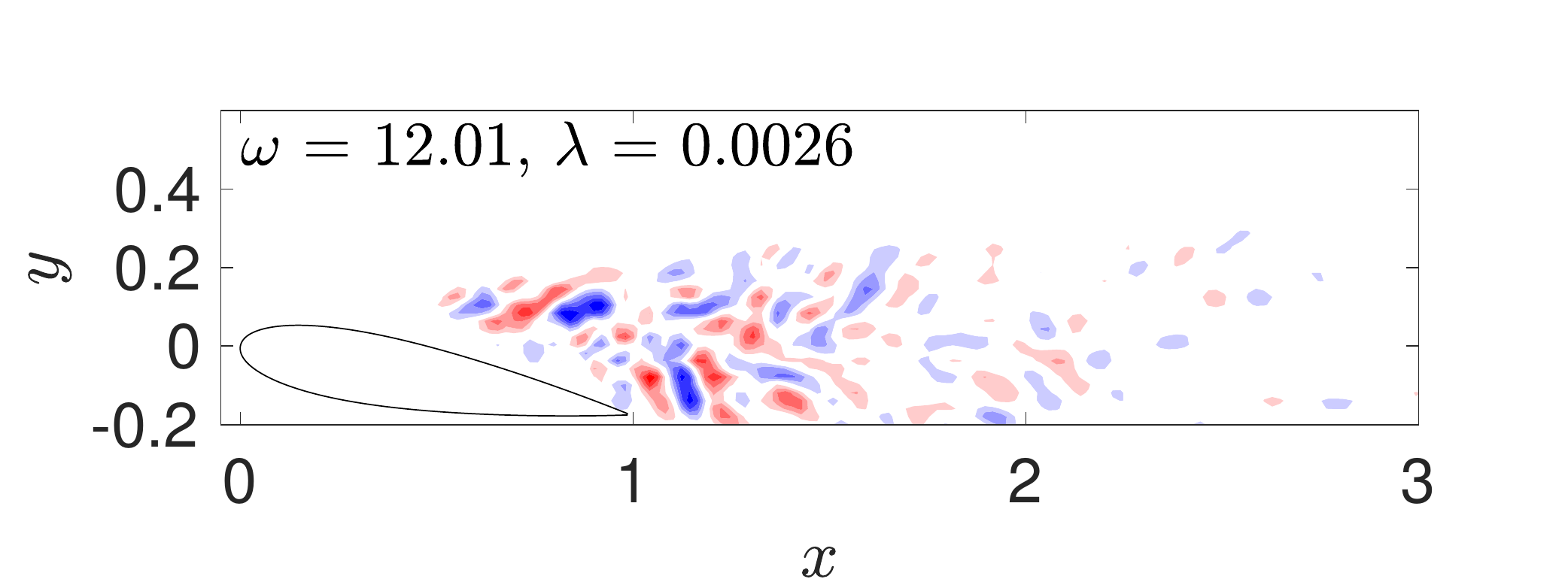}
	\includegraphics[scale=0.3]{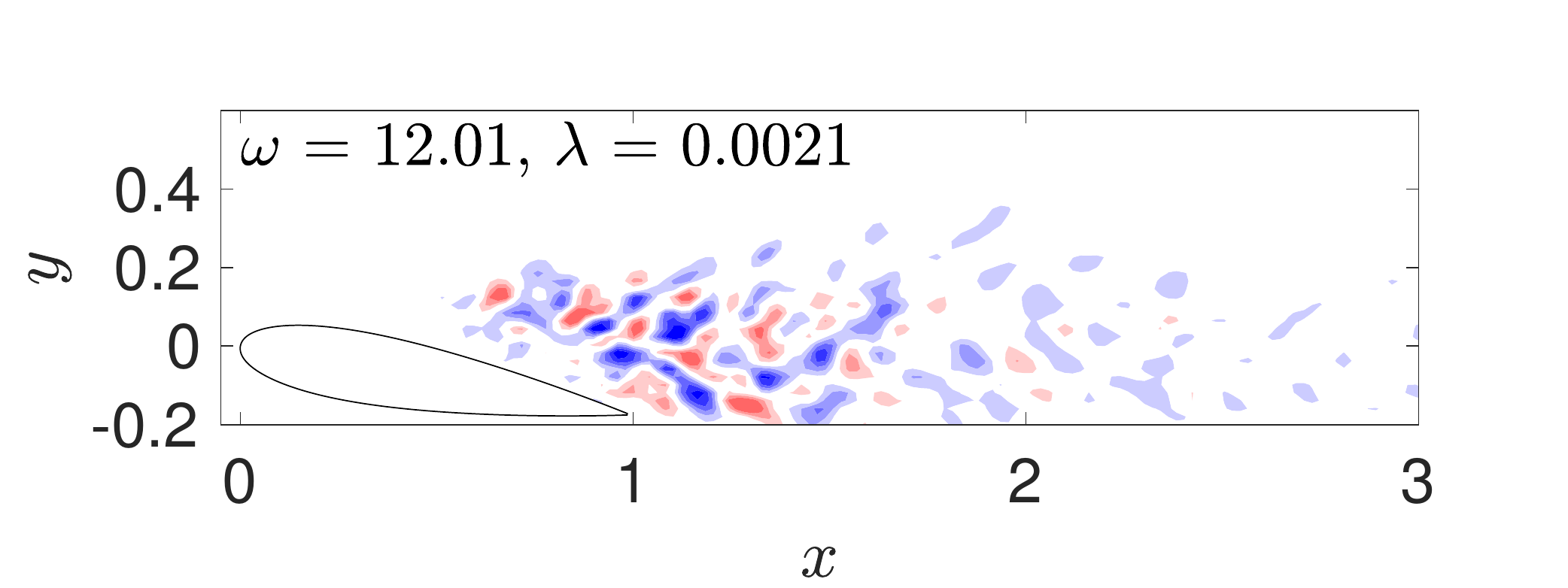} 
	
	\caption{The first (left column) and second (right column) SPOD mode for the nonlinear fluctuations of the A10 case.}\label{fig:NLFA10}
\end{figure}

The mode shapes for the A10 case are plotted in figure \ref{fig:NLFA0}. The results are reminiscent of those from the jet flow study of \cite{Towne15} in that there is a distinct lack of spatial coherence. While there may be some notion of a wavelength associated with these structures, it would be far more difficult to quantify in comparison to the results in figure \ref{fig:NLFA0}. There is also no noticeable difference between the spatial support of the lower frequency modes in figure \ref{fig:NLFA10} and that of the higher frequency mode even though the choice of frequency had an impact on the spatial location and distribution of the velocity fluctuations. 

\subsection{Discussion and classification of the two cases}

SPOD of the velocity fluctuations reveals low-rank behaviour for both the A0 and A10 cases although there are some major differences. The SPOD eigenspectra of A0 are highly peaked at the vortex shedding frequency and its harmonics. The A0 case, thus, can be considered sparse in that there are few energetic frequencies in the flow. The SPOD eigenspectra for A10 is more broadband, with separation between the first and second eigenvalues occurring at two distinct ranges of frequencies making the flow less sparse, i.e. more frequencies are required to model the fluctuating flow field. Within the lower range, the SPOD modes are concentrated in the wake whereas within the higher range, they are concentrated in the shear layer. Based on these trends alone, it seems that the A0 case resembles an oscillator-type flow while the A10 case behaves more like an amplifier since, in the case of an oscillator, the fluctuations are characterised by a well-defined frequency which can be readily identified in figure \ref{fig:SPODA0}. There is no clear peak in figure \ref{fig:SPODA10}.

Application of SPOD to the nonlinear forcing confirms this appraisal. In the case of A0, there is low-rank behaviour at all harmonics and the mode shapes are highly structured as they contain a well-defined wavelength and symmetry. This suggests they can be represented by a limited number of nonlinear interactions whereas in the A10 case, there is no apparent low-rank behaviour and the forcing field contains no coherent structure. These results are consistent with those of \cite{Rosenberg19}, who show that the forcing for the cylinder wake is highly structured at the shedding frequency and its harmonics, as well as \cite{Towne15}, who illustrate that the nonlinear forcing is comprised of incoherent turbulent motions for a turbulent jet. It is also in tune with the idea that the input to an amplifier is less important, as the linear dynamics have a preferential bias towards one particular structure. 

In theory, this makes modelling of amplifier flows easier as the linear dynamics will do the work for us while nonlinearity needs to be taken into account for oscillator flows. Consequently, amplifier flows are more amenable to stochastic forcing of an approximated linear operator. Alternatively, the modelling of oscillator flows may benefit from approximating the nonlinear forcing due to the emergence of parasitic modes. These ideas are tested in \S\ref{sec:modeling}.

\section{Resolvent analysis of the data-assimilated profiles}\label{sec:resolvent analysis}

For each angle of attack, two mean profiles are used as an input to resolvent analysis. The first is from the raw experimental mean, which is interpolated onto the same FreeFem mesh used in the data-assimilation algorithm. Due to the sparsity of data near the surface of the airfoil, the interpolated mean profile is not likely to satisfy the no-slip boundary condition, so this is manually enforced by setting the velocity at all points along the airfoil surface to zero. The points outside the PIV domain, where velocity vectors are not available, are set to uniform flow, i.e. $\overline{\boldsymbol{u}} = (1,~0)$. This may seem like an oversimplification, but it has no discernible impact on the shape of the resolvent modes in the region where the experimental mean is known. The second profile is the data-assimilated mean, whose results are compared to those of the interpolated experimental mean in order to reinforce the benefits of data-assimilation before performing resolvent analysis. This section is divided into three parts: the results for the A0 and A10 cases are discussed in \S\ref{sec:A0resolvent} and \S \ref{sec:A10resolvent}, respectively. In \S\ref{sec:comparison}, the results from resolvent analysis are compared to those from SPOD and we remark on the different behaviours observed for an oscillator versus an amplifier flow. 

\subsection{A0 case: a single linear mechanism} \label{sec:A0resolvent}

\begin{figure}
	\centering
	\includegraphics[scale=0.3]{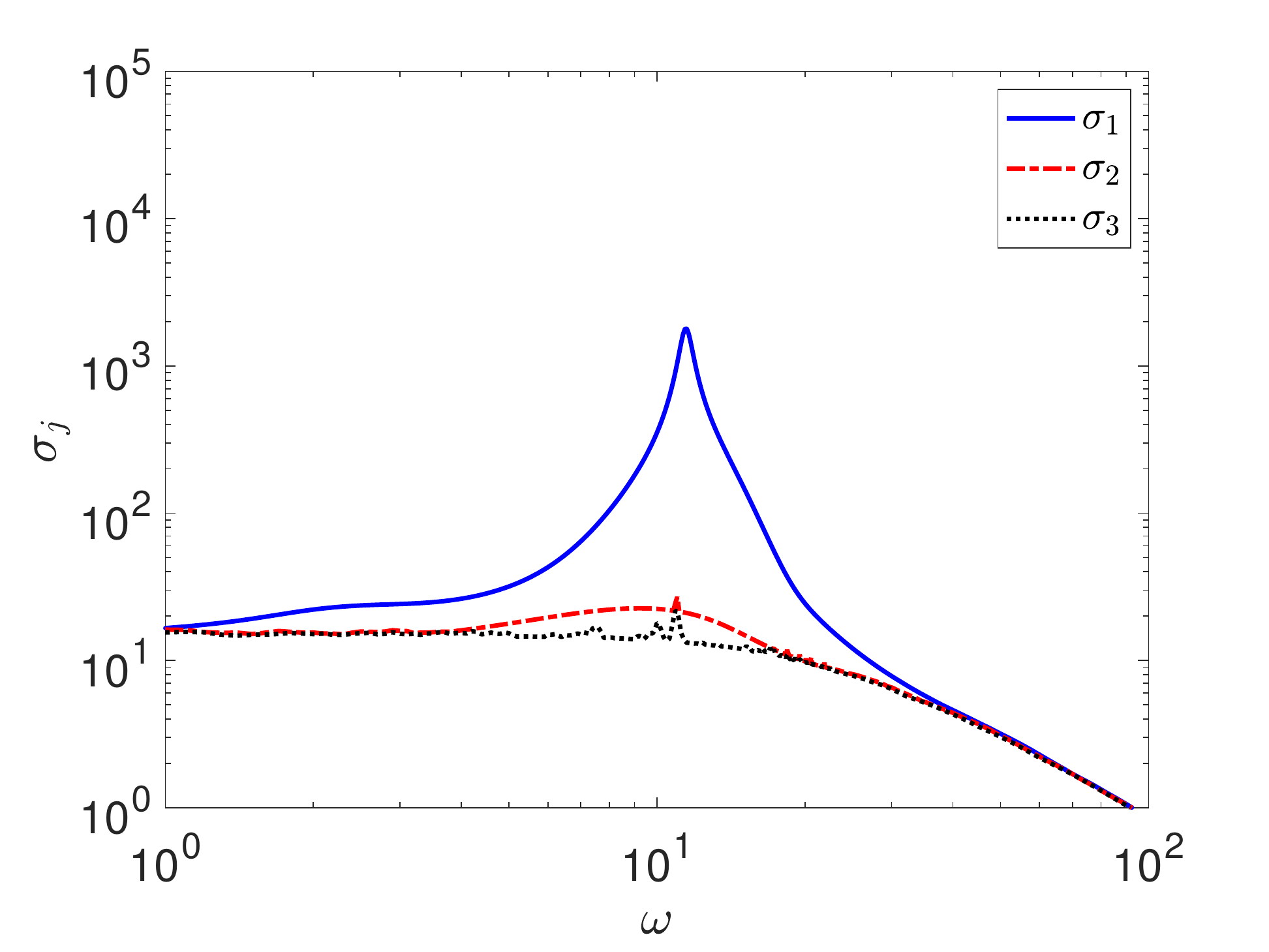}
	\includegraphics[scale=0.3]{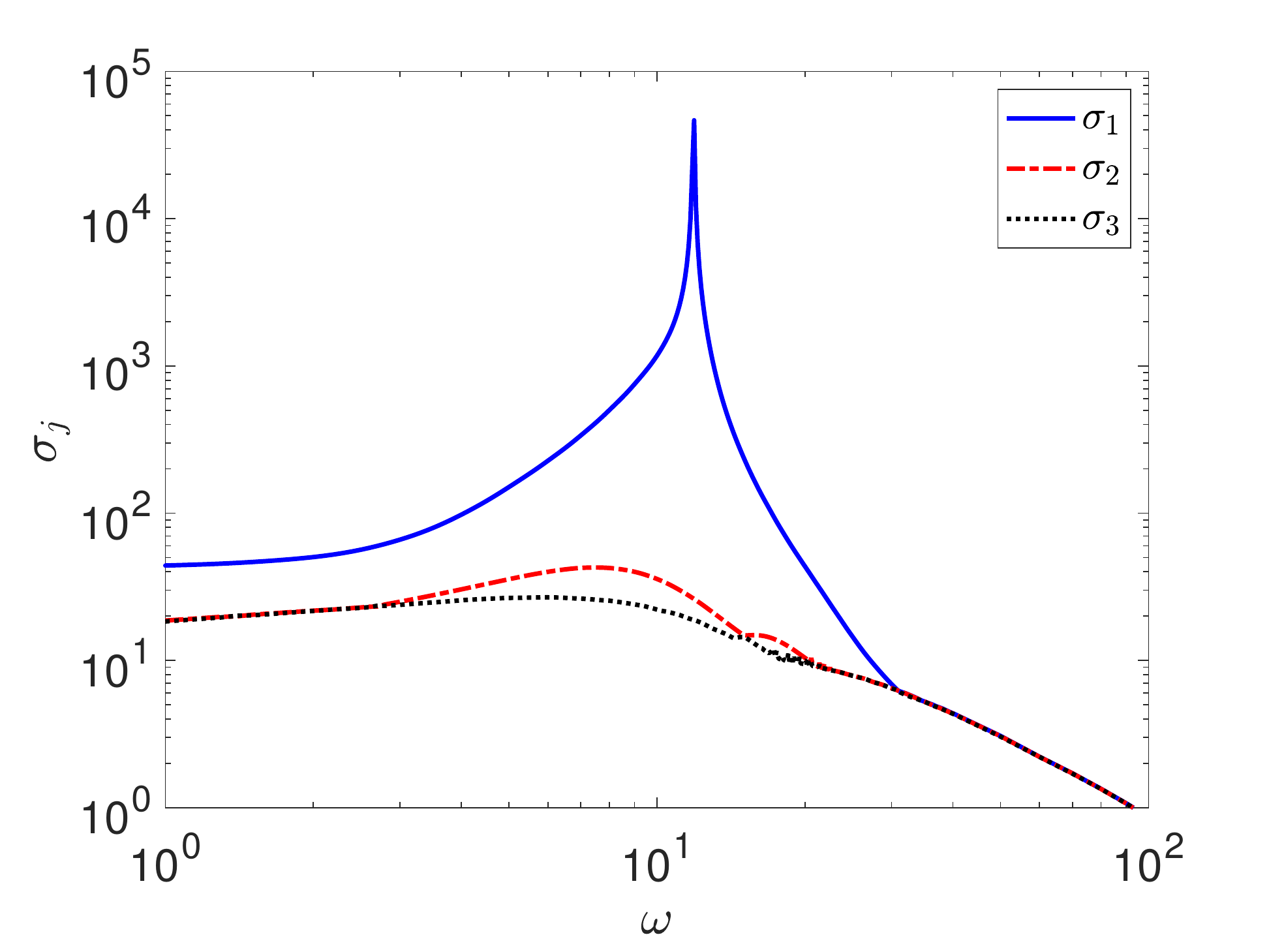}
	
	\caption{The first three singular values of the resolvent operator plotted as a function of $\omega$ for the experimental (left) and data-assimilated (right) mean profiles of the A0 case.}\label{fig:svalsA0}
\end{figure}

The first three singular values of the resolvent operator are plotted in figure \ref{fig:svalsA0} using both the interpolated experimental and data-assimilated means. The trends for both cases are quite similar to cylinder flow \citep{Symon18} in that there is a distinct range of frequencies where the first singular value is an order of magnitude higher than the second singular value. Data-assimilating the mean profile has a negligible impact on the absolute and relative magnitudes of the singular values with respect to one another. There is also no impact on the mode shapes \citep[see][]{Symon18b}. 

\subsection{A10 case: multiple linear mechanisms} \label{sec:A10resolvent}

\begin{figure}
	\centering
	\includegraphics[scale=0.3]{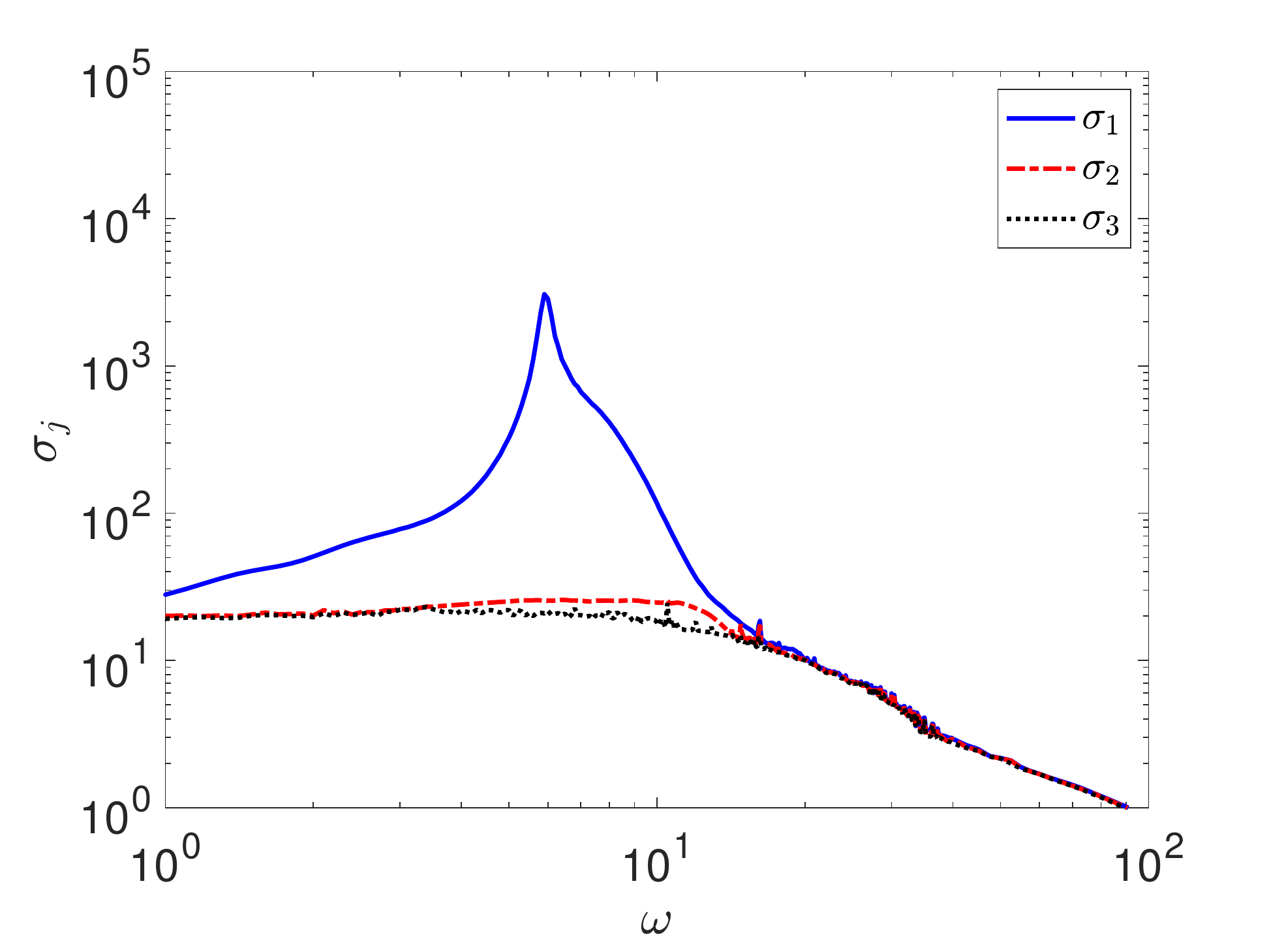}
	\includegraphics[scale=0.3]{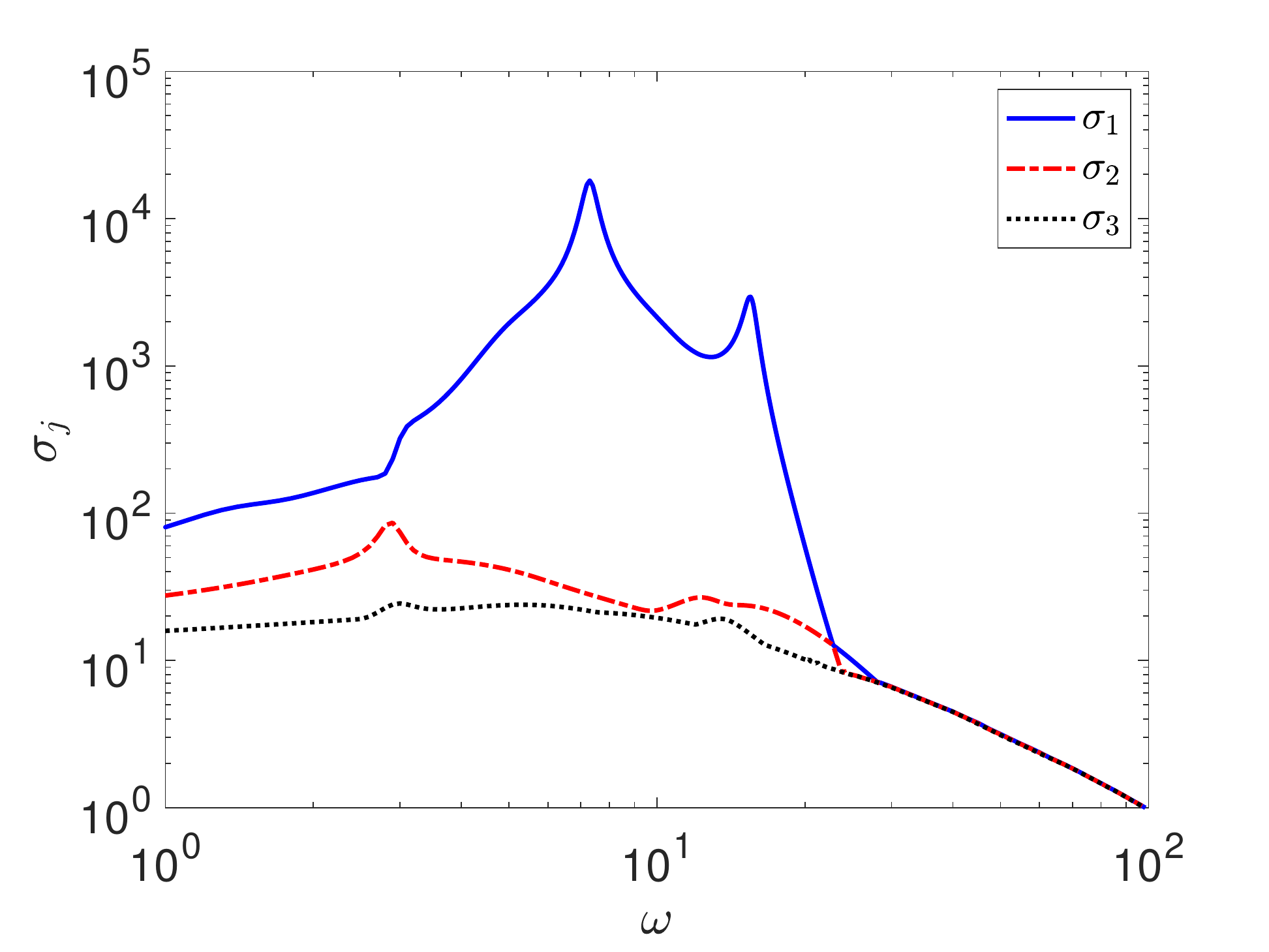}
	
	\caption{The first two singular values of the resolvent operator plotted as a function of $\omega$ for the experimental (left) and data-assimilated (right) mean profiles of the A10 case.}\label{fig:svalsA10}
\end{figure}

Unlike the A0 case, the singular values, illustrated in figure \ref{fig:svalsA10}, of the resolvent operator for the A10 case depend strongly on the mean profile. The interpolated mean has a single peak at $\omega = 5.9$ whereas the data-assimilated mean has two peaks at $\omega = 7.3$ and $\omega = 15.4$. The frequency range over which the resolvent is low-rank is also much wider for the data-assimilated mean. 

\begin{figure}
	\centering
	\includegraphics[scale=0.3]{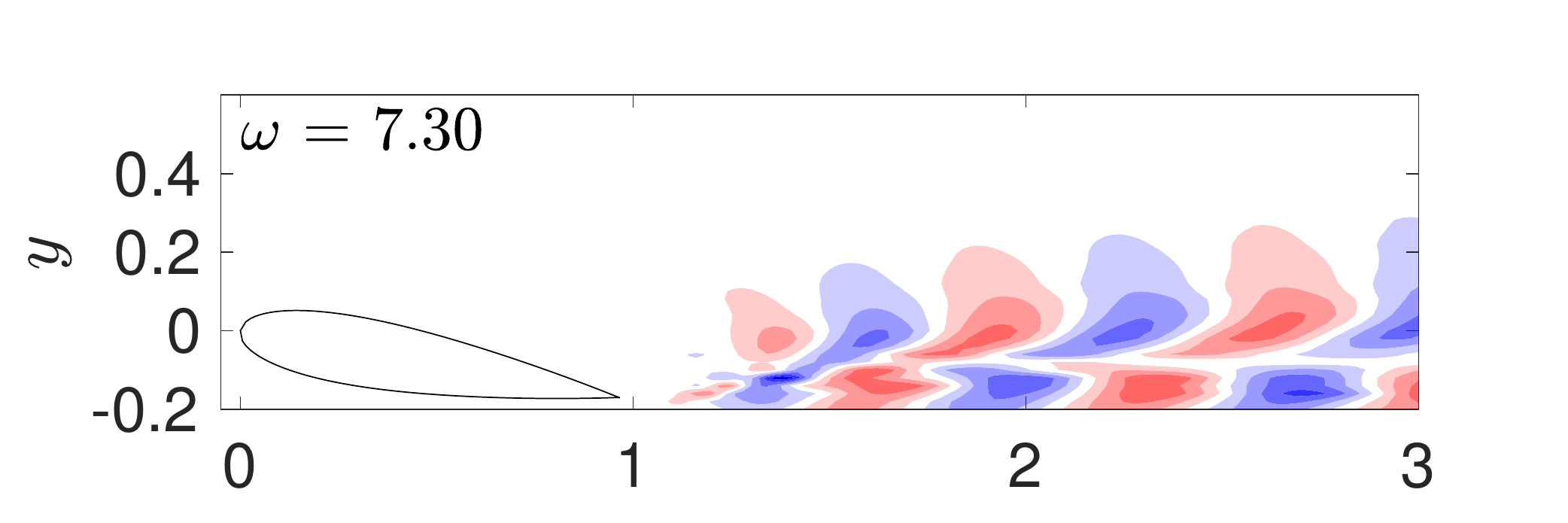}
	\includegraphics[scale=0.3]{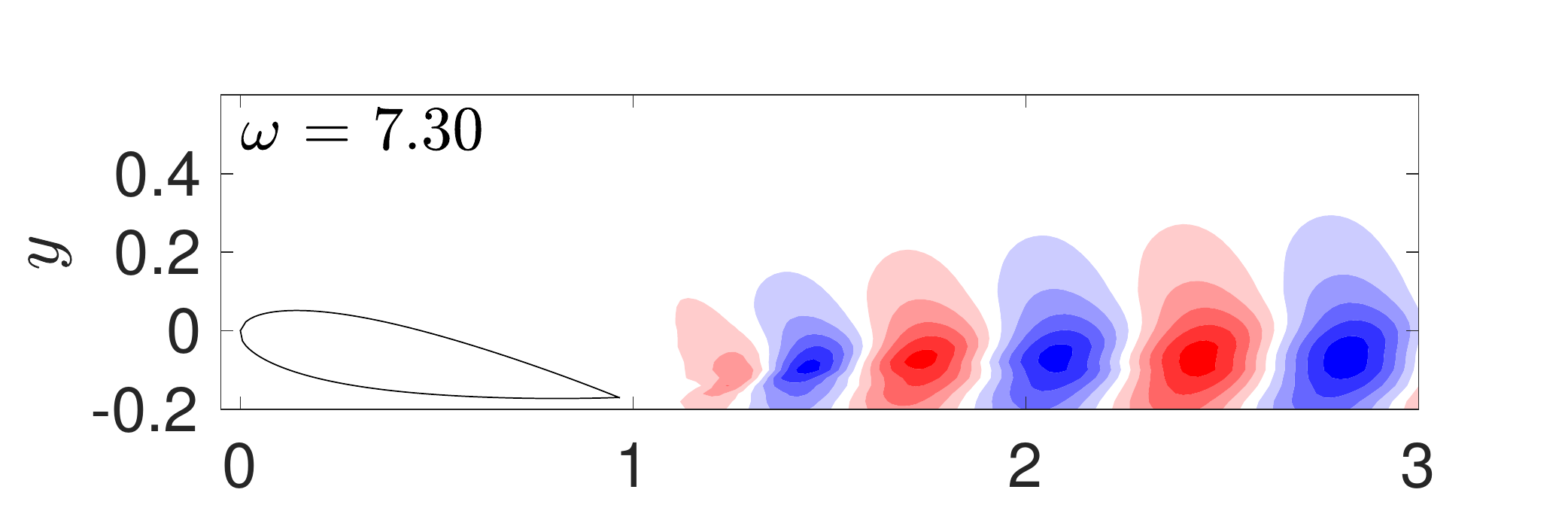}
	
	\includegraphics[scale=0.3]{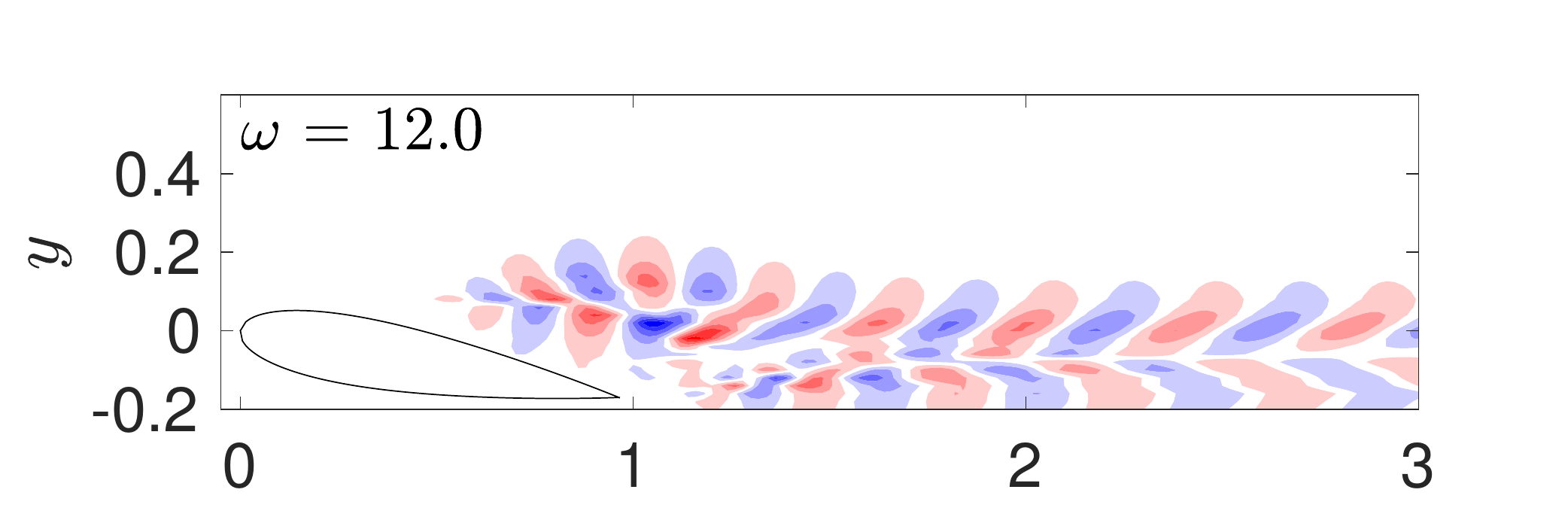}
	\includegraphics[scale=0.3]{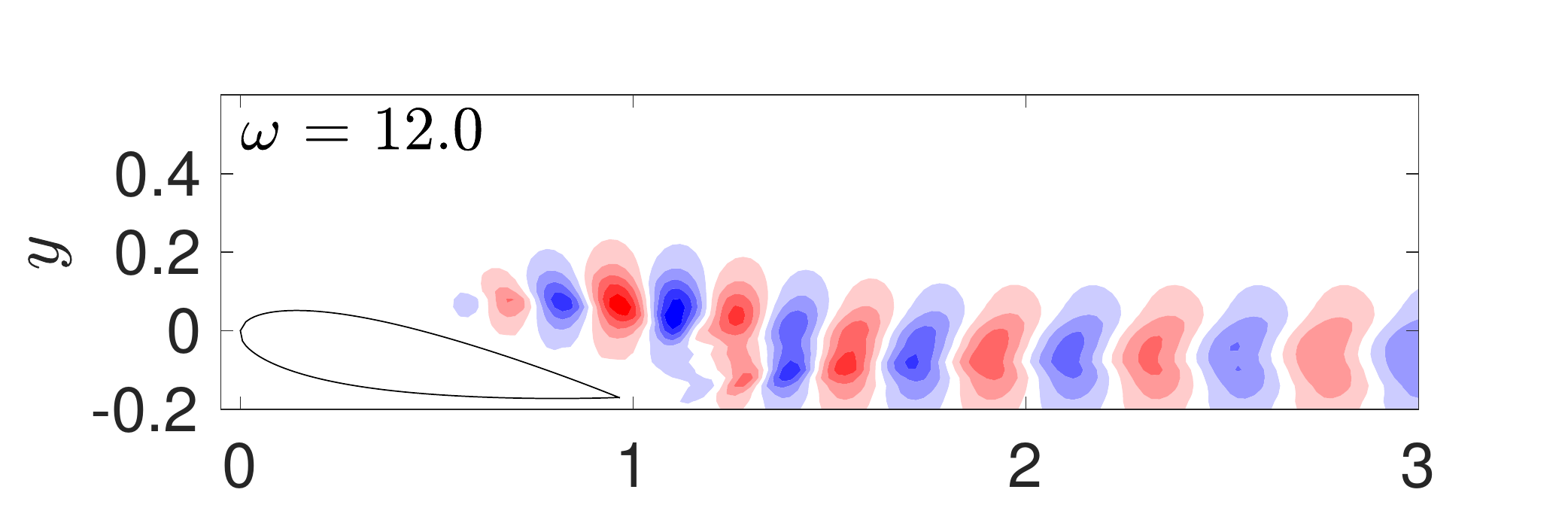}
	
	\includegraphics[scale=0.3]{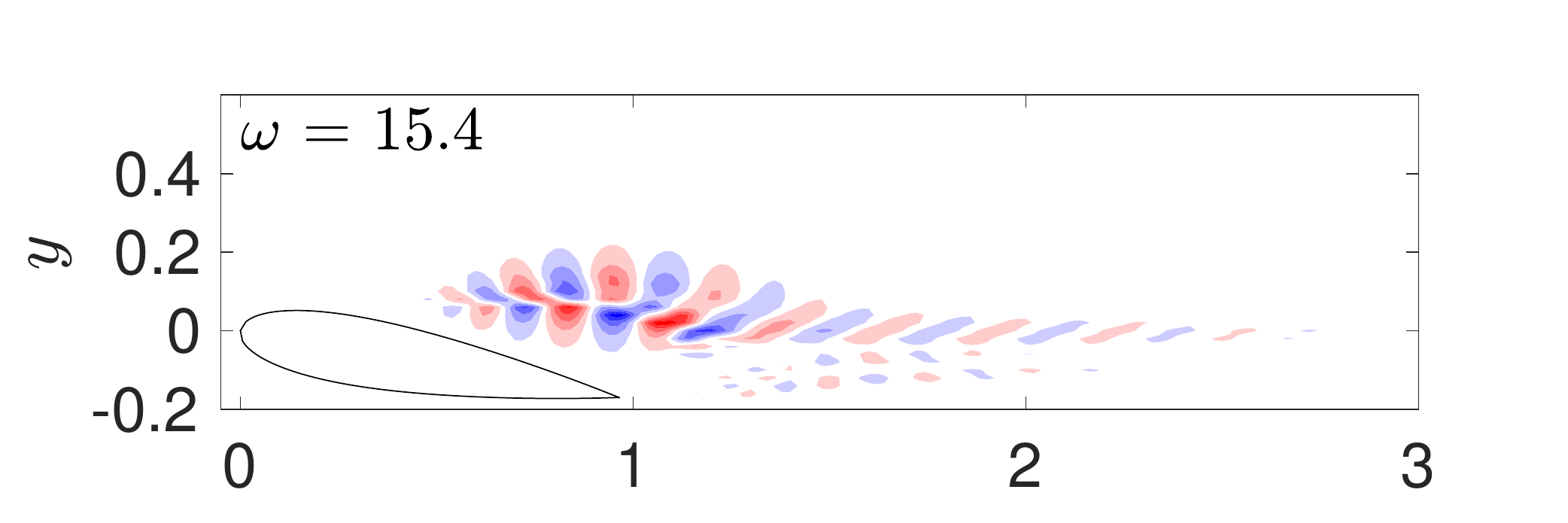}
	\includegraphics[scale=0.3]{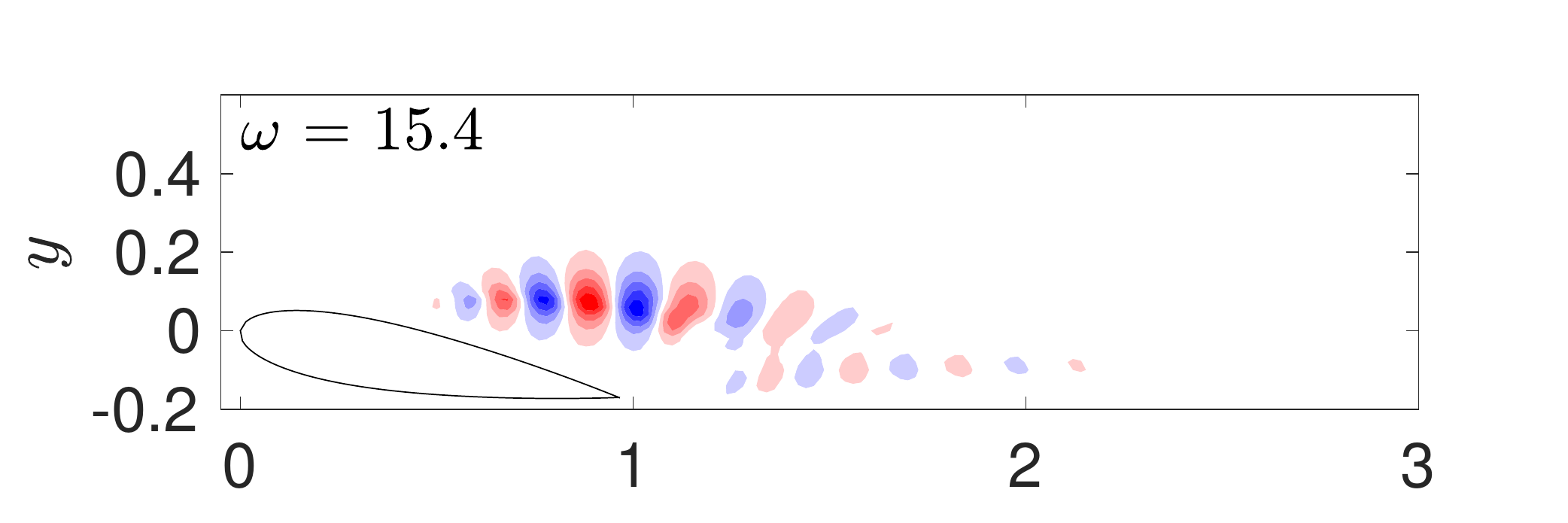}
	
	\caption{First optimal response mode $\boldsymbol{\psi}_1$ for various frequencies using the A10 data-assimilated mean profile. The streamwise and transverse velocity components are plotted in the left and right columns, respectively.}\label{fig:A10modeshapes}
\end{figure}

The identification of a secondary peak can be attributed to filling in data in the shear layer near the leading edge. As seen in figure \ref{fig:A10modeshapes}, which contains the most amplified resolvent mode at two frequencies corresponding to the peaks in figure \ref{fig:svalsA10} as well as one intermediate frequency ($\omega = 12.0$), two separate linear mechanisms are identified. The first corresponds to shedding- or wake-type modes which occur at lower temporal frequencies (top two rows of figure \ref{fig:A10modeshapes}). The second corresponds to a Kelvin-Helmholtz instability mechanism in the shear layer which occurs at higher temporal frequencies (bottom row of figure \ref{fig:A10modeshapes}.) The absence of PIV vectors near the leading edge results in resolvent analysis failing to capture the second linear mechanism when the interpolated experimental mean is analysed instead of the data-assimilated mean. The mode shapes for intermediate frequencies have a mix of both linear mechanisms as seen for $\omega = 12$ in figure \ref{fig:A10modeshapes}. These results are consistent with \citet{Thomareis18} and \citet{Yeh18}, who computed resolvent modes for a NACA 0012 airfoil at similar Reynolds numbers and angles of attack from time-averaged DNS and large-eddy simulation (LES) data, respectively. The latter study also identified two branches in the eigenspectrum of the LNS operator which were grouped into wake or shear layer modes. 

\subsection{Comparison to SPOD} \label{sec:comparison}

Even though data-assimilation adds no new insight for the A0 case, there are several noteworthy differences between the results of resolvent analysis and SPOD for this flow. The most significant one is that resolvent analysis identifies a single peak at the shedding frequency and low-rank activity is limited to nearby frequencies only. There is no low-rank behaviour at higher harmonics whereas in SPOD, there is an order of magnitude separation between the first and second eigenvalue of the first three harmonics. 

These observations can be explained by accounting for the nature of the nonlinear fluctuations. As in \citet{Towne18}, let us rewrite the cross-spectral density at an arbitrary frequency $\omega$ in terms of an expectation operator $\mathbb{E}(\cdot)$:
\begin{equation}
	\hat{\boldsymbol{S}}_{\hat{\boldsymbol{u}}\hat{\boldsymbol{u}}}(\omega) = \mathbb{E}\{\hat{\boldsymbol{u}}(\boldsymbol{x},\omega)\hat{\boldsymbol{u}}^*(\boldsymbol{x},\omega) \}. \label{eq:Suu}
\end{equation}
Noting that $\hat{\boldsymbol{u}}(\omega) = \mathcal{H}(\omega)\hat{\boldsymbol{f}}(\omega)$, (\ref{eq:Suu}) can be rewritten as
\begin{equation}
	\hat{\boldsymbol{S}}_{\hat{\boldsymbol{u}}\hat{\boldsymbol{u}}}(\omega) = \mathbb{E} \{\mathcal{H}(\omega)\hat{\boldsymbol{f}}(\boldsymbol{x},\omega)\hat{\boldsymbol{f}}^*(\boldsymbol{x},\omega)\mathcal{H}^*(\omega) \}.
\end{equation}
Instead of expressing $\mathcal{H}(\omega)$ in terms of its singular value decomposition, the resolvent operators are moved outside the expectation operator since they are deterministic to yield the following equation
\begin{equation} \label{eq:Sff}
	\hat{\boldsymbol{S}}_{\hat{\boldsymbol{u}}\hat{\boldsymbol{u}}}(\omega)=  \mathcal{H}(\omega)\hat{\boldsymbol{S}}_{\hat{\boldsymbol{f}}\hat{\boldsymbol{f}}}(\omega) \mathcal{H}^*(\omega),
\end{equation} 
where $\hat{\boldsymbol{S}}_{\hat{\boldsymbol{f}}\hat{\boldsymbol{f}}}(\omega) = \mathbb{E}\{\hat{\boldsymbol{f}}(\boldsymbol{x},\omega)\hat{\boldsymbol{f}}^*(\boldsymbol{x},\omega) \}$ is the cross-spectral density of the nonlinear forcing as in \cite{Towne19}. $\hat{\boldsymbol{S}}_{\hat{\boldsymbol{u}}\hat{\boldsymbol{u}}}(\omega)$ may be low-rank in two scenarios. 

The first is when the resolvent operator is low-rank, which occurs when there is a linear amplification mechanism. This is observed at the shedding frequency of the A0 case, frequencies corresponding to wake modes in the A10 case, and frequencies of the shear layer modes in the A10 case as summarised in rows 1, 4, and 5 in table \ref{tab:2 cases}. Both SPOD and resolvent analysis identify low-rank behaviour, similar to the case of the turbulent jet in \citet{Schmidt18} where the Kelvin-Helmholtz instability accounts for the low-rank nature of the cross-spectral density tensor. 

The second source of low-rank behavior can originate from the cross-spectral density of the nonlinear forcing, or $\hat{\boldsymbol{S}}_{\hat{\boldsymbol{f}}\hat{\boldsymbol{f}}}(\omega)$, being low-rank. Even if $\mathcal{H}(\omega)$ is nearly full-rank, as it is for higher harmonics of the shedding frequency, the left-hand side of (\ref{eq:Sff}) will be low-rank as long as at least one of the terms on the right-hand side is low-rank, i.e.
\begin{equation}
	 \text{rank}(\hat{\boldsymbol{S}}_{\hat{\boldsymbol{u}}\hat{\boldsymbol{u}}}(\omega)) \leq \text{min}\left[\text{rank}(\mathcal{H}(\omega)),\text{rank}(\hat{\boldsymbol{S}}_{\hat{f}\hat{f}}(\omega)) \right] .
\end{equation}
For the A0 case, the nonlinear forcing is very structured at the shedding frequency and its harmonics; consequently, SPOD of the velocity fluctuations reveals large spectral gaps at these frequencies as seen in rows 1 and 2 of table \ref{tab:2 cases}. We theorise that the nonlinear forcing can be estimated by limited triadic interactions of resolvent modes to educe the structure at these higher harmonics, a hypothesis which is tested in the next section. 

The SPOD eigenspectra of the nonlinear forcings in figures \ref{fig:NLFA0} and \ref{fig:NLFA10} verify that the cross-spectral density is low-rank at the shedding frequency and its harmonics in the A0 case and at no frequencies for the A10 case. The trends reinforce the notion that A0 behaves like an oscillator while A10 behaves like an amplifier. It also suggests that resolvent modes and SPOD modes will deviate substantially for the higher harmonics in the A0 case since the resolvent is not low-rank and the forcing is anything but white noise in space and time. This will be seen in \S\ref{sec:parasitic} for the second harmonic of the shedding frequency. The modes for the A10 case, on the other hand, can be expected to match at all frequencies where the the resolvent is low-rank since these coincide with the low-rank frequencies identified by SPOD.

\begin{figure}
	\centering	
	\includegraphics[scale=0.3]{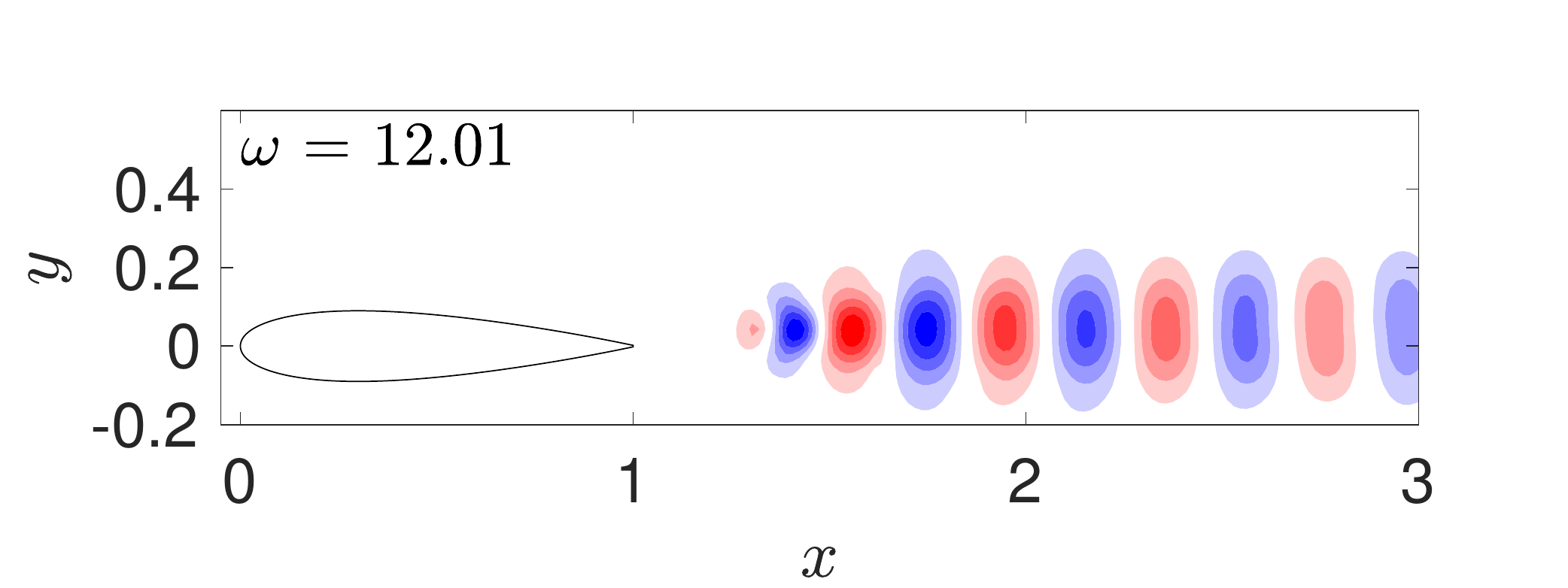}
	\includegraphics[scale=0.3]{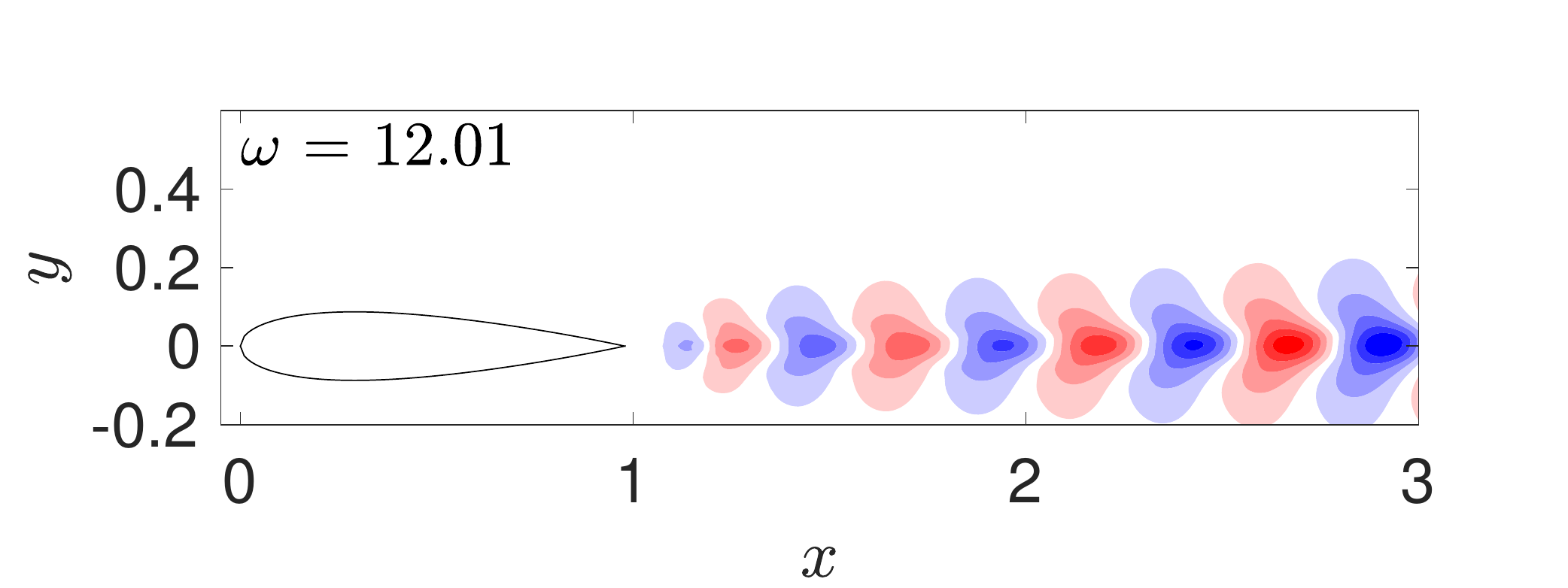}
	
	\includegraphics[scale=0.3]{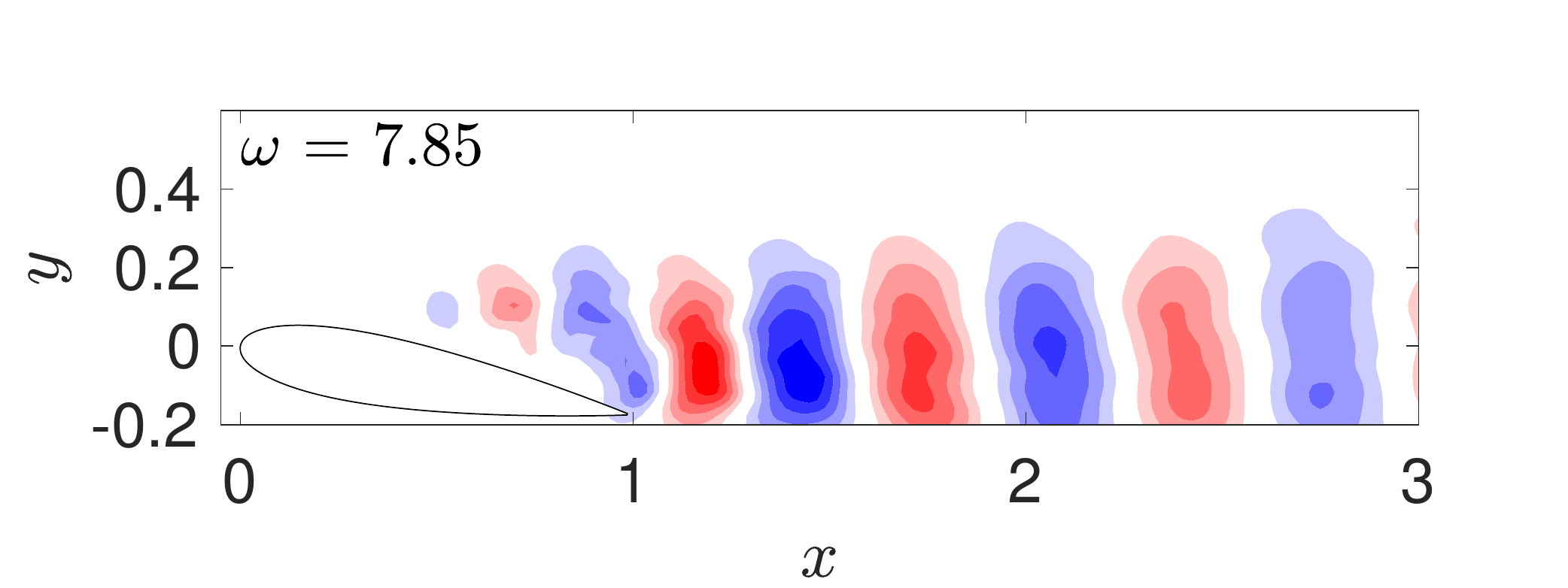}
	\includegraphics[scale=0.3]{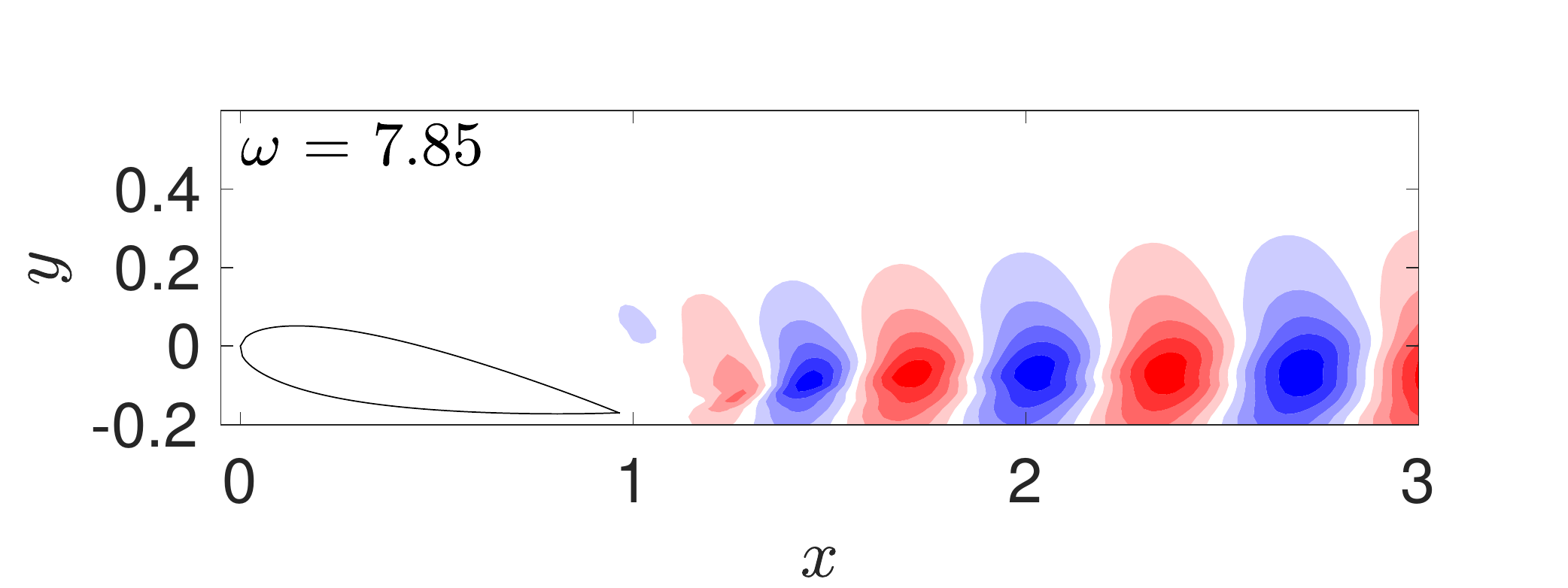}
	
	\includegraphics[scale=0.3]{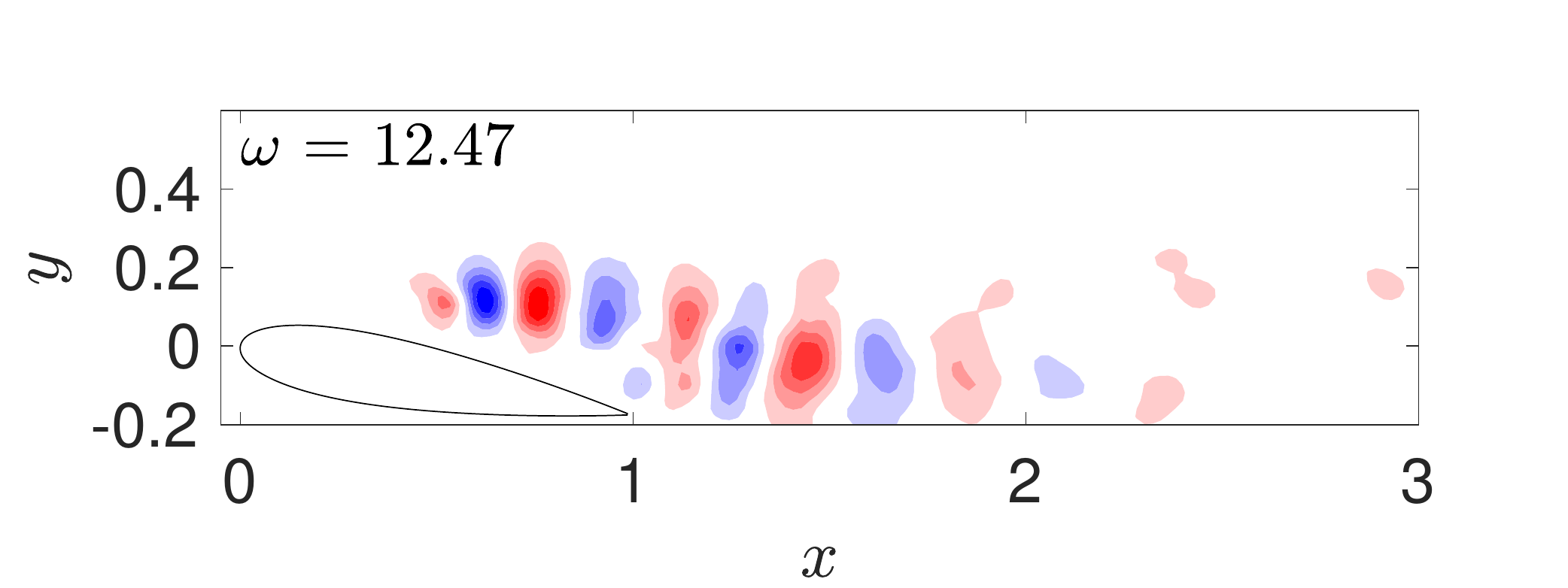}
	\includegraphics[scale=0.3]{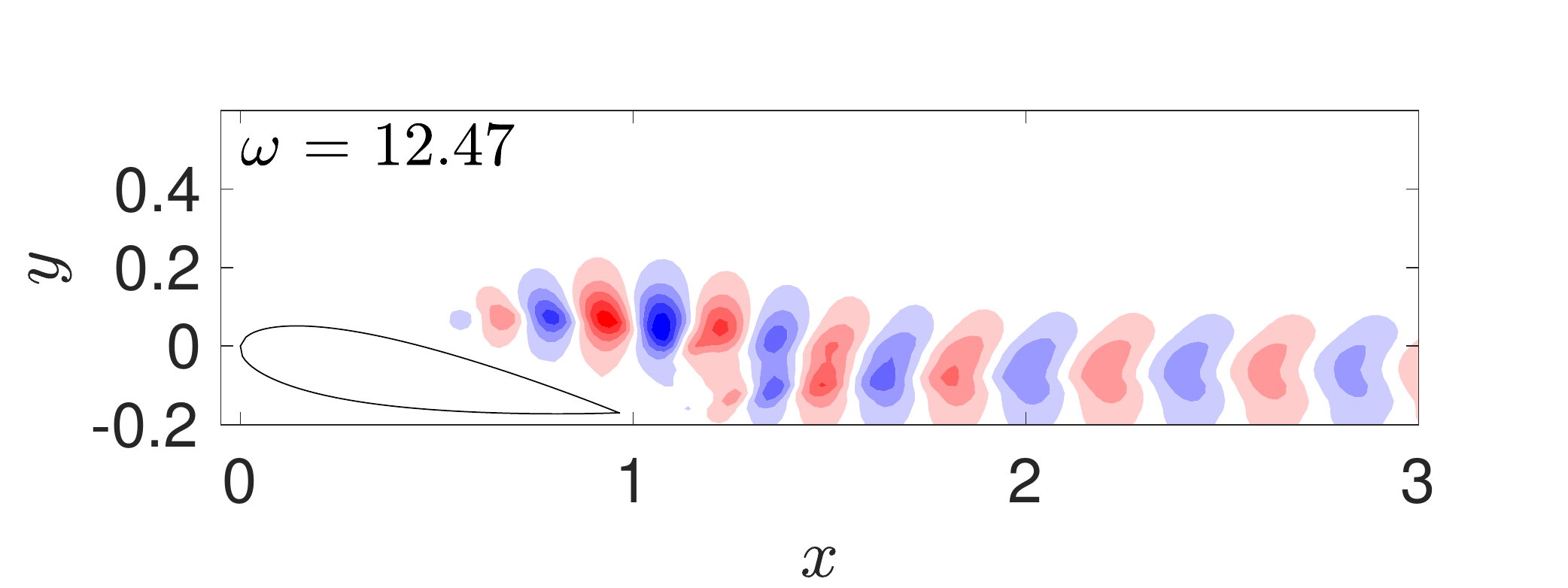} 
	
	\caption{A comparison between the dominant SPOD mode (left) and the optimal resolvent mode (right) at highly amplified frequencies.}\label{fig:compare modes}
\end{figure}

We compare the dominant SPOD mode to the optimal resolvent mode in figure \ref{fig:compare modes} to reinforce the notion that the first resolvent response mode is representative of the true fluctuations in the flow when the resolvent is low-rank. One difference, which is particularly notable for the A0 shedding mode, is that the amplitude of the resolvent mode increases as a function of $x$. By comparing the final red contour in the top row of figure \ref{fig:compare modes}, we note that the amplitude of the SPOD mode is quite weak at $x = 2.8$, whereas it is very strong for the resolvent mode. A similar phenomenon can be observed for the A10 wake mode. The shear layer mode, on the other hand, is a very close match. Despite these discrepancies, the wavelengths of the modes are in complete agreement, signifying that the resolvent modes correctly predict the convection velocities of the structures in the flow.

\begin{table}
	\begin{center}
		\begin{tabular}{c|c|c|c|c}
			\hline
			$\omega$-range    & $\boldsymbol{S_{\hat{u}\hat{u}}}(\omega)$ (SPOD) & $\mathcal{H}(\omega)$ (Resolvent) & $\boldsymbol{S_{\hat{f}\hat{f}}}(\omega)$ & modelling type \\
			\hline
			\multicolumn{5}{c}{A0 case (oscillator)}\\
			\hline
			1. near $\omega_s$      & low-rank   & low-rank   & low-rank  & first resolvent mode \\
			2. near harmonics of $\omega_s$       & low-rank   & high-rank & low-rank  & parasitic mode \\
			3. all other $\omega$      & high-rank  & high-rank & high-rank & not modelled   \\
			\hline
			\multicolumn{5}{c}{A10 case (amplifier)} \\
			\hline
			4. wake frequencies      & low-rank   & low-rank   & high-rank  & first resolvent mode \\
			5. shear layer frequencies       & low-rank   & low-rank & high-rank  & first resolvent mode \\
			6. all other $\omega$      & high-rank  & high-rank & high-rank & not modelled \\
			\hline
			
		\end{tabular}
	\end{center}
	\caption{Properties of the cross-spectral density tensors and resolvent operators for the A0 and A10 cases.} \label{tab:2 cases}
\end{table}


\section{Resolvent-based modelling of the velocity fluctuations}\label{sec:modeling}

We now move on to determine how much of the full fluctuating field can be deduced starting from only a limited mean profile and data at a few probe points. We search for an efficient basis for a reduced-order model of the energetic coherent structures using the data-assimilated mean and very limited time-resolved probes (i.e. knowledge of the velocity at one or two points). This section is divided into three parts: in \S\ref{sec:parasitic} we attempt to approximate the structure of the nonlinear forcing using limited triadic interactions at frequencies where the linear dynamics are not low-rank for the A0 case. The reduced-order models with and without these parasitic modes are compared. In \S\ref{sec:two}, we consider the A10 case and examine whether just one or multiple probes are needed to account for the presence of two linear mechanisms in the flow. Finally, we discuss our results in the context of experimental limitations encountered in \S\ref{sec:limitations}. 

\subsection{Parasitic modes and modelling of the A0 case}\label{sec:parasitic}

\subsubsection{Approximation of the nonlinear forcing}

To improve the basis for frequencies outside the range where the resolvent operator is low-rank, the nonlinear forcing is approximated using very limited triadic interactions. Let us consider two examples. The simplest is the second harmonic of the shedding frequency $\omega_s$, whose nonlinear forcing is approximated in exactly the same manner as that outlined in \cite{Rosenberg19}, i.e. 
\begin{equation} \label{eq:forcing 2ws}
	\hat{\boldsymbol{f}}_a(2 \omega_s) \sim \hat{\boldsymbol{\psi}}_1(\omega_s) \cdot \nabla \hat{\boldsymbol{\psi}}_1(\omega_s).
\end{equation}
Many other modes may contribute to the right-hand side of (\ref{eq:forcing 2ws}) although this approximation is adequate since the singular values associated with $\hat{\boldsymbol{\psi}}_1(\omega_s)$ are very large. Very low frequencies can also be recovered by postulating that they result from the beating of two highly amplified frequencies. The nonlinear forcing for the frequency $\omega = 3.70$, for example, is
\begin{equation} \label{eq:forcing3}
	\hat{\boldsymbol{f}}(\omega = 3.70) = - \sum_{\substack{\omega_a + \omega_b = 3.70 \\ \omega_a,\omega_b \neq 0}} \chi_1(\omega_a) \chi_1(\omega_b) \nabla \cdot (\hat{\boldsymbol{\psi}}_1(\omega_a) \hat{\boldsymbol{\psi}}^T_1(\omega_b)).   
\end{equation}
Since the singular value associated with the shedding frequency at $\omega_s = 12.0$ is very large, it can be reasoned that the magnitude of $\chi_1$ for this frequency will outweigh all other terms in (\ref{eq:forcing3}). To obtain the correct frequency, therefore, it only needs to interact with the frequency $\omega = -8.30$, where the resolvent operator is still approximately rank-1, to produce a mode at the desired frequency $\omega = +3.70$. (\ref{eq:forcing3}), therefore, reduces to
\begin{equation} \label{eq:forcing 3.7}
	\hat{\boldsymbol{f}}_a(\omega = 3.70) \sim \hat{\boldsymbol{\psi}}_1(\omega = 12.0) \cdot \nabla \hat{\boldsymbol{\psi}}_1(\omega = -8.30) + \hat{\boldsymbol{\psi}}_1(\omega = -8.30) \cdot \nabla \hat{\boldsymbol{\psi}}_1(\omega = 12.0),
\end{equation} 
where the $\chi_1$'s have been dropped since $\chi_1(\omega = 12.0)\chi_1(\omega = -8.3)$ appears before each term in (\ref{eq:forcing 3.7}) and, at this point, we are only interested in improving the basis for this particular frequency. If we were to include more triads in either (\ref{eq:forcing 2ws}) or (\ref{eq:forcing 3.7}), we would need to know the phase of the modes since $\chi_1$ appears before each mode in (\ref{eq:forcing3}). We argue here that the high amplifications associated with the shedding mode allow us to ignore the triadic interactions of suboptimal modes and frequencies which are less amplified. 

\begin{figure}
	\centering
	\includegraphics[scale=0.3]{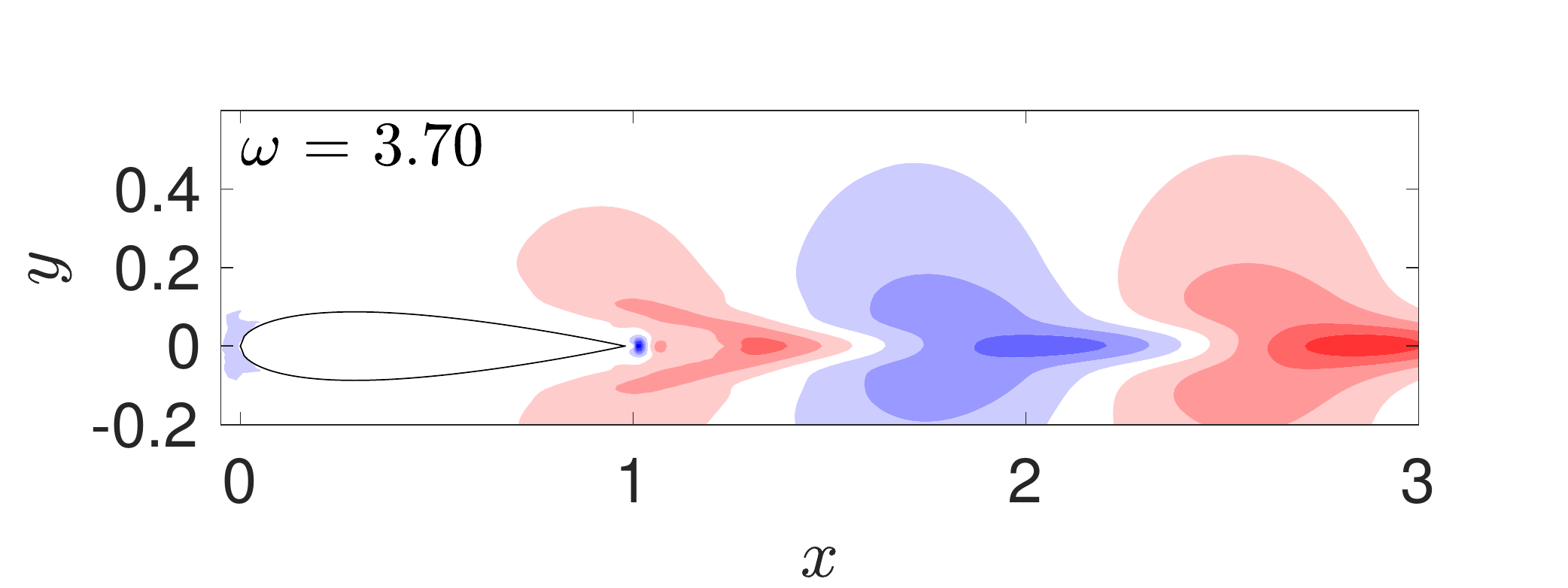}
	\includegraphics[scale=0.3]{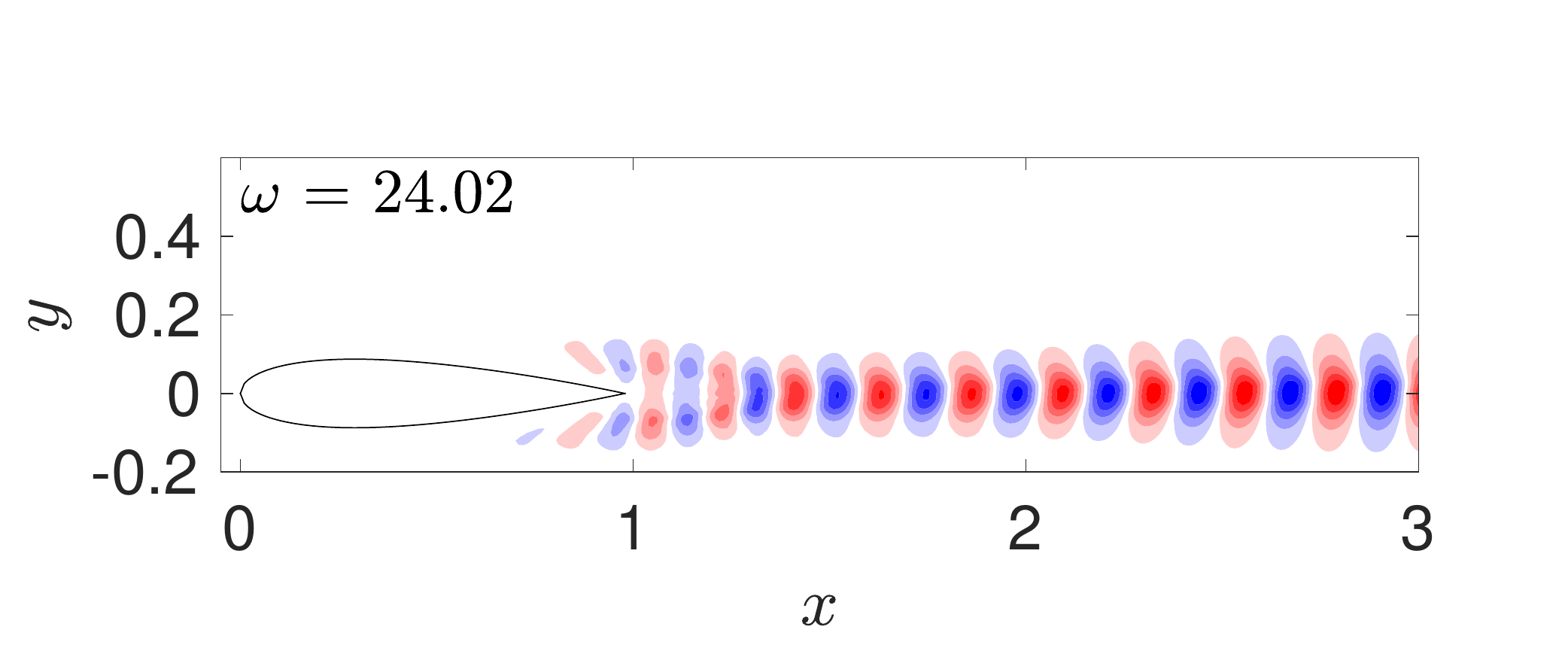}	
	\includegraphics[scale=0.3]{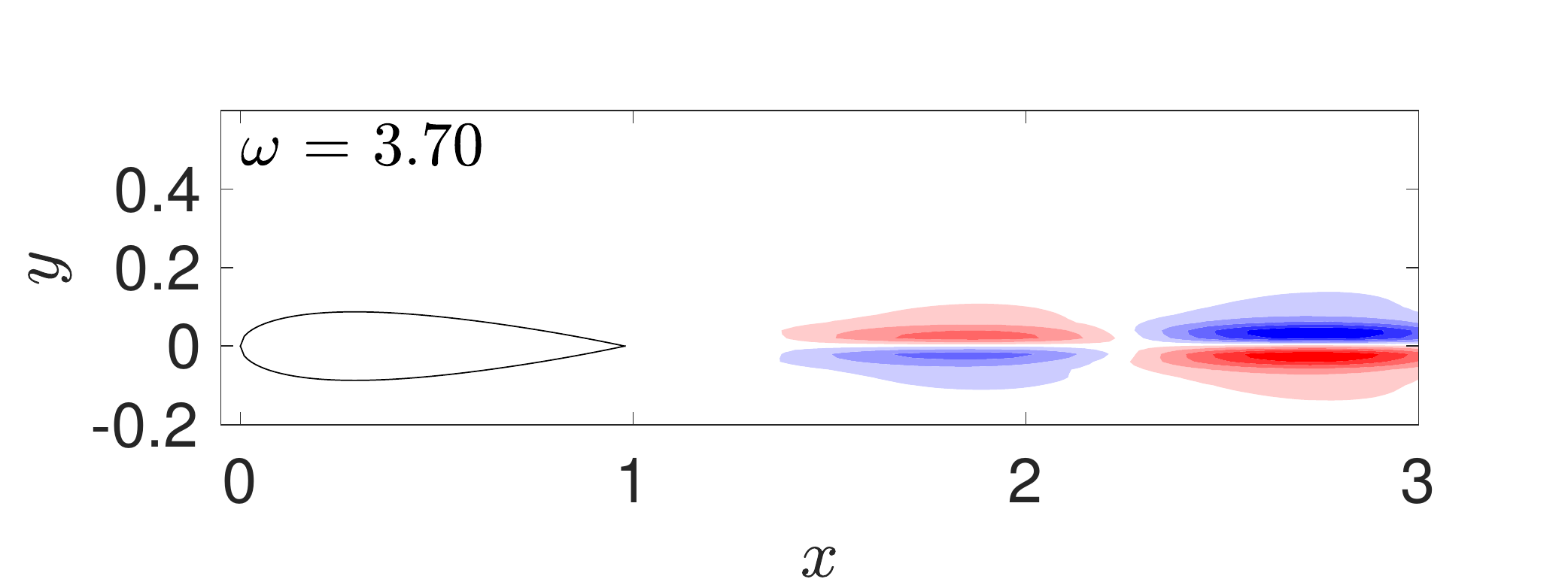}
	\includegraphics[scale=0.3]{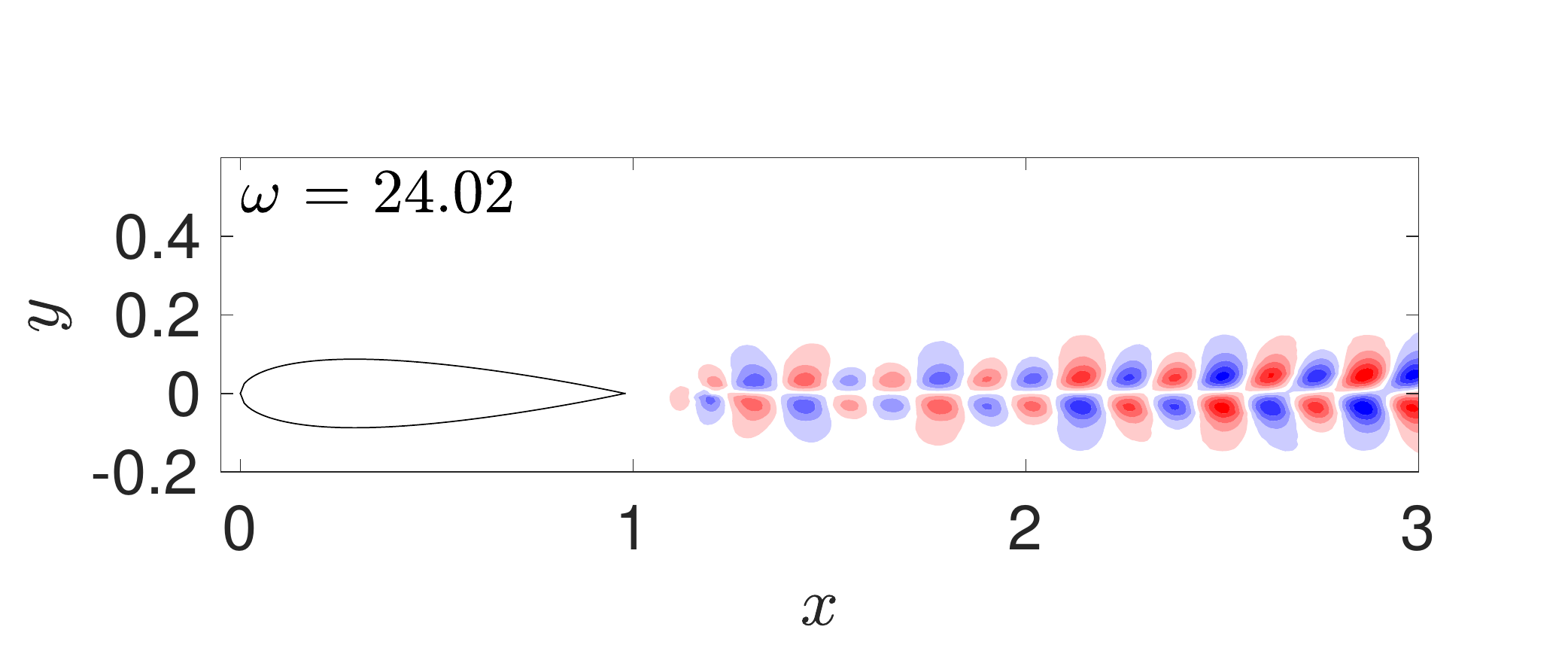}
	\includegraphics[scale=0.3]{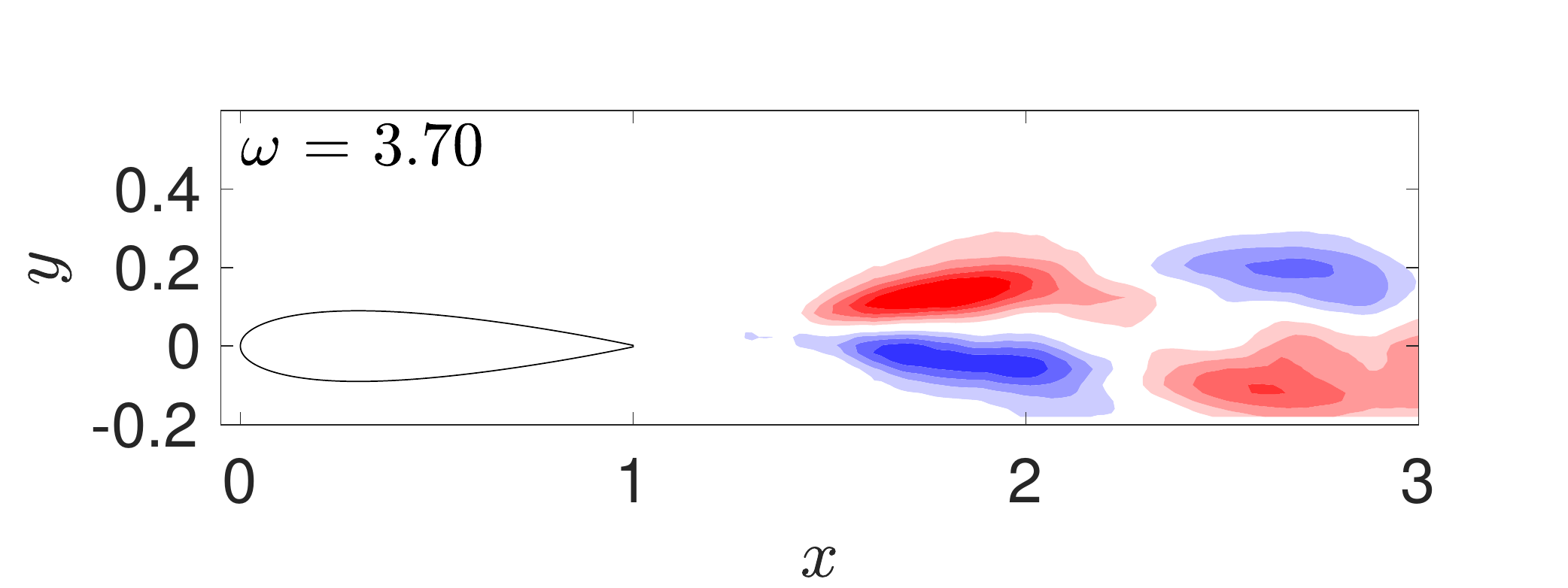}
	\includegraphics[scale=0.3]{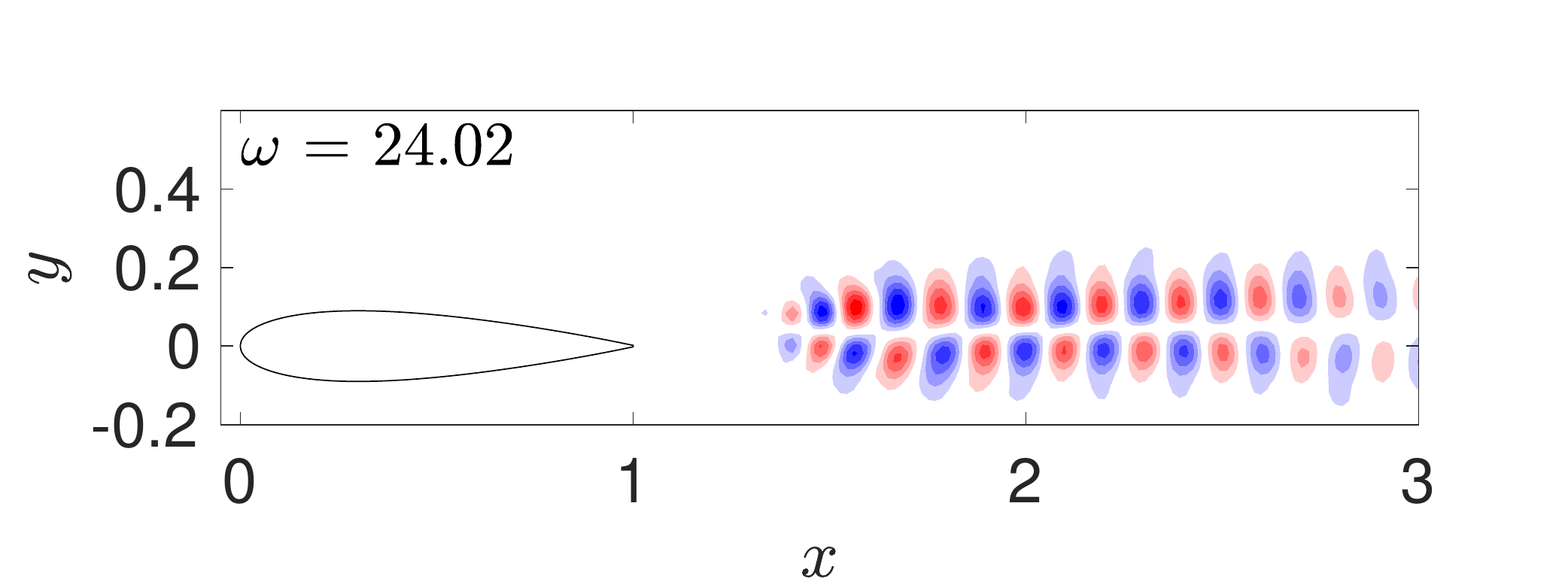}
	\caption{The optimal response mode (top row) compared to the parasitic mode (middle row) and the first SPOD mode (bottom row).}\label{fig:parasitic modes} 
\end{figure}

The parasitic modes, or `forced' resolvent modes (i.e. those obtained after right-multiplying the resolvent operator by the approximated nonlinear forcing), are compared to the optimal response mode $\hat{\boldsymbol{\psi}}_1$ from the singular value decomposition and the first SPOD mode in figure \ref{fig:parasitic modes}. 
Despite how simple the modelling for the nonlinear forcing is, the structure at all frequencies considered is a major improvement over the predicted structure from the first resolvent response mode. In particular, the parasitic mode obeys the correct symmetries while the optimal response mode resembles a stretched or compressed version of the vortex shedding mode. 

\subsubsection{Modelling preliminaries}

\begin{figure} 
	\centering
	\includegraphics[scale=0.3]{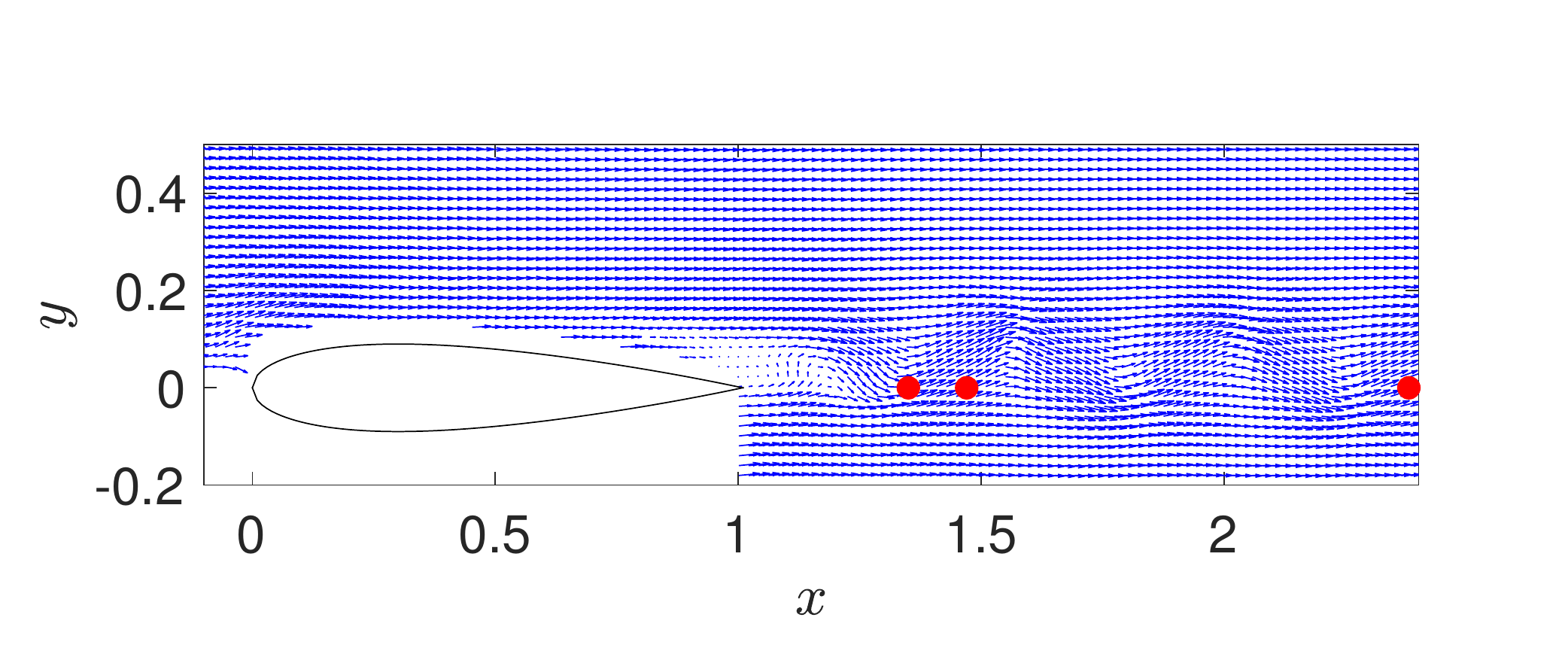}
	\includegraphics[scale=0.3]{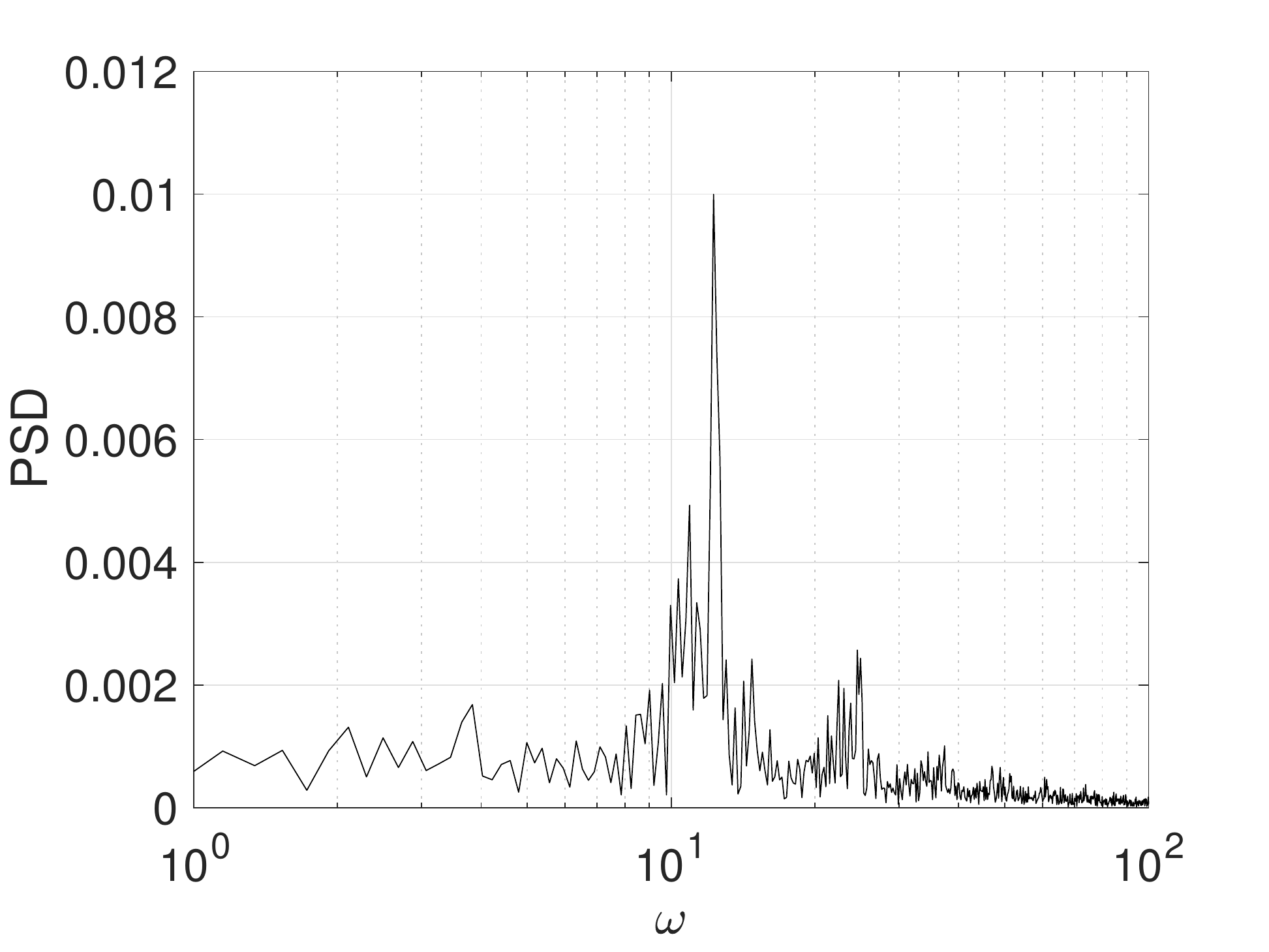}
	\caption{Candidate probe points (left) and the power spectrum of $\check{v}(t)$ for P2 (right).}\label{fig:A0calibration}
\end{figure}

Several candidate points to calibrate the complex weights of the resolvent modes are displayed in figure \ref{fig:A0calibration}. They are referred to as points 1, 2, and 3, and are ordered by their streamwise position, i.e. point 1 is the furthest upstream.
The power spectrum of P2 (P1 and P3 are very similar) for the transverse velocity component, plotted in figure \ref{fig:A0calibration}, contains a large peak at $\omega = 12.0$ and a secondary one at $\omega = 24.0$. These are the first and second harmonics of the shedding frequency, respectively. There is another discernible peak at $\omega = 10.7$ resulting from jitter in the shedding frequency. This is also present in the SPOD results for the leading eigenvalue, although it is less noticeable. The jitter does not influence the resolvent norm.

The unweighted resolvent modes are first normalised using the standard energy inner product based on the two velocity components. Once the resolvent modes are weighted using the calibration point, they should reflect the true magnitude of the fluctuations. To make a fair comparison between the model and snapshots obtained from PIV, the experimental data are filtered to exclude temporal frequencies outside the range used in the model. The filtering involves Fourier-transforming the raw PIV data in time, truncating Fourier modes at unwanted frequencies, and inverse Fourier-transforming back into the time domain. Another possible way to compare the resolvent model to the experiment is to project the raw data onto the first SPOD mode at all frequencies of interest. Generally these would be where SPOD is approximately rank-1. Since the frequency resolution of the SPOD modes is only $\Delta \omega \approx 0.5$, we choose to filter based on the first procedure.

\subsubsection{Results and quantification of error}

We begin by comparing the performance of each candidate point. The instantaneous error is computed between the model and PIV as follows 
\begin{equation}
	I_v(t) = \sqrt{\sum_{i=1}^N \left(v^i(t)-v_{PIV}^i(t) \right)^2},
\end{equation}
where $v$ denotes the transverse velocity component and $N$ is the number of PIV vectors. $I_v(t)$ is integrated in time to obtain a global error metric 
\begin{equation}
	E_v = \sqrt{\left(\frac{1}{T}\int_0^TI(t)^2dt\right)}.
\end{equation}
$E_v$ is normalised by the total kinetic energy contained in the filtered PIV and the results for each point are summarised in table \ref{tab:errorA0}. Even though the performance of each point is good at some time instants, the overall performance for all three is relatively poor although P3 is the best. If two points are considered, the performance is nearly an order of magnitude better. Furthermore, points which are situated further downstream in the flow yield better results, which is a consequence of the resolvent modes failing to capture the correct streamwise decay of the fluctuations. The combination of points P2 and P3 yields the lowest amount of error, so the following discussion considers this choice only.

\begin{table}  
	\centering
	\begin{tabular}{c|c|c|c}
		\hline 
		1 Point & Error & 2 Points & Error \\ \hline
		1 &  344\% & 1 + 2 & 165\% \\ \hline
		2 &  988\% & 1 + 3 & 36\% \\ \hline
		3 &  152\% & 2 + 3 & 32\% \\ \hline 
	\end{tabular}
	\caption{Error for the A0 case as a function of point selection.} \label{tab:errorA0}
\end{table} 

\begin{figure}
	\centering
	\includegraphics[scale=0.3]{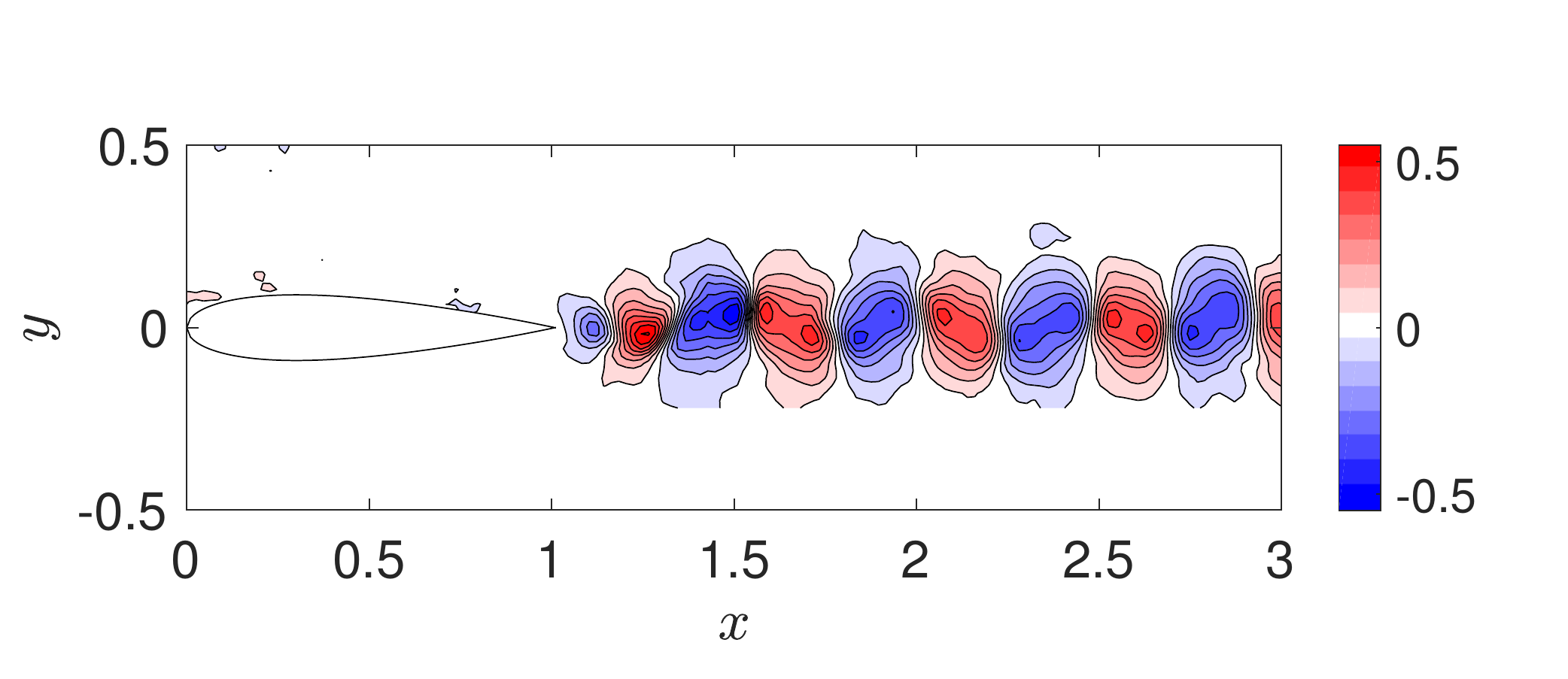}
	
	\includegraphics[scale=0.3]{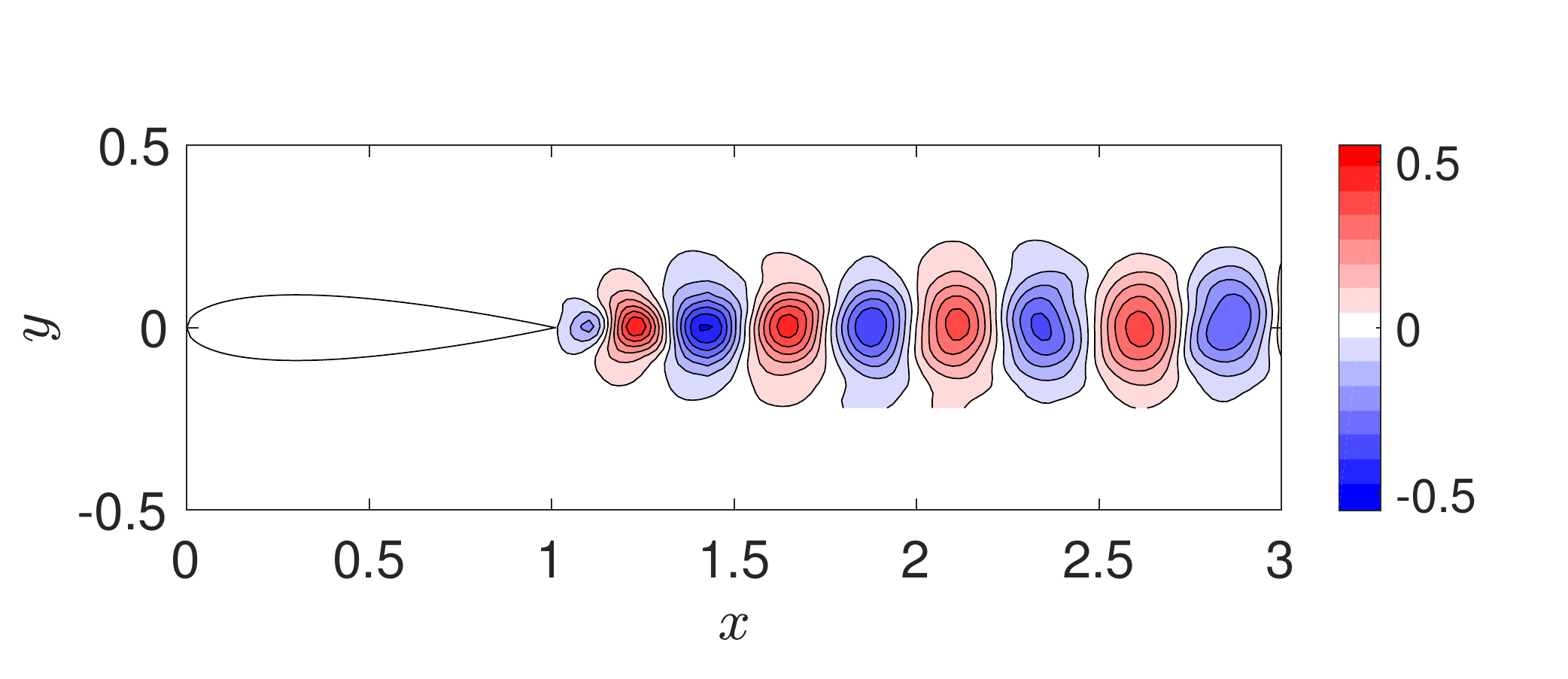}
	\includegraphics[scale=0.3]{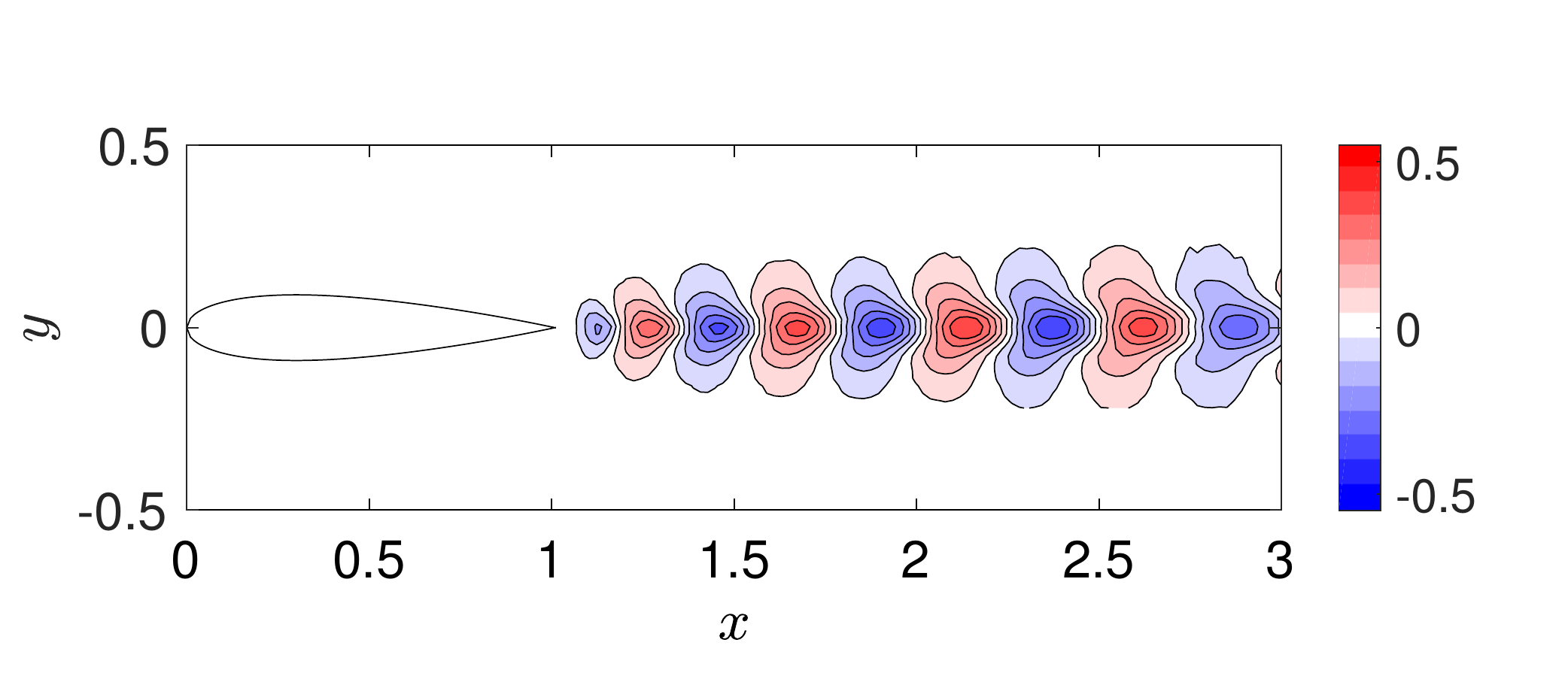}
	
    \includegraphics[scale=0.3]{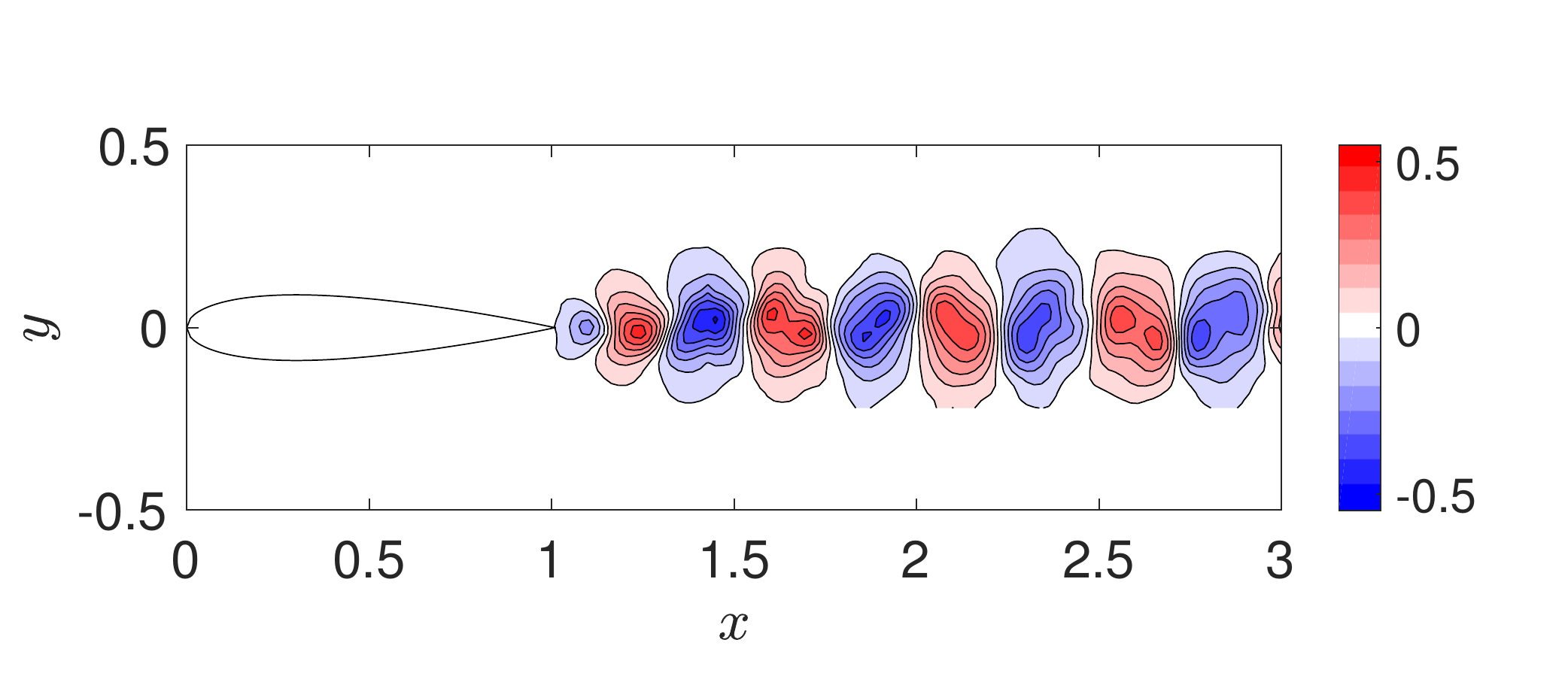}
	\includegraphics[scale=0.3]{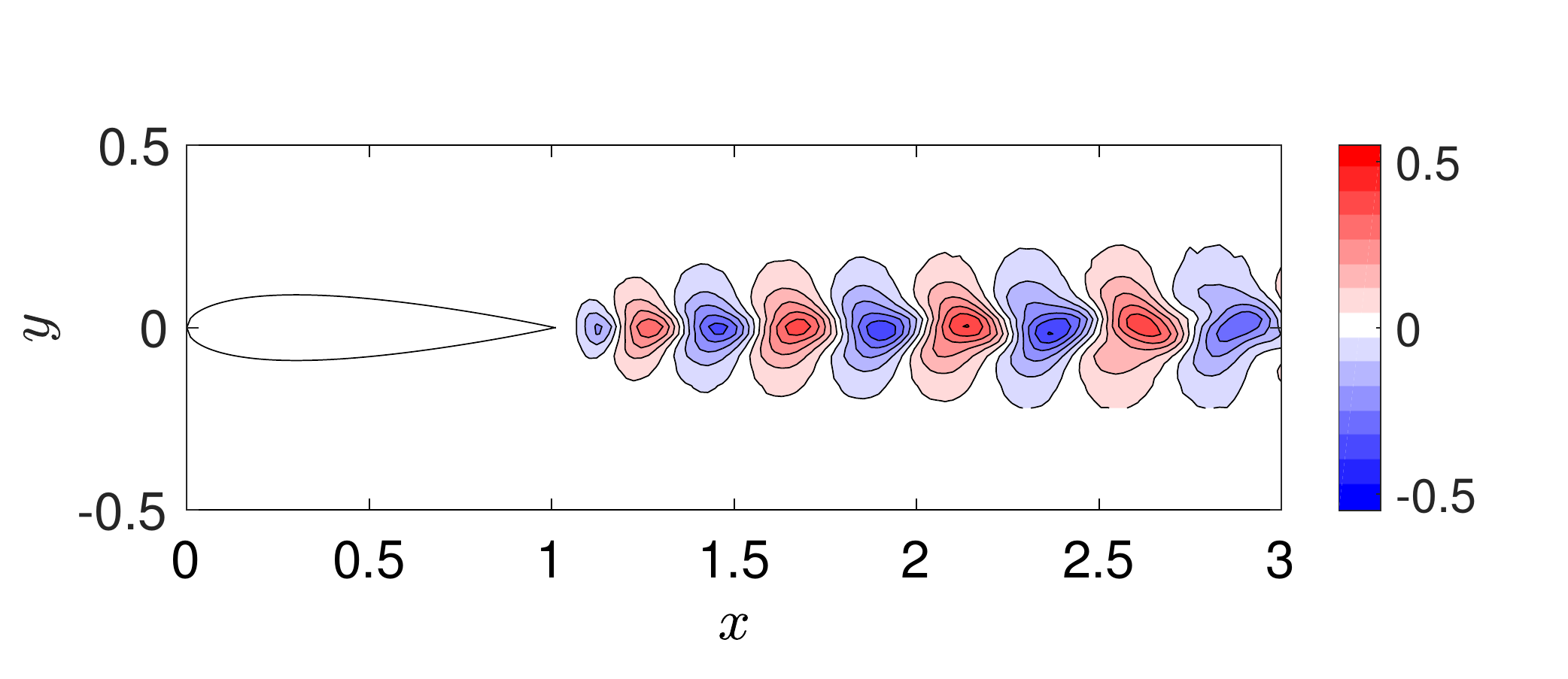}
	
	\caption{$T_1$ results: (a) The unfiltered PIV snapshot. (b) Filtered PIV containing frequencies $6.73 < \omega < 15.7$ compared to the (c) resolvent model. (d) Filtered PIV containing $\omega < 39$ compared to the (e) resolvent model.}\label{fig:T1A0}
\end{figure}

The values computed in table \ref{tab:errorA0} factored in all resolvent modes computed, i.e. both classical and parasitic. The first modelling attempt, however, only uses modes where the resolvent operator is nearly rank-1, i.e. $6.73 < \omega < 15.7$, henceforth referred to as rank-1 modes. These 41 modes are computed using the first singular response mode of the resolvent operator. Harmonics of the shedding frequency where SPOD reveals low-rank behaviour are excluded for now.  A representative snapshot is presented in figure \ref{fig:T1A0} and the flow resembles the traditional von K\'arm\'an vortex street behind a cylinder. The raw PIV data are contained in the top row while the filtered PIV are compared to the resolvent model using rank-1 modes only in the second row.

\begin{figure}
	\centering	
	\includegraphics[scale=0.3]{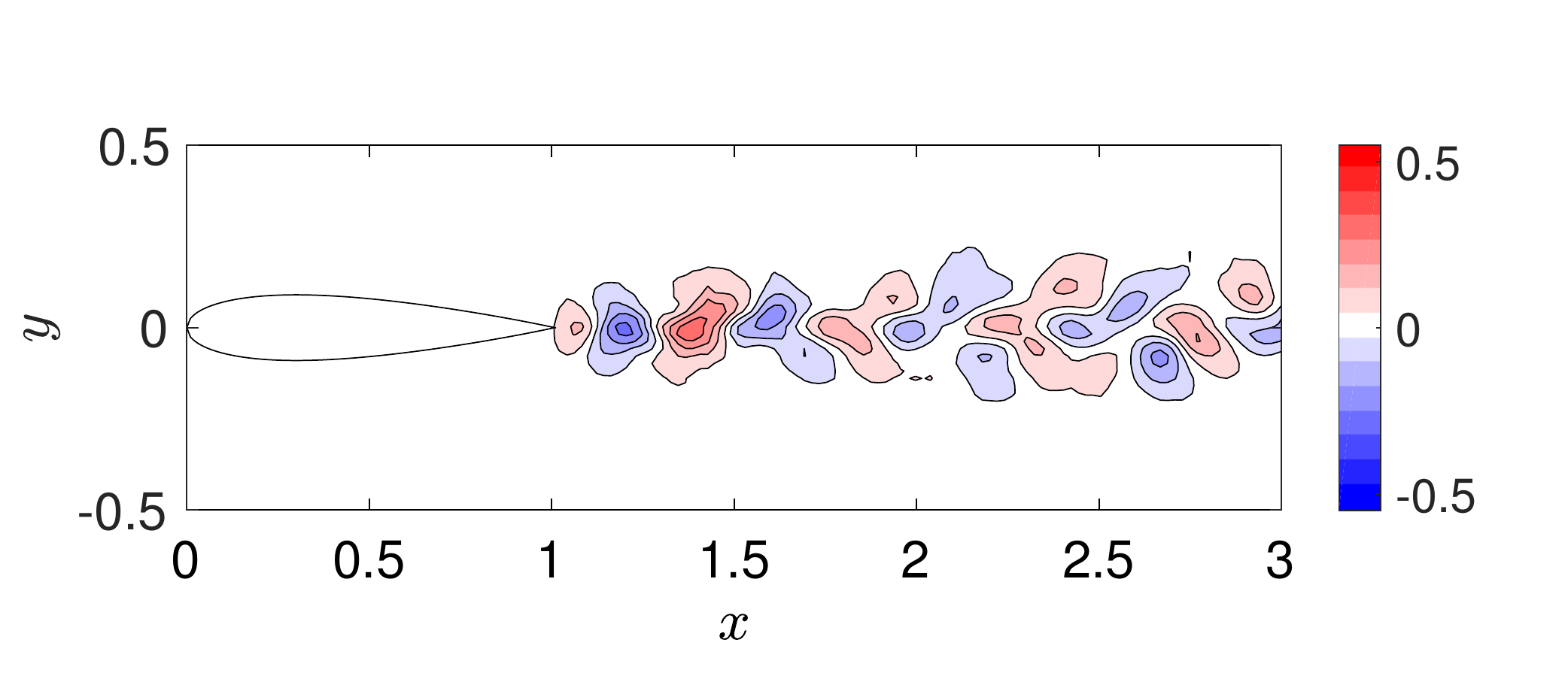}
	\includegraphics[scale=0.3]{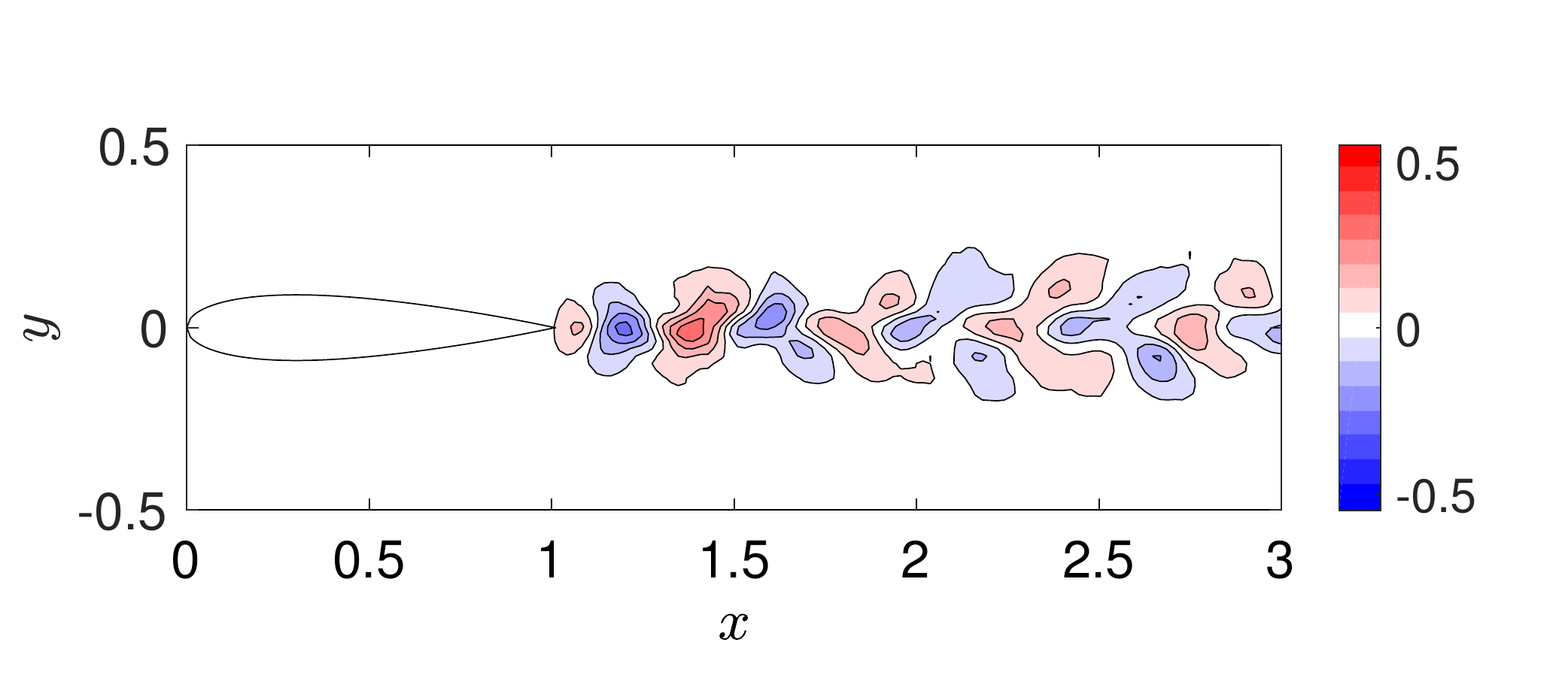}
	
	\caption{The discrepancy between the filtered PIV and the resolvent model using rank-1 frequencies only (left) and with the inclusion of parasitic modes (right).}\label{fig:A0error}
\end{figure}

The velocity fluctuations are symmetric with respect to the centreline since the most amplified structure predicted by resolvent analysis resembles the von K\'arm\'an vortex street at all rank-1 frequencies. Similar to the cylinder, there is only one linear mechanism, so modes at suboptimal frequencies end up being a stretched or compressed (in the streamwise direction) version of the shedding mode \citep[see][]{Symon18b, Symon19}. The true velocity fluctuations are not symmetric with respect to the centreline if all frequencies are considered. The difference between the filtered PIV and resolvent model is plotted in figure \ref{fig:A0error}. The largest discrepancies occur immediately behind the trailing edge ($1.0 < x < 1.5$), where the fluctuations are under predicted by the resolvent model. By calibrating at points P2 and P3, where the resolvent modes over predict the fluctuations, the resulting model has a harder time reconstructing the region further upstream.

When parasitic modes are included, the model is no longer constrained to be symmetric since the second harmonic and low frequency modes are antisymmetric with respect to the centreline. The flow is now modelled using 91 modes ranging from frequencies as low as $\omega = 2.72$ to frequencies as high as $\omega = 39$, in order to encompass the third harmonic. The resolvent-model is animated in time and compared to the filtered PIV in movies 1 and 2, respectively. Not all frequencies within this range are included since many of them are not energetic as indicated by the probe points. The model is compared to the filtered PIV in the bottom row of figure \ref{fig:T1A0}.

Because the modes outside the rank-1 range are not particularly energetic, the filtered PIV in the bottom row resembles the raw PIV snapshot in the top row fairly well (it contains 80\% of the kinetic energy). The addition of parasitic modes does not dramatically improve the model, but it does capture more detailed features of the flow field. Further downstream, the negative velocity perturbations tilt upstream while the positive velocity perturbations tilt downstream and this is consistent with the PIV data. The discrepancy between the improved model and the filtered PIV is illustrated in figure \ref{fig:A0error}. Small improvements can be seen in the region beyond $ x = 1.5$. For this particular snapshot, the addition of parasitic modes reduces the error from 22\% to 18\%.

\subsection{Two linear mechanisms and modelling of the A10 case} \label{sec:two}

\subsubsection{Selection of probe points}

\begin{figure}
	\centering
	\includegraphics[scale=0.4]{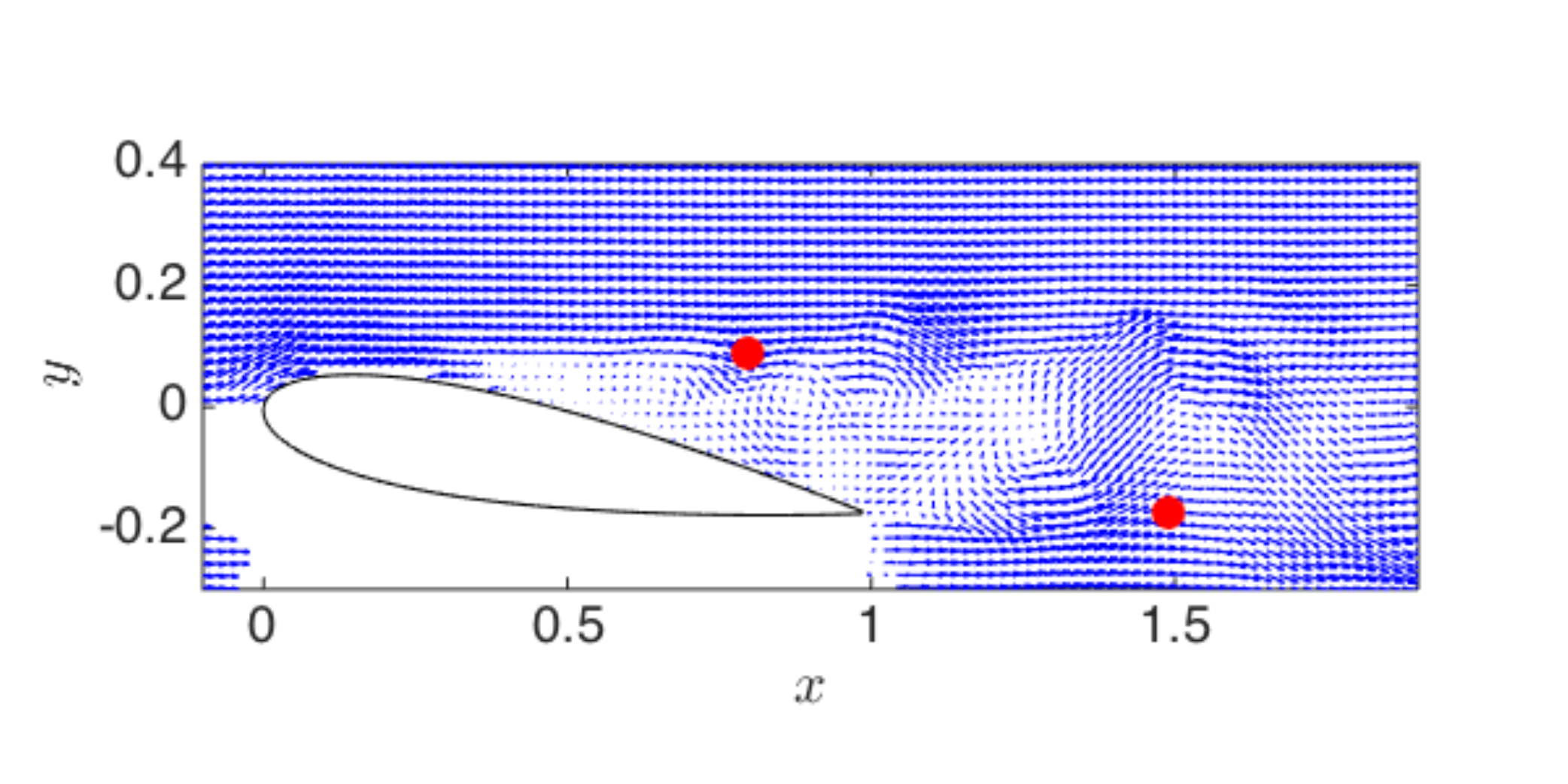}
	
	\includegraphics[scale=0.3]{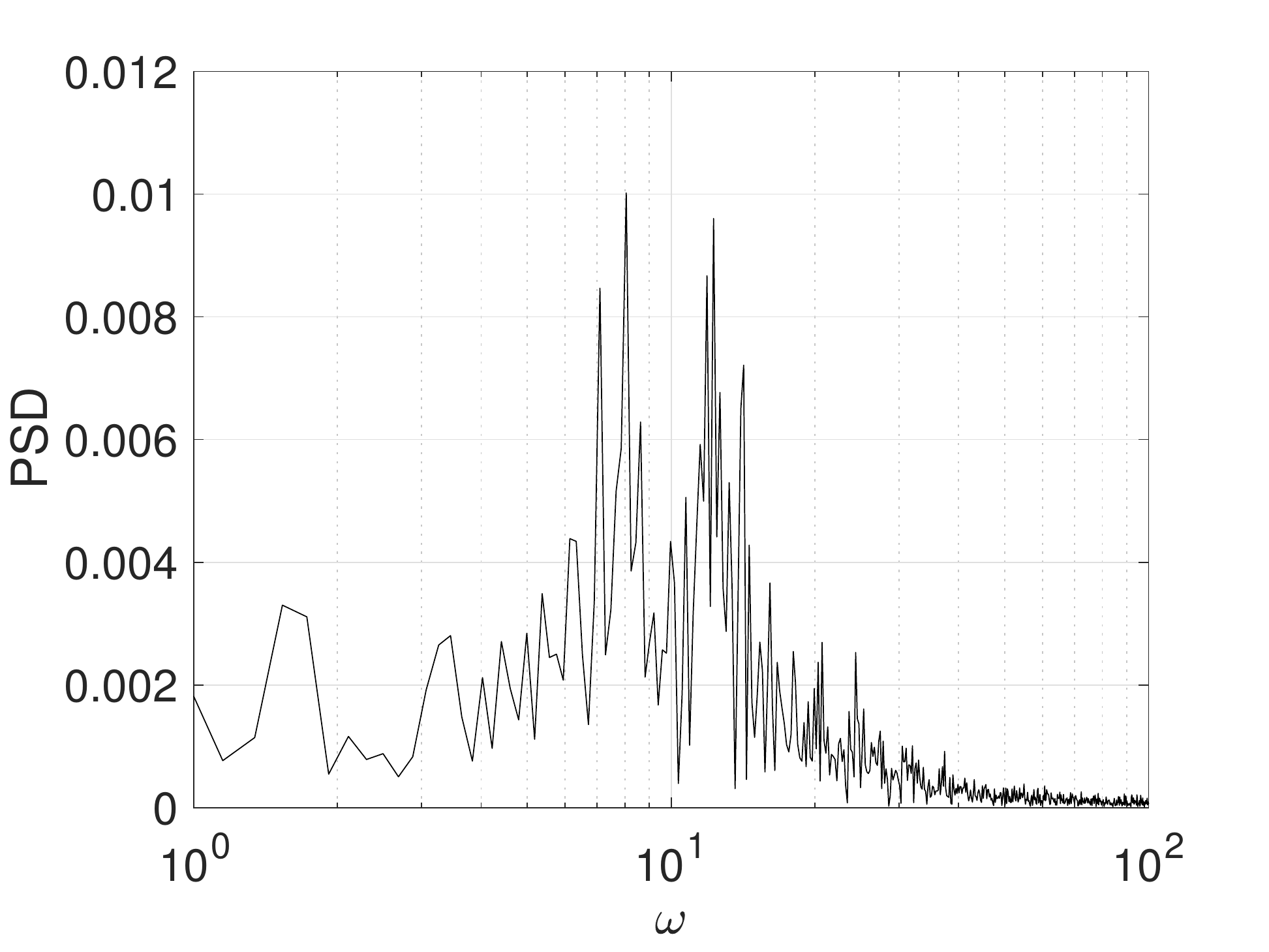}
	\includegraphics[scale=0.3]{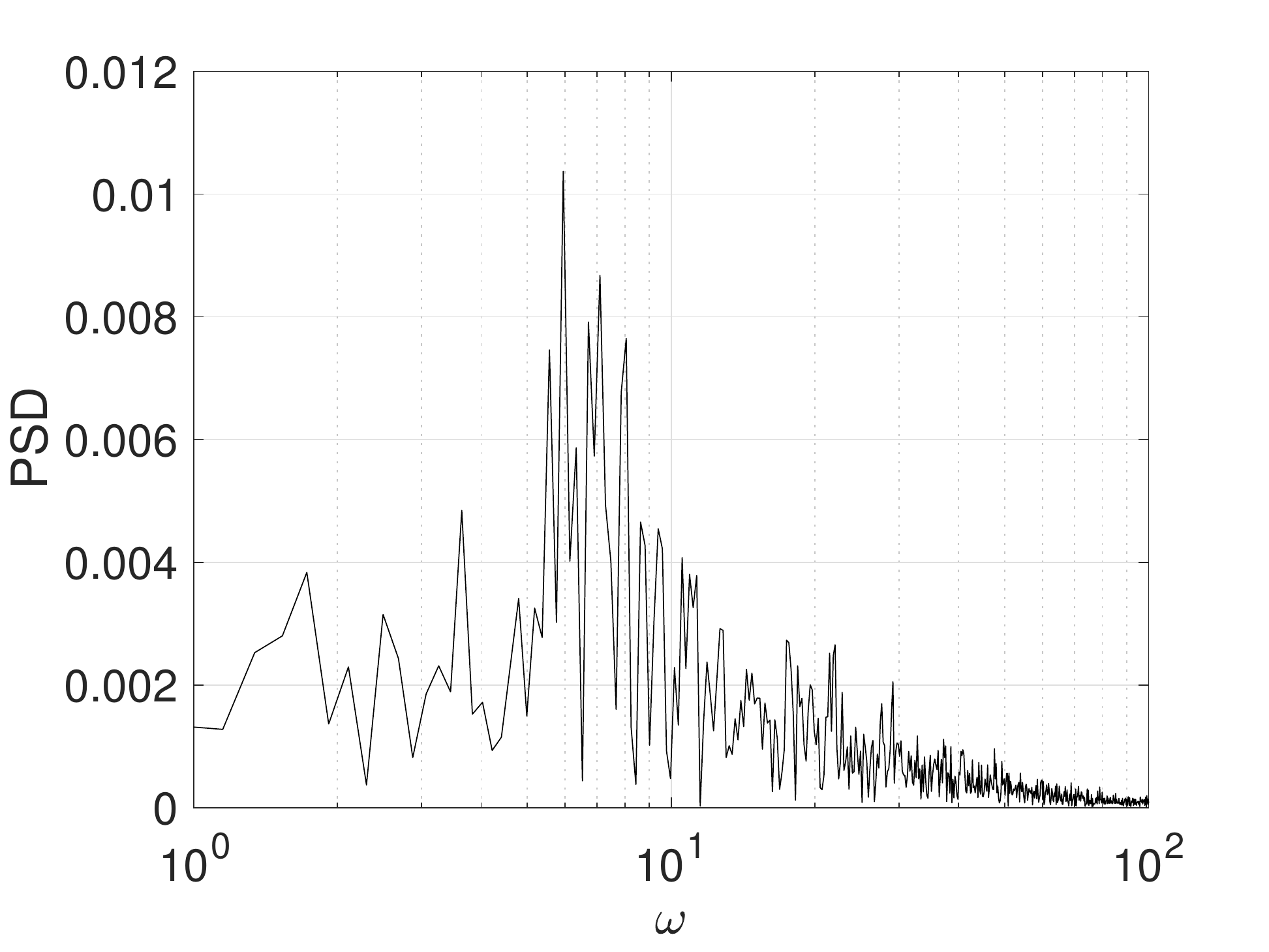}
	
	\caption{(Top) Probe points P4 and P5 in the shear layer and wake, respectively, and their power spectra in (left) and (right) for the A10 case.} \label{fig:probes}
\end{figure}

The challenge associated with the A0 case was to incorporate parasitic modes using nonlinear interactions. For the A10 case, the difficulty stems from being able to capture the two linear mechanisms present in the flow. Two candidate points are proposed in figure \ref{fig:probes}. P4 is located in the shear layer and P5 is in the airfoil wake. 
The power spectrum for P4 contains two broad peaks at frequencies similar to those identified by resolvent analysis and SPOD. This strengthens the argument that the resolvent tends to be low-rank at energetic frequencies as remarked by \cite{Moarref13} for wall-bounded turbulent flows. The power spectrum for P5, on the other hand, contains one broad peak centred around the wake frequencies. It seems reasonable, therefore, to suggest that calibrating the modes using this point alone would not successfully model the fluctuations in the shear layer. 

The performance using a single point or both points is summarised in table \ref{tab:error} using 66 modes in the range $3.14 < \omega < 17.7$. The reconstruction does very badly when only P4 is used as it overpredicts the fluctuations in the wake substantially. P5 is an improvement but it underpredicts the fluctuations in the shear layer. The model performs best when both points are used to calibrate the resolvent modes. 

\begin{table}  
	\centering
	\begin{tabular}{c|c|c}
		\hline
		Point 4& Point 5 & Points 4 $+$ 5 \\ \hline
		$\gg$ 100\%    & 956\%  &  47\% \\ \hline
	\end{tabular}
	\caption{Error for the A10 case as a function of point selection.} \label{tab:error}
\end{table}

\subsubsection{Modelling of the A10 velocity fluctuations}

\begin{figure}
	\centering
	\includegraphics[scale=0.3]{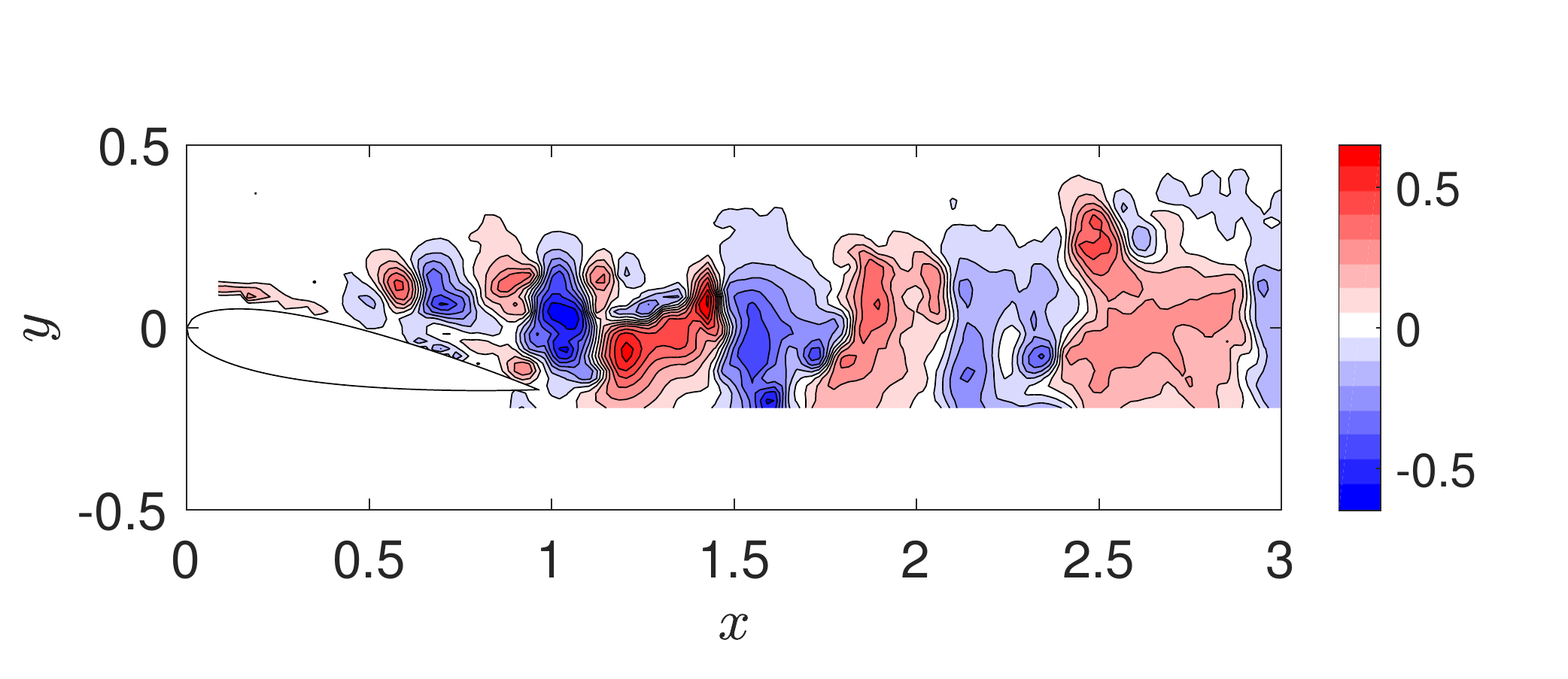}
	
	\includegraphics[scale=0.3]{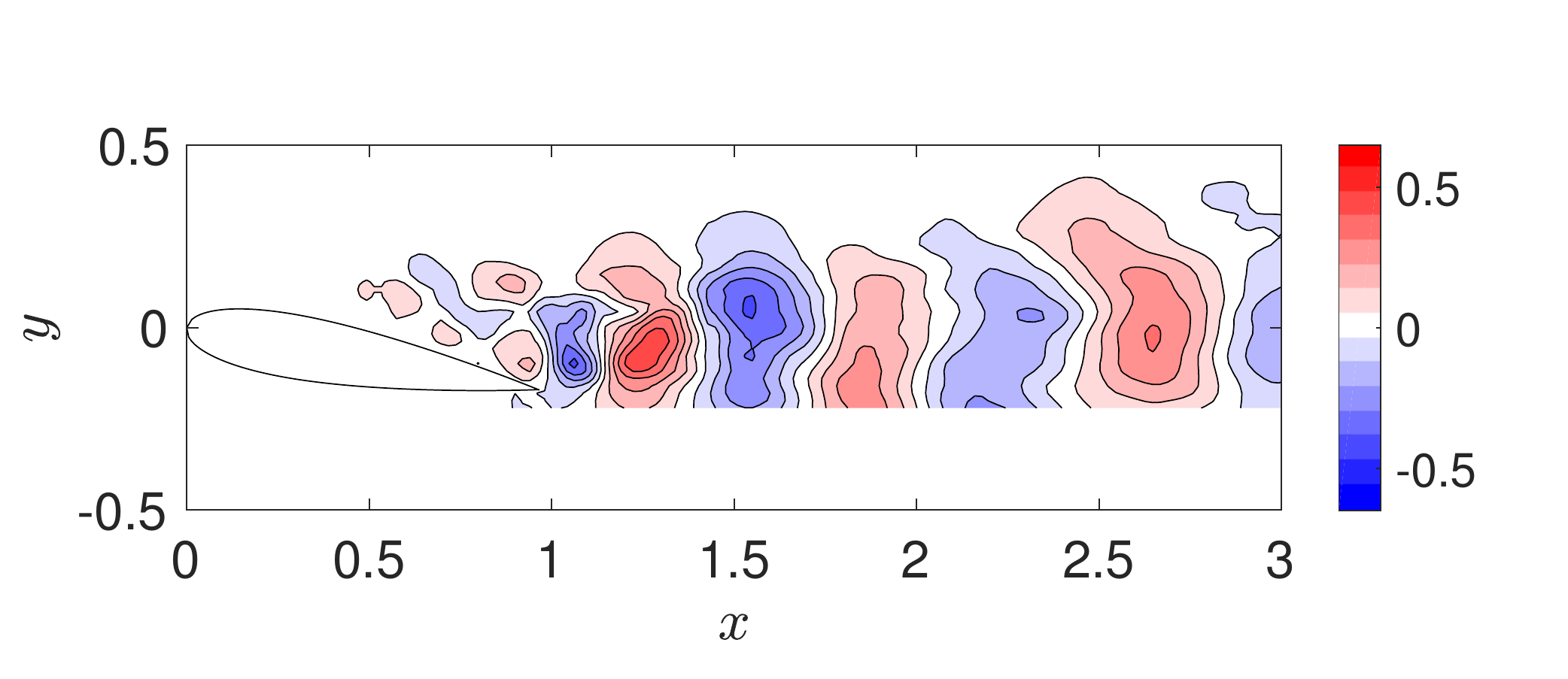}
	\includegraphics[scale=0.3]{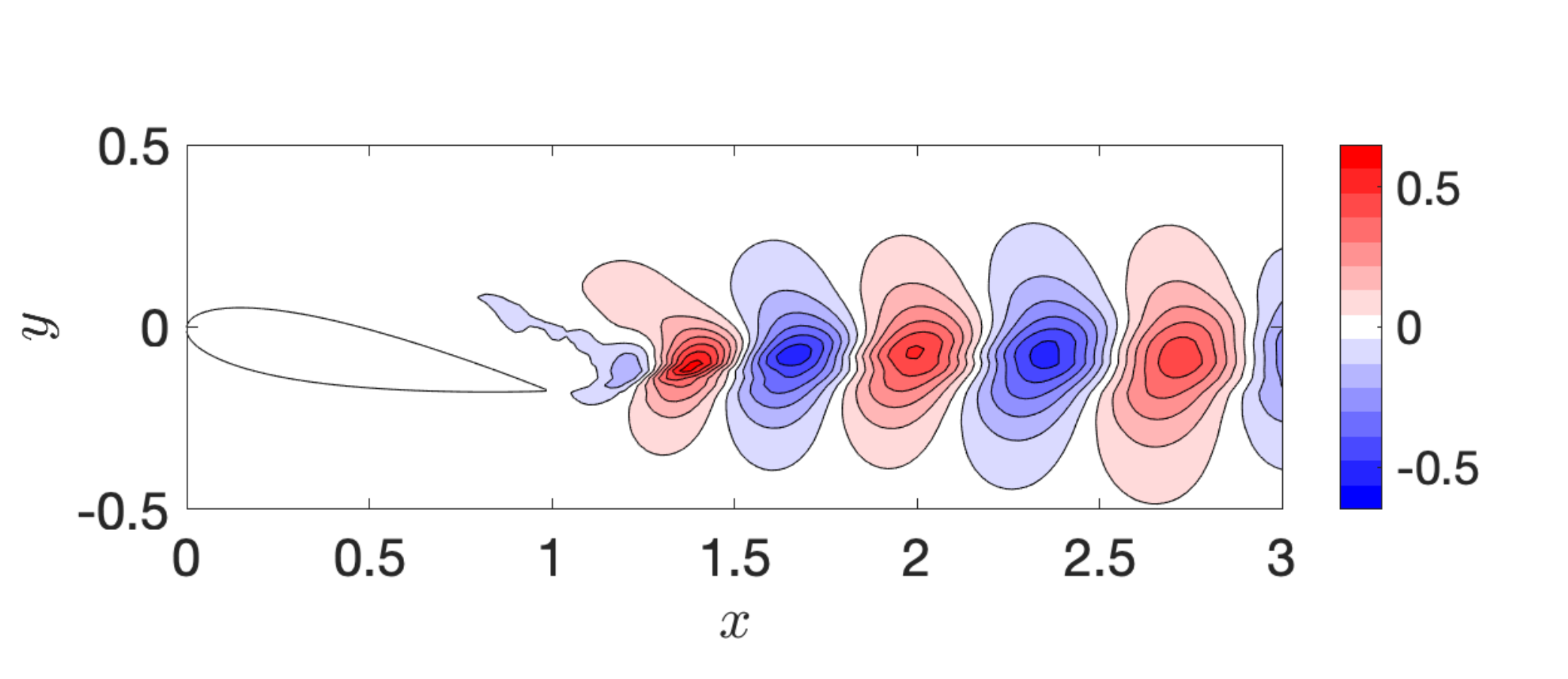}
	
	\includegraphics[scale=0.3]{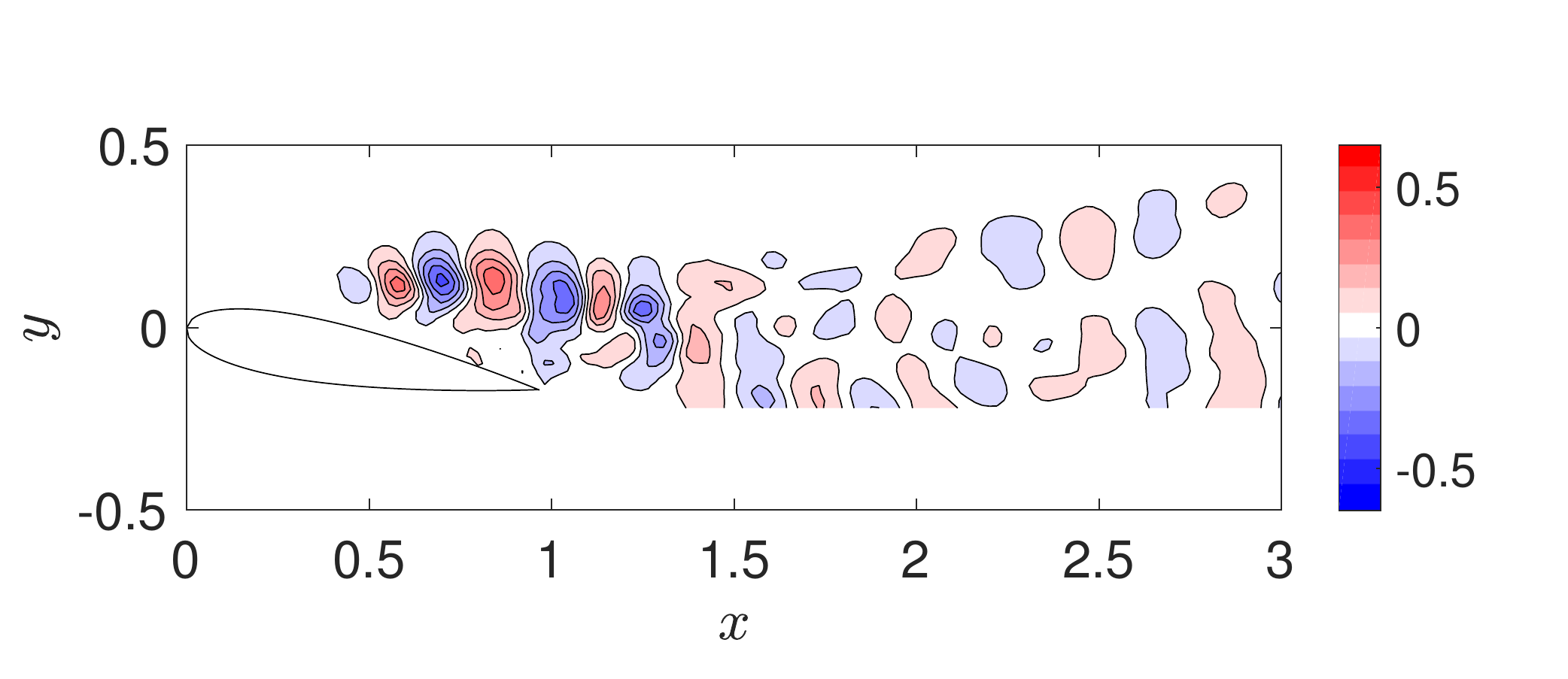}
	\includegraphics[scale=0.3]{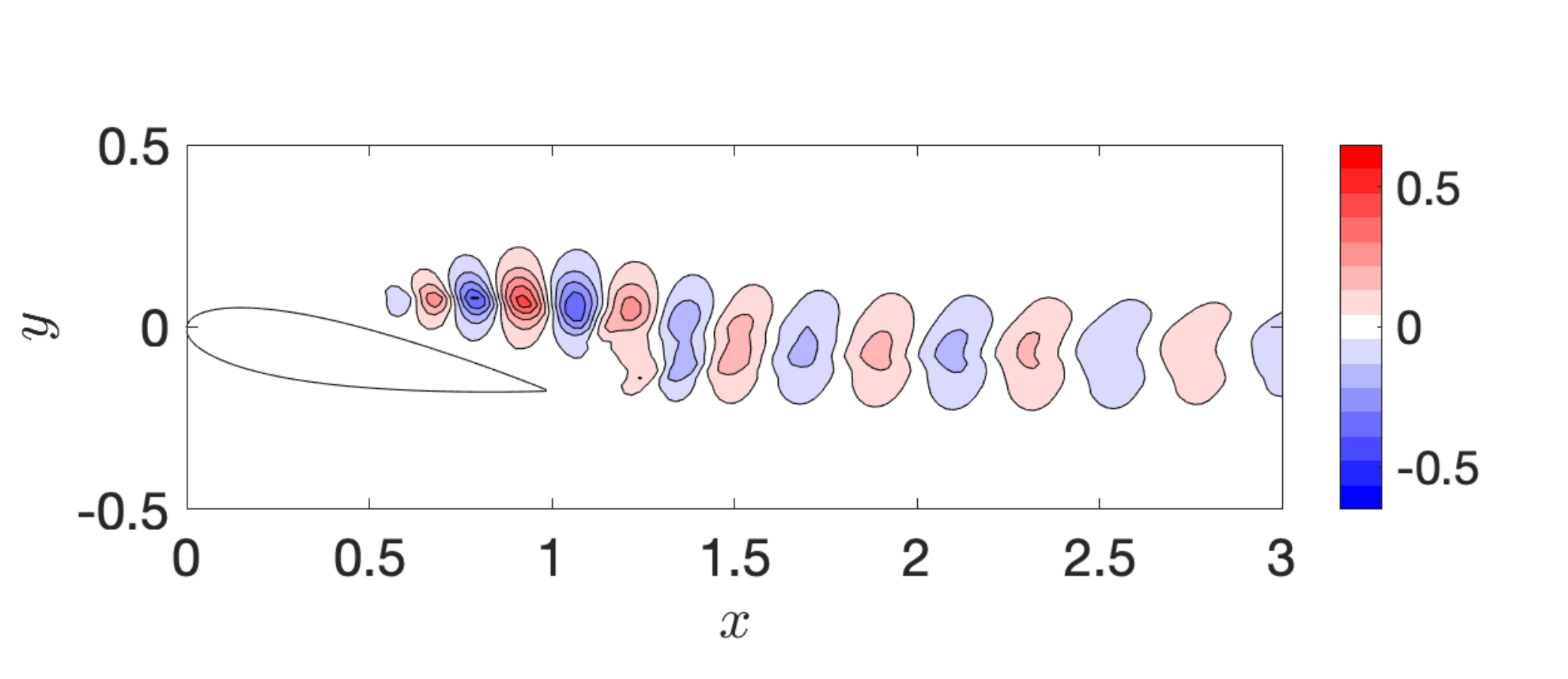}
	
	\includegraphics[scale=0.3]{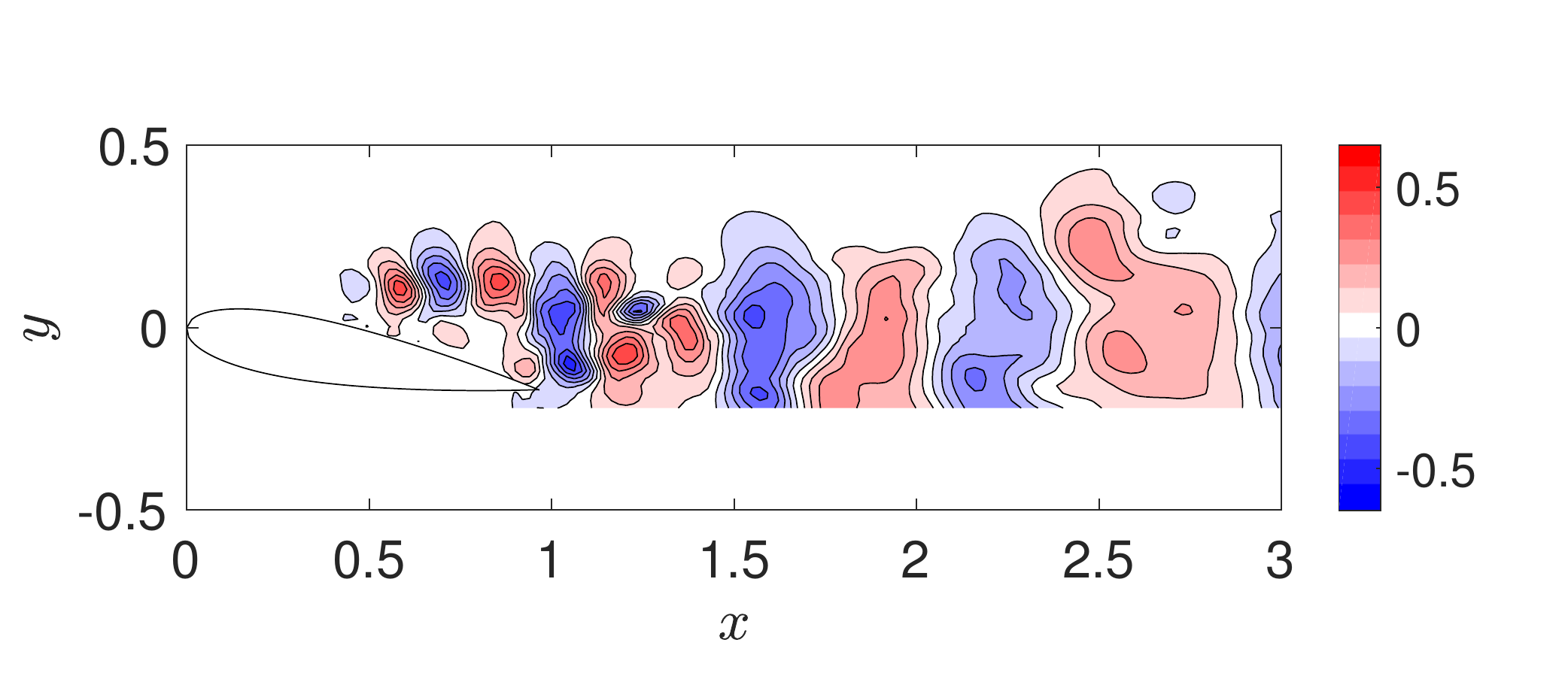}
	\includegraphics[scale=0.3]{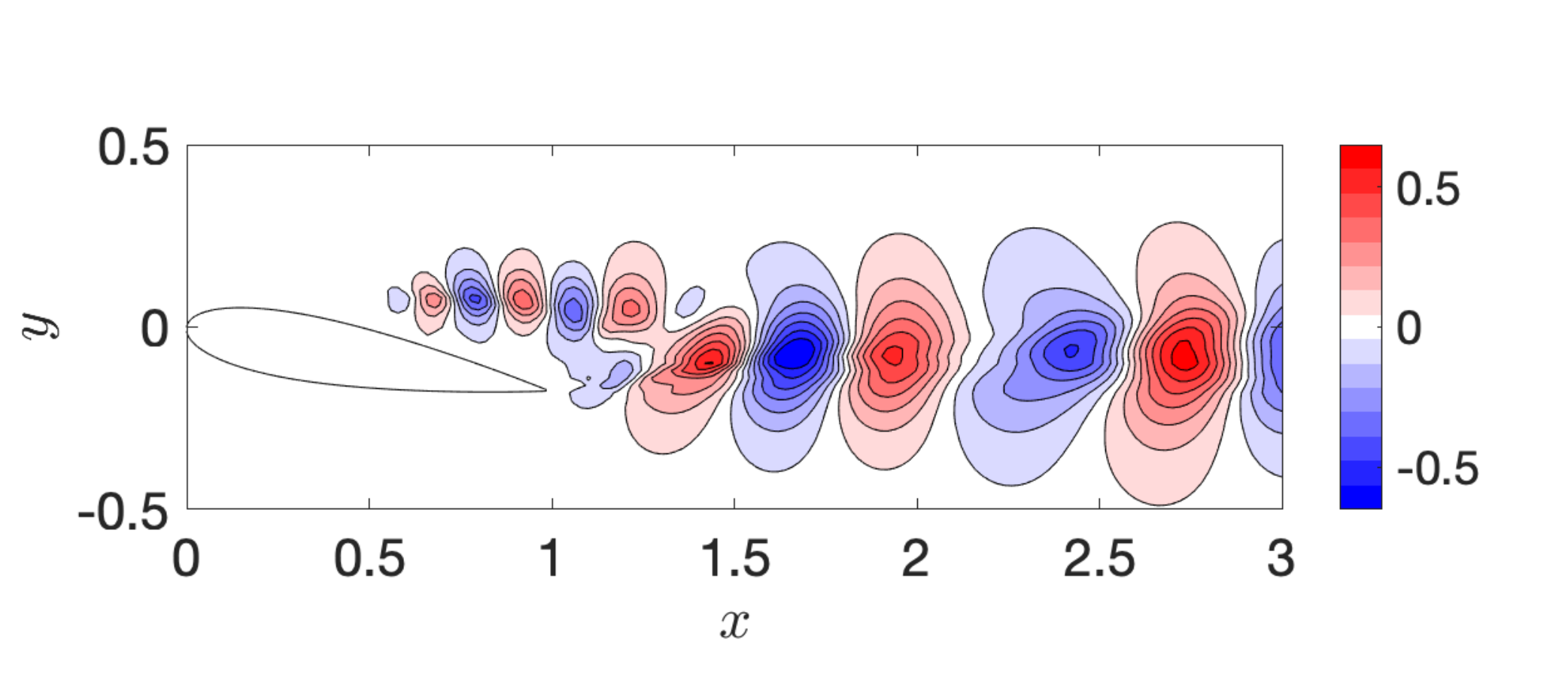}
	
	\caption{A10 results: (Top row) The unfiltered PIV snapshot. (Second row) Filtered PIV containing frequencies $3.14 < \omega < 9.88$ compared to the resolvent model. (Third row) Filtered PIV containing $11.0 < \omega < 17.7$ compared to the resolvent model. (Fourth row) Filtered PIV containing $3.14 < \omega < 17.7$ compared to the resolvent model.}\label{fig:A10ROM}
\end{figure}

For the sake of brevity, only results for the use of both points are summarised in figure \ref{fig:A10ROM} as they help explain why a single point is not sufficient to obtain a good model of the flow. The raw PIV data are visualised in the top row. It is fairly evident that the flow is more complicated for this angle of attack as the presence of both linear mechanisms can be observed in relatively distinct regions of the flow. There is also an area immediately behind the trailing edge where they overlap leading to more complex flow structures. 

Similar to the A0 case, the PIV data are frequency-filtered. We begin by considering 31 frequencies where roughly wake modes only are present, i.e. $3.15 < \omega < 9.65$, as seen in row 2 of figure \ref{fig:A10ROM}. Even though there is some activity in the shear layer, it is small in comparison to the unfiltered case. The resolvent model is plotted alongside it. While the agreement between the wavelengths of the structures is excellent, the model overpredicts the intensity of the fluctuations as far upstream as $x = 1.5$. One potential method for combating this problem is to consider a discounted resolvent, i.e. 
\begin{equation}
	\mathcal{H}(\omega) = \boldsymbol{C}^T(s \boldsymbol{B} - \boldsymbol{L})^{-1}\boldsymbol{C},
\end{equation}
where $s$ is a complex number with nonzero real part. \citet{Yeh18} applied the discounted resolvent to flow around a NACA 0012 airfoil since they encountered unstable modes. Since we do not find unstable eigenmodes near the wake frequencies, we opt to not apply this procedure. 

The PIV data are next filtered to retain 31 frequencies corresponding to the shear layer dynamics, i.e. $11.0 < \omega < 17.7$, and are compared to the resolvent model in row 3 of figure \ref{fig:A10ROM}. The model is very good, both in terms of the wavelengths of the structures predicted and their amplitudes. The signature of these modes at P5 is almost negligible, however, so the model misses them altogether when only P5 is considered. This explains why the use of two points fares significantly better than just one.

Finally, the model is compared to the filtered PIV for $3.14 < \omega < 17.7$ (all 66 frequencies) in the bottom row of figure \ref{fig:A10ROM}. Consistent with earlier results, the agreement is best in the shear layer while the performance of the model degrades further downstream. The model does capture some of the complex behaviour immediately behind the airfoil and the inclination of the fluctuations is correct overall. Despite poor performance further downstream, potential reasons for which are discussed in the next subsection, the resolvent model does quite well at reproducing the coherent structures of the flow. While it is beyond the scope of this study, the very good agreement in the shear layer for the A10 case suggests that the fluctuating forces could be predicted by the resolvent modes. Furthermore, these could be tested by measuring the fluctuating forces on the airfoil with a force sensor. 

\subsubsection{Dynamics and convection velocities of the structures}

Before discussing experimental limitations of the procedure, it is worth analysing the convection velocities of the structures in the resolvent model. To aid the reader, both the filtered PIV fluctuations ($3.14 < \omega < 17.7$) and resolvent model are animated and uploaded as supplementary movies 3 and 4, respectively, to visualise the evolution of the flow in time. The movies contain 1000 snapshots and show that there is a coupling between the shear layer and wake modes in the region of the flow where the recirculation bubble is situated. Despite jitter in the flow's behaviour, the vortices which form in the shear layer meet those which emerge from the trailing edge and, in some cases, they merge. The convection velocity of the fluctuations, consequently, is relatively constant as a function of $x$ although the perturbations begin to slow down further downstream. The reduced-order model from resolvent analysis is able to capture these key features despite poor predictions beyond $x = 2$.

\subsection{Experimental limitations} \label{sec:limitations}

Before we conclude, it is worth discussing how the results could be improved and why this is difficult to achieve due to the limitations of the current experimental data sets. Based on the SPOD results in this study and previous literature, it is likely that A0 and A10 have separate solutions.

The main problem associated with the A0 case is that the resolvent modes do not correctly model the streamwise decay of the fluctuations, a phenomenon also remarked upon by \citet{Rosenberg19} for cylinder flow. In order to combat this problem, \citet{Rosenberg19} approximated the nonlinear forcing for the shedding frequency using the triadic interaction between the shedding mode, predicted by the singular value decomposition, and the parasitic mode at the second harmonic, i.e.
\begin{equation}
	\hat{\boldsymbol{f}}_a(\omega_s) \sim \hat{\boldsymbol{\psi}}_1(-\omega_s) \cdot \nabla \hat{\boldsymbol{u}}_a(2\omega_s) + \hat{\boldsymbol{u}}_a(2\omega_s) \cdot \nabla \hat{\boldsymbol{\psi}}_1(-\omega_s).
\end{equation}
We attempted to do the same for the A0 case, since we have a good prediction of the fluctuation at the shedding frequency from the optimal resolvent mode and the second harmonic by approximating its nonlinear forcing. SPOD of the nonlinear forcing, furthermore, strongly suggests that it is necessary to account for this feedback since there is separation between the first two SPOD eigenvalues and the first mode is very structured. The approximated nonlinear forcing for the airfoil at the shedding frequency, however, did not improve the mode shape. 

A possible reason for why this failed might be related to the enormous amplification at the shedding frequency due to the proximity of a stable eigenvalue to the neutral axis. Similar to cylinder flow, it is necessary to obtain a very accurate estimate of the nonlinear forcing in order to obtain the correct mode since the resolvent operator is so biased in favour of the first response mode. \cite{Lesshafft18} also remarked that spurious numerical noise can be sufficient to trigger the optimal mode and thus overwhelm suboptimal modes. The data-assimilated mean profile is not completely free of noise so the resolvent modes and their triadic interactions are not sufficiently accurate to alter the mode shape predicted by the resolvent at the shedding frequency. It would be worth investigating low Reynolds number DNS data, which are uncontaminated by noise, to understand the exact source of the deviations and how to alleviate these problems when using experimental data.  

The main region of disagreement for the A10 case is in the far wake, where the model overestimates the velocity fluctuations. Since the nonlinear forcing is not structured at this angle of attack, it suggests the root cause to be three-dimensional effects which the model does not incorporate. Three-dimensionality results from the formation of stall cells, the number of which is given by $n_s = AR/2.28$ \citep{Weihs83}, resulting in $n_s \approx 2$ for this setup. We observed earlier that the mean spanwise velocity component and its spanwise gradient are approximately zero. The mean flow, consequently, is nearly two-dimensional, hence the success of the data-assimilation (particularly with respect to the captured Reynolds stress gradients). In order to consider spanwise wavenumbers, however, we would need to obtain the spanwise-averaged mean profile \citep[similar to][]{Beneddine16}, which is inaccessible from planar PIV. By considering spanwise wavenumbers other than $k_z = 0$, which is more suitable for the region of the flow dominated by the Kelvin-Helmholtz instability, it may be possible to obtain resolvent modes which more accurately depict the far wake dynamics. The investigation of these proposed solutions is an area of current study.

\section{Conclusions}\label{sec:conclusion}

The flows around an airfoil at two angles of attack, $\alpha = 0^{\circ}$ and $\alpha = 10^{\circ}$, have been modelled from resolvent analysis using the procedure summarised in figure \ref{fig:flowchart}. The experimental measurements which were necessary included a crude mean profile obtained by PIV and no more than two probe measurements. The experimental data were first analysed using SPOD, which suggested that the angle of attack resulted in significant changes in the flow. SPOD of the velocity fluctuations indicated low-rank behaviour at the shedding frequency and its harmonics for the A0 case. The first eigenvalue of the cross-spectral density tensor was significantly higher than the second eigenvalue, resulting in large but narrow peaks. The A10 case, on the other hand, did not contain well-defined peaks although there was separation between the first two eigenvalues around two ranges of frequencies. The mode shapes within these ranges resembled wake and shear layer modes, respectively. 

We also used SPOD to analyse the nonlinear fluctuations, which were directly computed from the PIV data. In the A0 case, the cross-spectral density of the nonlinear forcing exhibited similar trends to the velocity fluctuations in that a spectral gap existed at the shedding frequency and its harmonics. No such similarity existed for the A10 case as there was no low-rank behaviour at any frequency. The mode shapes observed for the A10 case, consequently, were highly unstructured, suggesting they did not emerge from a small number of modal interactions. As expected, the mode shapes for the A0 case were very structured for the shedding frequency and its harmonics, suggesting that the flow behaves like an oscillator with intrinsic dynamics which are insensitive to background noise. It followed that the A10 case behaved more like an amplifier.

In order to confirm this interpretation, the flows were analysed from a resolvent point of view. Before using the mean profile as an input, they were data-assimilated to decrease noise, fill in missing data, and correct artificial errors which resulted from stitching together multiple mean profiles (in the A10 case). The effect of data-assimilation was far more evident for the A10 case as it managed to identify two linear mechanisms which peaked at frequencies very similar to those identified by SPOD. When the mean profile was interpolated onto the CFD mesh, it was unable to identify the shear layer peak due to missing data near the separation point. Both the data-assimilated and interpolated means only identified a single linear mechanism for the A0 case at the shedding frequency, confirming our earlier interpretation that the A0 case behaved like an oscillator which revolved around the shedding mode while the A10 case behaved like an amplifier. 

By expanding the cross-spectral density tensor into the product of the resolvent operator, its Hermitian transpose, and the cross-spectral density tensor of the nonlinear forcing \citep[similar to][]{Towne18}, we noted that low-rank behaviour may emerge when either the resolvent is low-rank (a linear amplification mechanism) or the cross-spectral density tensor of the nonlinear fluctuations is low-rank (structured nonlinear forcing due to modal interactions). The instances of low-rank behaviour due to either cause were summarised in table \ref{tab:2 cases} for both airfoils. Notably in the A0 case, we educed the structure of the fluctuations at higher harmonics, where the resolvent operator is not low-rank, by considering limited triadic interactions of highly amplified resolvent modes. 

When it came to resolvent-based modelling of the A0 flow, we no longer had to rely exclusively on the resolvent operator being low-rank to predict fluctuations. Indeed when the model was limited to rank-1 modes, we only recovered fluctuations which are symmetric with respect to the centreline since all rank-1 modes resemble the main instability mechanism. By adding parasitic modes, we were able to recover asymmetric fluctuations and capture finer details of the velocity fluctuations. Parasitic modes were not featured in the reconstruction of the A10 case. Instead, we determined that two probe points were necessary to capture both linear mechanisms in the flow and prevent overestimation of the fluctuations in the far wake region. 

The combination of data-assimilation and resolvent analysis have shown their potential to fill in data from incomplete measurements and it is for this reason that they are a topic of ongoing study. The governing equations were exploited to obtain a mean velocity field which obeyed prescribed boundary conditions, removed experimental noise, and increased resolution. The performance of interpolation is clearly inferior as seen in the A10 case. Moreover, resolvent analysis produces a good basis which can be calibrated from limited time-resolved measurements.

It can also be concluded that the combination of SPOD and resolvent analysis can shed light on the nature of the nonlinear forcing in complex flows. In particular, they highlight when the nonlinear forcing is structured and the computation of parasitic modes in this study from experimental data suggest that this process can be leveraged for reduced-order modelling in the future. In cases where the nonlinear forcing is less structured, e.g. the A10 case, it could be possible to devise new strategies for predicting their statistics, e.g. \citet{Towne19}, or modelling their effect on the large-scale structures, e.g. \citet{Illingworth18}. It may even suffice to rely on linear mechanisms only and treat the forcing as stochastic. The results from the paper also confirm the potential of resolvent analysis to predict acoustic fields, e.g. \cite{Schmidt18}, or the forces acting on a solid body, e.g. \citet{Gomez16b}. The A10 model, in particular, is quite promising in the vicinity of the airfoil so it would be possible to verify whether the predicted pressure field is valid by comparing the forces to those measured experimentally using a force sensor. 

\section{Acknowledgments}

The support of the Army Research Office (ARO) through grant number W911NF-17-1-0306 and the Office of Naval Research (ONR) through grant number N00014-17-1-3022 is gratefully acknowledged.

\bibliographystyle{jfm}
\bibliography{DAResSPOD}

\begin{thebibliography}{48}
\expandafter\ifx\csname natexlab\endcsname\relax\def\natexlab#1{#1}\fi
\def\au#1{#1} \def\ed#1{#1} \def\yr#1{#1}\def\at#1{#1}\def\jt#1{\textit{#1}}
  \def\bt#1{#1}\def\bvol#1{\textbf{#1}} \def\vol#1{#1} \def\pg#1{#1}
  \def\publ#1{#1}\def\arxiv#1{#1}\def\org#1{#1}\def\st#1{\textit{#1}}

\bibitem[Amestoy {\em et~al.\/}(2001)Amestoy, Duff, L'{E}xcellent \&
  Koster]{Amestoy01}
{\sc \au{Amestoy, P.~{R}.}, \au{Duff, I.~{S}.}, \au{L'{E}xcellent, J.-{Y}.} \&
  \au{Koster, J.}} \yr{2001}  \at{A fully asynchronous multifrontal solver
  using distributed dynamic shedding}.  \jt{SIAM J. Matrix Anal. Appl.}
  \bvol{23}~(1),  \pg{15--41}.

\bibitem[Beneddine {\em et~al.\/}(2016)Beneddine, Sipp, Arnault, Dandois \&
  Lesshafft]{Beneddine16}
{\sc \au{Beneddine, S.}, \au{Sipp, D.}, \au{Arnault, A.}, \au{Dandois, J.} \&
  \au{Lesshafft, L.}} \yr{2016}  \at{Conditions for validity of mean flow
  stability analysis}.  \jt{J. Fluid Mech.}  \bvol{798},  \pg{485--504}.

\bibitem[Beneddine {\em et~al.\/}(2017)Beneddine, Yegavian, Sipp \&
  Leclaire]{Beneddine17}
{\sc \au{Beneddine, S.}, \au{Yegavian, R.}, \au{Sipp, D.} \& \au{Leclaire, B.}}
  \yr{2017}  \at{Unsteady flow dynamics reconstruction from mean flow and point
  sensors: an experimental study}.  \jt{J. Fluid Mech.}  \bvol{824},
  \pg{174--201}.

\bibitem[Dergham {\em et~al.\/}(2013)Dergham, Sipp \& Robinet]{Dergham13}
{\sc \au{Dergham, G.}, \au{Sipp, D.} \& \au{Robinet, J.-{C}h.}} \yr{2013}
  \at{Stochastic dynamics and model reduction of amplifier flows: the backward
  facing step flow}.  \jt{J. Fluid Mech.}  \bvol{719},  \pg{406--430}.

\bibitem[Dunne(2016)]{Dunne16}
{\sc \au{Dunne, R.}} \yr{2016}  \at{Dynamic stall on vertical axis wind turbine
  blades}. PhD thesis, California Institute of Technology.

\bibitem[Foures {\em et~al.\/}(2014)Foures, Dovetta, Sipp \& Schmid]{Foures14}
{\sc \au{Foures, D. P.~G.}, \au{Dovetta, N.}, \au{Sipp, D.} \& \au{Schmid,
  P.~J.}} \yr{2014}  \at{A data-assimilation method for {R}eynolds-averaged
  {N}avier-{S}tokes-driven mean flow reconstruction}.  \jt{J. Fluid Mech.}
  \bvol{759},  \pg{404--431}.

\bibitem[G{\'o}mez {\em et~al.\/}(2014)G{\'o}mez, Blackburn, Rudman, McKeon,
  Luhar, Moarref \& Sharma]{Gomez14}
{\sc \au{G{\'o}mez, F.}, \au{Blackburn, H.~{M}.}, \au{Rudman, M.}, \au{McKeon,
  B.~J.}, \au{Luhar, M.}, \au{Moarref, R.} \& \au{Sharma, A.~S.}} \yr{2014}
  \at{On the origin of frequency sparsity in direct numerical simulations of
  turbulent pipe flow}.  \jt{Phys. Fluids}  \bvol{26},  \pg{101703}.

\bibitem[G{\'o}mez {\em et~al.\/}(2016{\natexlab{{\em a\/}}})G{\'o}mez,
  Blackburn, Rudman, Sharma \& McKeon]{Gomez16a}
{\sc \au{G{\'o}mez, F.}, \au{Blackburn, H.~M.}, \au{Rudman, M.}, \au{Sharma,
  A.~S.} \& \au{McKeon, B.~J.}} \yr{2016{\natexlab{{\em a\/}}}}  \at{A
  reduced-order model of three-dimensional unsteady flow in a cavity based on
  the resolvent operator}.  \jt{J. Fluid Mech.}  \bvol{798},  \pg{R2}.

\bibitem[G{\'o}mez {\em et~al.\/}(2016{\natexlab{{\em b\/}}})G{\'o}mez, Sharma
  \& Blackburn]{Gomez16b}
{\sc \au{G{\'o}mez, F.}, \au{Sharma, A.~S.} \& \au{Blackburn, H.~M.}}
  \yr{2016{\natexlab{{\em b\/}}}}  \at{Estimation of unsteady aerodynamic
  forces using pointwise velocity data}.  \jt{J. Fluid Mech.}  \bvol{804},
  \pg{R4}.

\bibitem[Hayase(2015)]{Hayase15}
{\sc \au{Hayase, T.}} \yr{2015}  \at{Numerical simulation of real-world flows}.
   \jt{Fluid Dyn. Res.}  \bvol{47}~(051201).

\bibitem[He {\em et~al.\/}(2019)He, Liu, Gan \& Lesshafft]{He19}
{\sc \au{He, C.}, \au{Liu, Y.}, \au{Gan, L.} \& \au{Lesshafft, L.}} \yr{2019}
  \at{Data assimilation and resolvent analysis of turbulent flow behind a
  wall-proximity rib}.  \jt{Phys. Fluids}  \bvol{31},  \pg{025118}.

\bibitem[Hecht(2012)]{Hecht12}
{\sc \au{Hecht, F.}} \yr{2012}  \at{New development in {F}ree{F}em++}.  \jt{J.
  Numer. Math.}  \bvol{20},  \pg{251--265}.

\bibitem[Huerre \& Rossi(1998)]{Huerre98}
{\sc \au{Huerre, P.} \& \au{Rossi, M.}} \yr{1998} {\em Hydrodynamics and
  nonlinear instabilities\/},  \pg{pp. 81--294}.  \publ{Cambridge University
  Press}.

\bibitem[Illingworth {\em et~al.\/}(2018)Illingworth, Monty \&
  Marusic]{Illingworth18}
{\sc \au{Illingworth, S.~{J.}}, \au{Monty, J.~{P.}} \& \au{Marusic, I.}}
  \yr{2018}  \at{Estimating large-scale structures in wall turbulence using
  linear models}.  \jt{J. Fluid Mech.}  \bvol{842},  \pg{146--162}.

\bibitem[Lehoucq \& Sorensen(1996)]{Lehoucq96}
{\sc \au{Lehoucq, R.~{B}.} \& \au{Sorensen, D.~{C}.}} \yr{1996}  \at{Deflation
  techniques for an implicitly restarted {A}rnoldi iteration}.  \jt{SIAM J.
  Matrix Anal. Appl.}  \bvol{17}~(4),  \pg{789--821}.

\bibitem[Lesshafft {\em et~al.\/}(2018)Lesshafft, Semeraro, Jaunet, Cavalieri
  \& Jordan]{Lesshafft18}
{\sc \au{Lesshafft, L.}, \au{Semeraro, O.}, \au{Jaunet, V.}, \au{Cavalieri, A.
  {V.}~{G.}} \& \au{Jordan, P.}} \yr{2018}  \at{Resolvent-based modelling of
  coherent wavepackets in a turbulent jet}.  \jt{arXiv preprint
  arXiv:1810.09340} .

\bibitem[Lumley(1967)]{Lumley67}
{\sc \au{Lumley, J.~L.}} \yr{1967} The structure of inhomogeneous turbulent
  flows.  \bt{In {\em Atmospheric Turbulence and Radio Propagation\/} (ed.
  \ed{A.~M. Yaglom \& V.~I. Tatarski})}.

\bibitem[Lumley(1970)]{Lumley70}
{\sc \au{Lumley, J.~L.}} \yr{1970} {\em Stochastic tools in turbulence\/}.
  \publ{Academic Press}.

\bibitem[McKeon(2017)]{McKeon17}
{\sc \au{McKeon, B.~J.}} \yr{2017}  \at{The engine behind (wall) turbulence:
  perspectives on scale interactions}.  \jt{J. Fluid Mech.}  \bvol{817},
  \pg{P1}.

\bibitem[McKeon \& Sharma(2010)]{McKeon10}
{\sc \au{McKeon, B.~J.} \& \au{Sharma, A.~S.}} \yr{2010}  \at{A critical-layer
  framework for turbulent pipe flow}.  \jt{J. Fluid Mech.}  \bvol{658},
  \pg{336--382}.

\bibitem[McKeon {\em et~al.\/}(2013)McKeon, Sharma \& Jacobi]{McKeon13}
{\sc \au{McKeon, B.~J.}, \au{Sharma, A.~S.} \& \au{Jacobi, I.}} \yr{2013}
  \at{Experimental manipulation of wall turbulence: {A} systems approach}.
  \jt{Phys. Fluids}  \bvol{25},  \pg{031301}.

\bibitem[Moarref {\em et~al.\/}(2013)Moarref, Sharma, Tropp \&
  McKeon]{Moarref13}
{\sc \au{Moarref, R.}, \au{Sharma, A.~{S}.}, \au{Tropp, J.~{A.}} \& \au{McKeon,
  B.~{J}.}} \yr{2013}  \at{Model-based scaling of the streamwise energy density
  in high-{R}eynolds-number turbulent channels}.  \jt{J. Fluid Mech.}
  \bvol{734},  \pg{275--316}.

\bibitem[Raffel {\em et~al.\/}(2018)Raffel, Willert, Scarano, K\"ahler, Wereley
  \& Kompenhans]{Raffel18}
{\sc \au{Raffel, M.}, \au{Willert, {C.}~{E.}}, \au{Scarano, F.}, \au{K\"ahler,
  C.~J.}, \au{Wereley, S.~T.} \& \au{Kompenhans, J.}} \yr{2018} {\em Particle
  Image Velocimetry - A practical guide\/}.  \publ{Springer}.

\bibitem[Rosenberg {\em et~al.\/}(2019)Rosenberg, Symon \& McKeon]{Rosenberg19}
{\sc \au{Rosenberg, K.}, \au{Symon, S.} \& \au{McKeon, B.~J.}} \yr{2019}
  \at{The role of parasitic modes in nonlinear closure via the resolvent
  feedback loop}.  \jt{Physical Review Fluids}  \bvol{in press}.

\bibitem[Rowley {\em et~al.\/}(2009)Rowley, Mezi\'c, Bagheri, Schlatter \&
  Henningson]{Rowley09}
{\sc \au{Rowley, C.~{W.}}, \au{Mezi\'c, I.}, \au{Bagheri, S.}, \au{Schlatter,
  P.} \& \au{Henningson, D.~{S}.}} \yr{2009}  \at{Spectral analysis of
  nonlinear flows}.  \jt{J. Fluid Mech.}  \bvol{641},  \pg{115--127}.

\bibitem[Sasaki {\em et~al.\/}(2017)Sasaki, Piantanida, Cavalieri \&
  Jordan]{Sasaki17}
{\sc \au{Sasaki, K.}, \au{Piantanida, S.}, \au{Cavalieri, A. {V.}~{G.}} \&
  \au{Jordan, P.}} \yr{2017}  \at{Real-time modelling of wavepackets in
  turbulent jets}.  \jt{J. Fluid Mech.}  \bvol{821},  \pg{458--481}.

\bibitem[Schmid(2010)]{Schmid10}
{\sc \au{Schmid, P.~{J.}}} \yr{2010}  \at{Dynamic mode decomposition of
  numerical and experimental data}.  \jt{J. Fluid Mech.}  \bvol{656},
  \pg{5--28}.

\bibitem[Schmidt \& Towne(2019)]{Schmidt18b}
{\sc \au{Schmidt, O.~{T.}} \& \au{Towne, A.}} \yr{2019}  \at{An efficient
  streaming algorithm for spectral proper orthogonal decomposition}.
  \jt{Comput. Phys. Commun.}  \bvol{237},  \pg{98--109}.

\bibitem[Schmidt {\em et~al.\/}(2018)Schmidt, Towne, Rigas, Colonius \&
  Br{\`e}s]{Schmidt18}
{\sc \au{Schmidt, O.~{T.}}, \au{Towne, A.}, \au{Rigas, G.}, \au{Colonius, T.}
  \& \au{Br{\`e}s, G.~{A.}}} \yr{2018}  \at{Spectral analysis of jet
  turbulence}.  \jt{J. Fluid Mech.}  \bvol{855},  \pg{953--982}.

\bibitem[Sharma {\em et~al.\/}(2016)Sharma, Mezi\'c \& McKeon]{Sharma16}
{\sc \au{Sharma, A.~{S}.}, \au{Mezi\'c, I.} \& \au{McKeon, B.~{J}.}} \yr{2016}
  \at{Correspondence between {K}oopman mode decomposition, resolvent mode
  decomposition, and invariant solutions of the {N}avier-{S}tokes equations}.
  \jt{Phys. Rev. Fluids}  \bvol{1},  \pg{032402(R)}.

\bibitem[da~Silva(2019)]{daSilva19}
{\sc \au{da~Silva, A. F.~C.}} \yr{2019}  \at{An {E}n{KF}-based flow estimator
  for aerodynamic flows}. PhD thesis, California Institute of Technology.

\bibitem[Sipp \& Lebedev(2007)]{Sipp07}
{\sc \au{Sipp, D.} \& \au{Lebedev, A.}} \yr{2007}  \at{Global stability of base
  and mean flows: a general approach and its applications to cylinder and open
  cavity flows}.  \jt{J. Fluid Mech.}  \bvol{593},  \pg{333--358}.

\bibitem[Sipp \& Marquet(2013)]{Sipp13}
{\sc \au{Sipp, D.} \& \au{Marquet, O.}} \yr{2013}  \at{Characterization of
  noise amplifiers with global singular modes: the case of the leading-edge
  flat-plate boundary layer}.  \jt{Theor. Comput. Fluid Dyn.}  \bvol{27},
  \pg{617--635}.

\bibitem[Sipp {\em et~al.\/}(2010)Sipp, Marquet, Meliga \& Barbagallo]{Sipp10}
{\sc \au{Sipp, D.}, \au{Marquet, O.}, \au{Meliga, P.} \& \au{Barbagallo, A.}}
  \yr{2010}  \at{Dynamics and control of instabilities in open-flows: a
  linearized approach}.  \jt{Appl. Mech. Rev.}  \bvol{63},  \pg{030801}.

\bibitem[Symon(2018)]{Symon18b}
{\sc \au{Symon, S.}} \yr{2018}  \at{Reconstruction and estimation of flows
  using resolvent analysis and data-assimilation}. PhD thesis, California
  Institute of Technology.

\bibitem[Symon {\em et~al.\/}(2017)Symon, Dovetta, McKeon, Sipp \&
  Schmid]{Symon17}
{\sc \au{Symon, S.}, \au{Dovetta, N.}, \au{McKeon, B.~J.}, \au{Sipp, D.} \&
  \au{Schmid, P.~J.}} \yr{2017}  \at{Data assimilation of mean velocity from
  2{D} {PIV} measurements of flow over an idealized airfoil}.  \jt{Exp. Fluids}
   \bvol{5}~(61).

\bibitem[Symon \& McKeon(2018)]{Symon18c}
{\sc \au{Symon, S.} \& \au{McKeon, B.~J.}} \yr{2018} Experimental flow
  reconstruction using resolvent analysis and data-assimilation.  \bt{In {\em
  21st Australasian Fluid Mechanics Conference, Adelaide, Australia\/}}.

\bibitem[Symon {\em et~al.\/}(2018)Symon, Rosenberg, Dawson \& McKeon]{Symon18}
{\sc \au{Symon, S.}, \au{Rosenberg, K.}, \au{Dawson, S. T.~M.} \& \au{McKeon,
  B.~J.}} \yr{2018}  \at{Non-normality and classification of amplification
  mechanisms in stability and resolvent analysis}.  \jt{Phys. Rev. Fluids}
  \bvol{3},  \pg{053902}.

\bibitem[Symon {\em et~al.\/}(2019)Symon, Sipp, Schmid \& McKeon]{Symon19}
{\sc \au{Symon, S.}, \au{Sipp, D.}, \au{Schmid, P.~J.} \& \au{McKeon, B.~J.}}
  \yr{2019}  \at{Mean and unsteady flow reconstruction using data-assimilation
  and resolvent analysis}.  \jt{AIAA J.}  \bvol{in press}.

\bibitem[Thomareis \& Papadakis(2018)]{Thomareis18}
{\sc \au{Thomareis, N.} \& \au{Papadakis, G.}} \yr{2018}  \at{Resolvent
  analysis of separated and attached flows around an airfoil at transitional
  {R}eynolds number}.  \jt{Phys. Rev. Fluids}  \bvol{3},  \pg{073901}.

\bibitem[Towne(2016)]{Towne16}
{\sc \au{Towne, A.}} \yr{2016}  \at{Advancements in jet turbulence and noise
  modeling: accurate one-way solutions and empirical evaluation of the
  nonlinear forcing of wavepackets}. PhD thesis, California Institute of
  Technology.

\bibitem[Towne {\em et~al.\/}(2015)Towne, Colonius, Jordan, Cavalieri \&
  Br{\`e}s]{Towne15}
{\sc \au{Towne, A.}, \au{Colonius, T.}, \au{Jordan, P.}, \au{Cavalieri, A.
  {V.}~{G.}} \& \au{Br{\`e}s, G.~{A.}}} \yr{2015}  \at{Stochastic and nonlinear
  forcing of wavepackets in a {M}ach 0.9 jet}.  \jt{21st AIAA/CEAS Aeroacustics
  Conf.} .

\bibitem[Towne {\em et~al.\/}(2019)Towne, Lozano-Dur{\'a}n \& Yang]{Towne19}
{\sc \au{Towne, A.}, \au{Lozano-Dur{\'a}n, A.} \& \au{Yang, X.}} \yr{2019}
  \at{Resolvent-based estimation of space-time flow statistics}.  \jt{arXiv
  preprint arXiv:1901:07478} .

\bibitem[Towne {\em et~al.\/}(2018)Towne, Schmidt \& Colonius]{Towne18}
{\sc \au{Towne, A.}, \au{Schmidt, O.~{T.}} \& \au{Colonius, T.}} \yr{2018}
  \at{Spectral proper orthogonal decomposition and its relationship to dynamic
  mode decomposition and resolvent analysis}.  \jt{J. Fluid Mech.}  \bvol{847},
   \pg{821--867}.

\bibitem[Weihs \& Katz(1983)]{Weihs83}
{\sc \au{Weihs, D.} \& \au{Katz, J.}} \yr{1983}  \at{Cellular patterns in
  poststall flow over unswept wings}.  \jt{AIAA J.}  \bvol{21}~(12),
  \pg{1757--1759}.

\bibitem[Welch(1967)]{Welch67}
{\sc \au{Welch, P.}} \yr{1967}  \at{The use of fast {F}ourier transform for the
  estimation of power spectra: a method based on time averaging over short,
  modified periodograms}.  \jt{IEEE Trans. Audio Electroacoust.}  \bvol{2},
  \pg{70--73}.

\bibitem[Westerweel \& Scarano(2005)]{Westerweel05}
{\sc \au{Westerweel, J} \& \au{Scarano, F}} \yr{2005}  \at{Universal outlier
  detection for {P}{I}{V} data}.  \jt{Exp. Fluids}  \bvol{39}~(6),
  \pg{1096--1100}.

\bibitem[Yeh \& Taira(2019)]{Yeh18}
{\sc \au{Yeh, C.~A.} \& \au{Taira, K.}} \yr{2019}  \at{Resolvent-analysis-based
  design of airfoil separation control}.  \jt{J. Fluid Mech.}  \bvol{867},
  \pg{572--610}.

\end{thebibliography}

\end{document}